\documentclass[a4paper,english]{article}
\usepackage[T1]{fontenc}
\usepackage{amsmath}
\usepackage{amssymb}
\usepackage[dvips]{graphicx}
\makeatletter

\providecommand{\LyX}{L\kern-.1667em\lower.25em\hbox{Y}\kern-.125emX\@}
\newcommand{\noun}[1]{\textsc{#1}}

 \newcommand{\lyxaddress}[1]{
   \par {\raggedright #1 
   \vspace{1.4em}
   \noindent\par}
 }

\usepackage{babel}
\makeatother
\begin{document}

\title{Unfashionable observations \\
about $3$ dimensional gravity}

\author{Buffenoir E.%
\footnote{e-mail: buffenoi@lpm.univ-montp2.fr%
}, Noui K.%
\footnote{e-mail:knoui@lpm.univ-montp2.fr%
}}

\maketitle

\lyxaddress{UMR 5825 du C.N.R.S. Laboratoire de Physique Mathématique et théorique,
Université Montpellier 2, 34000 Montpellier, France}

\begin{abstract}
It is commonly accepted that the study of 2+1 dimensional quantum
gravity could teach us something about the 3+1 dimensional case. The
non-perturbative methods developed in this case share, as basic ingredient,
a reformulation of gravity as a gauge field theory. However, these
methods suffer many problems. Firstly, this perspective abandon the
non-degeneracy of the metric and causality as fundamental principles, hoping
to recover them in a certain low-energy limit. Then, it is not clear
how these combinatorial techniques could be used in the case where
matter fields are added, which are however the essential ingredients
in order to produce non trivial observables in a generally covariant
approach. Endly, considering the status of the observer in these approaches, 
it is not clear at all if they really could
produce a completely covariant description of quantum gravity. We propose
to re-analyse carefully these points. This study leads us to a really
covariant description of a set of self-gravitating point masses in
a closed universe. This approach is based on a set of observables
associated to the measurements accessible to a participant-observer,
they manage to capture the whole dynamic in Chern-Simons gravity
as well as in true gravity. The Dirac algebra of these observables
can be explicitely computed, and exhibits interesting algebraic features
related to Poisson-Lie groupoids theory. 
\end{abstract}

\section{Introduction}

Despite of the lack of local degrees of freedom in pure 2+1 dimensional
gravity some basic facts have motivated intensive studies of this
theory during last twenty years: the asymptotic structures associated
to event horizons are sufficiently rich to produce non trivial thermodynamic
properties as the Bekenstein-Hawking law and non-perturbative techniques
have been imported from Topological and Conformal field theories to
try to obtain a complete understanding of this theory. These results
have inspired the development of similar approaches in the 3+1 dimensional
situation. The aim is clearly to obtain a completely covariant description
of quantum gravity and these approaches share, as basic ingredient,
a reformulation of gravity as a gauge field theory. This perspective
abandon the non-degeneracy of the metric and causality as fundamental
principles, hoping to recover them in a certain low-energy limit. Recent
numerical simulations \cite{Ambj} have shed doubt on this point of view. The problem
can be traced back to some misunderstandings about large gauge transformations
in general relativity. And it seems necessary to rethink to the status
of these symmetries. On another part, it is not clear how these combinatorial
techniques could be used in the case where matter fields are added.
Matter degrees of freedom are nevertheless the essential ingredients
in order to produce non trivial observables, solving the frozen dynamic 
problem, in a generally covariant approach, and it is then not clear
at all if these approaches could produce a really covariant description
of quantum gravity. It seems difficult for the present time to consider
the canonical descriptions of 3+1 quantum gravity as falsifiable theories
and to check these points in this situation. We propose to analyse
them in the 2+1 dimensional case. The section 2 is devoted to a re-analysis
of the second order and first order formalism of deSitter gravity
in vacuo, and in particular of the differences between Chern-Simons
gravity and true gravity. We profit of this study to emphasize certain
misunderstandings concerning the status of general invariance and
the choices usually made concerning the status of the observer. In
section 3, we present results concerning dynamics of free particles
on a fixed deSitter spacetime in a new algebraic fashion exhibiting
interesting algebraic features related to dynamical classical Yang-Baxter
equation, we then discuss the coupling of particles to gravity using
Witten's proposal of a minimal coupling and give, for the first time
the complete Dirac treatment of the gravity+matter action. In section
3, we exhibit for Chern-Simons gravity as well as true gravity, certain
classes of non-local observables associated to measurements made by
a localized participant-observer, which permit us to reconstruct the
apparent sky seen by this observer (distances of stars, angles between
light rays coming from them,...) and its dynamics. As strange as it
is, the Dirac bracket of this complete set of observables can be computed,
leading to a beautiful combinatorial description of the phase space
of gravity generalizing a previous study about constants of motion
subalgebra \cite{key-31}. This paper is an attempt to give physical 
foundations to the combinatorial description program and aims to 
develop a description of gravity where the spectator is really internal
to the universe and not just a spectator placed in an asymptotic region 
of spacetime.

\section{Gravity in vacuo}

\subsection{Basic notions on 2+1 deSitter pure gravity}

\subsubsection{\emph{basic notations}}

We will, for the rest of the paper, define our fields on an oriented
smooth 3-manifold $\mathcal{M}$ whose set of local coordinates will
be denoted $I_{\mathcal{M}}=\left\{ x^{0},x^{1},x^{2}\right\} $,
and the Minkowski space $\mathbf{M}_{3}$, whose set of coordinates
will be denoted $I_{\mathbf{M}}=\left\{ 0,1,2\right\} $, will be
equipped with the flat metric $(\eta _{ab})_{a,b\in I_{\mathbf{M}}}$
with $\eta _{ab}=\delta _{a,b}-2\delta _{a,b}\delta _{a,0}$(the light
velocity will be chosen to be $c=1$) . A soldering frame at $x\in \mathcal{M}$
is an \emph{isomorphism} $(e_{\mu }^{a})_{\mu \in I_{\mathcal{M}},a\in I_{\mathbf{M}}}$
from the tangent space $T_{x}\mathcal{M}$ to the Minkowski space
$\mathbf{M}_{3}$, $(\, ^{\star }\! e_{a}^{\mu })_{\mu \in I_{\mathcal{M}},a\in I_{\mathbf{M}}}$
will denote its inverse (we will assume that they are smooth functions
of $x$). 

The metric field $(g_{\mu \nu })_{\mu ,\nu =0,1,2}$ will be build
from the soldering form as\begin{equation}
g_{\mu \nu }=\eta _{ab}e_{\mu }^{a}e_{\nu }^{b}.\label{metric}\end{equation}
 The metric $\eta $ (resp. g) and its inverse will be used to lower
and raise internal coordinates $I_{\mathbf{M}}$ (resp. base spacetime
coordinates $I_{\mathcal{M}}$) indices written with latin (resp.
greek) letters $a,b,c...$(resp. $\alpha ,\beta ,\gamma ...$), the
Einstein convention will be adopted. We will use the following conventions
for $\varepsilon -$symbols: $(\epsilon ^{\alpha \beta \gamma })_{\alpha ,\beta ,\gamma \in I_{\mathcal{M}}}$
and $(\varepsilon _{abc})_{a,b,c\in I_{\mathbf{M}}}$ are totally
antisymmetric and $\epsilon ^{x^{0}x^{1}x^{2}}=1,$ $\varepsilon _{012}=1$,
and we will denote $\mathsf{g}=\frac{1}{6}\epsilon ^{\alpha \beta \gamma }\epsilon ^{\rho \sigma \varpi }g_{\alpha \rho }g_{\beta \sigma }g_{\gamma \varpi }$
and $\mathsf{e}=\frac{1}{6}\varepsilon _{abc}\epsilon ^{\alpha \beta \gamma }e_{\alpha }^{a}e_{\beta }^{b}e_{\gamma }^{c}$.

The affine connection $(\Gamma _{\alpha \beta }^{\gamma })_{\alpha ,\beta ,\gamma \in I_{\mathcal{M}}}$
(with zero-torsion, i.e. $\Gamma _{[\alpha \beta ]}^{\gamma }=0$
) and the spin-connection $(\varpi _{\gamma a}{}^{b})_{a,b\in I_{\mathbf{M}},\gamma \in I_{\mathcal{M}}}$
( $so(2,1)-$valued, i.e. $\varpi _{\gamma }^{(ab)}=0$) are defined
as the unique solutions to the equations $\mathbb{D}_{\alpha }e_{\beta }^{a}:=\partial _{\alpha }e_{\beta }^{a}+\varpi _{\alpha }{}^{a}{}_{b}e_{\beta }^{b}-\Gamma _{\alpha \beta }^{\gamma }e_{\gamma }^{a}=0$
(the spin-connection can also be seen as the unique solution of the
equations $\mathbb{D}_{[\alpha }e_{\beta ]}^{a}=0$). As a remark,
we have $\Gamma _{\alpha \beta }^{\gamma }=\frac{1}{2}g^{\gamma \delta }(\partial _{\alpha }g_{\beta \delta }+\partial _{\beta }g_{\alpha \delta }-\partial _{\delta }g_{\alpha \beta })$.
It will be convenient to introduce variables $(\varpi _{\alpha }{}^{c})_{\alpha \in I_{\mathcal{M}},c\in I_{\mathbf{M}}}$
defined such that $\varpi _{\alpha }{}^{a}{}_{b}=\varepsilon ^{a}{}_{bc}\varpi _{\alpha }{}^{c}.$ 

The Riemann tensor $(R_{\alpha \beta \gamma }{}^{\delta })_{\alpha ,\beta ,\gamma ,\delta \in I_{\mathcal{M}}}$
is defined such that for any vector field $(v_{\alpha })_{\alpha =0,1,2}$
we have $\mathbb{D}_{[\alpha }\mathbb{D}_{\beta ]}v_{\gamma }=R_{\alpha \beta \gamma }{}^{\delta }\, v_{\delta }$.
It can be expressed in terms of the strength of the spin-connection
$(R_{\alpha \beta c}{}^{d})_{c,d\in I_{\mathbf{M}},\alpha ,\beta \in I_{\mathcal{M}}}$,
defined by $R_{\alpha \beta c}{}^{d}\equiv \partial _{[\alpha }\varpi _{\beta ]c}{}^{d}+\varpi _{[\alpha c}{}^{e}\varpi _{\beta ]e}{}^{d}$
, as $R_{\alpha \beta \gamma }{}^{\delta }=R_{\alpha \beta c}{}^{d}e_{\gamma }^{c}e_{d}^{\delta }$.
The Ricci curvature and the scalar curvature are defined as usual
by $R_{\alpha \beta }=R_{\alpha \beta \gamma }{}^{\beta }$ and $R=R_{\alpha }^{\alpha }$.

We will often choose to perform a $2+1$ split of the base manifold
\cite{key-13}. We assume that $M$ can be foliated by a family of
non-overlapping spacelike Cauchy surfaces $(\Sigma _{t})_{t\in [t_{1},t_{2}]}$
indexed by a dumb label $t$ (all diffeomorphic to a fixed $2-$dimensional
compact manifold without boundary $\Sigma $). We will denote by $(n^{\alpha })_{\alpha \in I_{\mathcal{M}}}$
the unit vector normal to spacelike slices. The spatial projection
of the metric, denoted $(h_{\alpha \beta })_{\alpha ,\beta }$, will
be given on each slice, as $h_{\alpha \beta }=g_{\alpha \beta }+n_{\alpha }n_{\beta }$
. To complete the splitting, we need to describe the physical correspondence
between points belonging to neighbour surfaces. To this aim, it suffices
to choose a future-pointing timelike vector field $t^{\alpha }$ verifying
$t^{\alpha }\partial _{\alpha }t=1$ (the evolution will be given
by the Lie derivative associated to this vector field). It can be
decomposed, in terms of the lapse function $N^{t}$ and the shift
vector $(N^{i})_{i}$ tangent to spatial slices, as $t^{\alpha }\equiv N^{t}n^{\alpha }+N^{\alpha }$.
Since $N^{\alpha }n_{\alpha }=n^{\alpha }h_{\alpha \beta }=0,$ it
is then clear that we can uniquely recover the space-time tensor fields
from these datas. We will denote by $K$ the trace of the extrinsic
curvature $K_{\alpha \beta }\equiv h_{\alpha }^{\mu }h_{\beta }^{\nu }\mathbb{D}_{\mu }n_{\nu }$
on the slices $\Sigma _{t}$. 

In the following, we will choose the local coordinates $\left\{ x^{0},x^{1},x^{2}\right\} $
as being the coordinates associated to the spacetime splitting (i.e.
$(t^{\alpha })_{\alpha }=(1,0,0),\: (n_{\alpha })_{\alpha }=(-N,0,0)$
and $(n^{\alpha })_{\alpha }=(\frac{1}{N},\frac{-N^{i}}{N})$). We
define the space base coordinates to be elements of $I_{\Sigma }=\left\{ x^{1},x^{2}\right\} ,$
and will be indexed by latin letters $i,j,k...$ We will introduce
the spatial metric $(q_{ij})_{i,j\in I_{\Sigma }}$ such that $h_{\alpha \beta }=h_{\alpha }^{i}h_{\beta }^{j}q_{ij}$,
here we have $h_{\alpha }^{i}=(1-\delta _{\alpha }^{x^{0}})\delta _{\alpha }^{i}$
which projects onto spatial indexes ($\mathsf{q}$ will denote the
determinant of the spatial metric). $(q^{ij})_{i,j\in I_{\Sigma }}$
will denote coefficients of the inverse metric, and spatial indexes
will be raised or lowered by this spatial metric. In this context,
the expression of the metric is given by \[
g_{\alpha \beta }=\left(\begin{array}{cc}
 -N^{2}+N^{i}N^{j}q_{ij} & q_{il}N^{l}\\
 q_{lj}N^{l} & q_{ij}\end{array}
\right)\]

and the extrinsic curvature is given by $K_{ij}=\frac{1}{2N}(\dot{q}\, _{ij}-\nabla _{j}N_{i}-\nabla _{i}N_{j})$

\subsubsection{\emph{equations of motion and action principles}}

The Einstein's equations for pure gravity with a positive cosmological
constant $\frac{1}{l_{c}^{2}}$ are\begin{equation}
R_{\alpha \beta }-\frac{1}{2}g_{\alpha \beta }R-\frac{1}{l_{c}^{2}}g_{\alpha \beta }=0.\label{einsteineq}\end{equation}

They describe the kinematics and dynamics of the gravitational field.
They are equivalent to the following set of equations

\begin{eqnarray}
T_{\alpha \beta }^{a}(e,\varpi )\equiv \partial _{[\alpha }e_{\beta ]}^{a}+\varepsilon ^{a}{}_{bc}\varpi _{[\alpha }^{b}e_{\beta ]}^{c} & = & 0\label{eqmove1}\\
C_{\alpha \beta }^{a}(e,\varpi )\equiv \partial _{[\alpha }\varpi _{\beta ]}^{a}+\varepsilon ^{a}{}_{bc}\varpi _{\alpha }^{b}\varpi _{\beta }^{c}-\frac{1}{l_{c}^{2}}\varepsilon ^{a}{}_{bc}e_{\alpha }^{b}e_{\beta }^{c} & = & 0\label{eqmove2}
\end{eqnarray}

in the sense that (\ref{eqmove1}) can be used to express uniquely
the spin-connection in terms of the soldering form and its inverse
(the invertibility is then crucial), this expression being reinjected
in (\ref{eqmove2}), we recover Einstein's equation for the metric
field (\ref{metric}). 

The previous classical dynamic can be obtained from the extremization
of the Einstein-Hilbert's action principle. The aim of the present
section is obviously to prepare the study of the physical phase space
of gravity. We will then always put relevant actions into their canonical
forms in order to use Dirac formalism. 

The action principle generally choosen in the metric formalism to
generate Einstein's dynamics is the ``trace K'' action (we will denote $\vartheta =8\pi \mathcal{G},$
and we have to notice that, for $c=1,$ the Newton's constant $\mathcal{G}$
has the unit of an inverse mass, $l_{c}$ is a length, and $l_{P}\equiv \hbar \mathcal{G}$
is also a length ):

\begin{equation}
S_{EH}[g]=\frac{1}{\vartheta }\int \! \! \! \! \! \int \! \! \! \! \! \int _{\mathcal{M}}d^{3}x\, \sqrt{-\mathsf{g}}\left(R[g]-\frac{2}{l_{c}^{2}}\right)+\frac{2}{\vartheta }\int \! \! \! \! \! \int _{\Sigma _{t_{2}}-\Sigma _{t_{1}}}d^{2}x\, \sqrt{\mathsf{q}}K,\label{einsteinhilbertaction}\end{equation}

where $\int \! \! \! \int _{\Sigma _{t_{2}}-\Sigma _{t_{1}}}$ is
a shortcut for $\int \! \! \! \int _{\Sigma _{t_{2}}}\! \! \! \! \! -\int \! \! \! \int _{\Sigma _{t_{1}}}.$
It can be reexpressed, in terms of the datas associated to the splitting,
as\[
S_{EH}[h,N]=\frac{1}{\vartheta }\int dx^{0}\int \! \! \! \! \! \int _{\Sigma _{x^{0}}}d^{2}x\, N^{x^{0}}\sqrt{\mathsf{h}}\left(\mathcal{R}+K_{ij}K^{ij}-K^{2}-\frac{2}{l_{c}^{2}}\right)\]

where $\mathcal{R}$ is the scalar curvature of the unique torsion
free spatial derivative operator compatible with the induced metric
on spatial slices $\Sigma _{t}$ denoted $\nabla $. 

The main properties of this action \cite{key-9} is that its extremization,
under variations of the metric compatible with fixed values of the
induced metric on boundaries $\Sigma _{t_{1}},\Sigma _{t_{2}}$, gives
equations of motion (\ref{einsteineq}). The canonical form of this
action is obtained as follows. The Legendre transform yields the canonical
momenta $(\pi ^{ij})_{i,j\in \left\{ x^{1},x^{2}\right\} }$ of the
spatial metric $(q_{ij})_{i,j\in \left\{ x^{1},x^{2}\right\} }$,
$\pi ^{ij}=-\vartheta \sqrt{\mathsf{q}}\left(K^{ij}-Kq^{ij}\right)$.
The primary constraints are $\rho _{\alpha }\approx 0$, where $\rho _{\alpha }$
are the canonical momenta conjugated to the lapse and the shift variables.
Their conservation under time evolution generates the secondary constraints
\begin{eqnarray*}
H_{t}\equiv  & \frac{\vartheta }{2\sqrt{\mathsf{q}}}\left(\pi ^{ij}\pi _{ij}-\frac{1}{2}(\pi _{i}^{i})^{2}\right)-\frac{\sqrt{\mathsf{q}}}{\vartheta }\left(\mathcal{R}+\frac{2}{l_{c}^{2}}\right) & \approx 0\\
H_{i}\equiv  & -\nabla _{j}\pi _{i}^{j} & \approx 0.
\end{eqnarray*}

All these constraints are first-class. And the canonical hamiltonian
density can be written in terms of them as\[
H_{can}[h,\pi ,N,\rho ]=\frac{1}{\vartheta }\int \! \! \! \! \! \int _{\Sigma _{t}}d^{2}x\, N^{\alpha }H_{\alpha }.\]

The total hamiltonian is obtained by adding a combination of primary
constraints smeared by arbitrary functions:\[
H_{EH}^{tot}[h,\pi ,N,\rho ,\lambda ]=\frac{1}{\vartheta }\int \! \! \! \! \! \int _{\Sigma _{t}}d^{2}x\, N^{\alpha }H_{\alpha }+\lambda ^{\alpha }\rho _{\alpha }.\]

And the first order problem is described by the total action $S_{EH}^{tot}[h,\pi ,N,\rho ,\lambda ]=\int dx^{0}\int \! \! \! \! \! \int _{\Sigma _{x^{0}}}d^{2}x\, \pi ^{ij}\dot{q}\, _{ij}+\rho _{\alpha }\dot{N}\, ^{\alpha }-\int dx^{0}H_{tot}[h,\pi ,N,\rho ,\lambda ].$

It is important to remark that, in $2+1$ dimensional gravity, there
is no local dynamical degrees of freedom ($12$ canonical variables,
$6$ first class constraints and then $6$ gauge fixings).

Let us now discuss the Einstein-Palatini's action principle which
extremization gives the first order description of general relativity. 

Forgetting for a while the boundary terms, we choose as a bulk action
the Hilbert-Palatini action:\begin{eqnarray}
S_{HP}[e,\varpi ] & = & \frac{1}{\vartheta }\int \! \! \! \! \! \int \! \! \! \! \! \int _{\mathcal{M}}\! \! d^{3}x\, \eta _{ab}\epsilon ^{\alpha \beta \gamma }e_{\alpha }^{a}\left(\partial _{[\beta }\varpi _{\gamma ]}^{b}+(\varpi _{\beta }^{c}\varpi _{\gamma }^{d}-\frac{1}{3l_{c}^{2}}e_{\beta }^{c}e_{\gamma }^{d})\varepsilon _{cd}{}^{b}\right).\label{hilbertpalatiniaction}
\end{eqnarray}

We will perform the same $2+1$ splitting as before. We will denote
by $E_{i}^{a}\equiv e_{i}^{a},\: \Omega _{i}^{a}\equiv \varpi _{i}^{a}\: \forall a\in I_{\mathbf{M}},i\in I_{\Sigma }$
the spatial parts of dynamical fields.%
\footnote{Let us notice that we have $q_{ij}=E_{i}^{a}E_{j}^{b}\eta _{ab}$.
Then, if we introduce the quantities $\mathcal{N}_{a}$$\equiv $$\frac{\varepsilon _{abc}\epsilon ^{ij}E_{i}^{b}E_{j}^{c}}{2\sqrt{\mathsf{q}}}$
($\epsilon ^{ij}$ is a shortword for $\epsilon ^{x^{0}ij}$, and
$\mathsf{q}$ is the determinant of the spatial metric) and $\, ^{\star }\! E_{a}^{i}\equiv \eta _{ab}q^{ij}E_{j}^{b},$
which verify the properties $\mathcal{N}_{a}E_{i}^{a}=0,\: \mathcal{N}_{a}\mathcal{N}^{a}=-1,$
$\, ^{\star }\! E_{a}^{i}E_{j}^{a}=\delta _{j}^{i},\: \, ^{\star }\! E_{a}^{i}E_{i}^{b}=\delta _{a}^{b}+\mathcal{N}_{a}\mathcal{N}^{b}$
and $\, ^{\star }\! E_{a}^{i}\mathcal{N}^{a}=0,$ we can obtain a
nice and general parametrization associated to the splitting. Indeed
we have $e_{x^{0}}^{a}=N\mathcal{N}^{a}+N^{i}E_{i}^{a},\: e_{i}^{a}=E_{i}^{a}$,
and $\, ^{\star }\! e_{a}^{x^{0}}=\frac{-\mathcal{N}_{a}}{N},\: \, ^{\star }\! e_{a}^{i}=\, ^{\star }\! E_{a}^{i}+\frac{N^{i}\mathcal{N}_{a}}{N}.$%
} The Hilbert-Palatini action can be recast into the following canonical
form:\begin{eqnarray}
S_{HP}[e,\varpi ] & = & \frac{2}{\vartheta }\int _{x_{1}^{0}}^{x_{2}^{0}}dx^{0}\int \! \! \! \! \! \int _{\Sigma }\! \! d^{2}x\, \eta _{ab}\epsilon ^{ij}\left(-E_{i}^{a}\partial _{x^{0}}\Omega _{j}^{b}+\frac{1}{2}e_{x^{0}}^{a}C_{ij}^{b}(E,\Omega )+\right.\nonumber \\
 &  & \left.+\frac{1}{2}\varpi _{x^{0}}^{a}T_{ij}^{b}(E,\Omega )\right).\label{actioncanonHP}
\end{eqnarray}

The simplicity of the bulk part of the action is a consequence of
the dimensionality of spacetime. Indeed, we have\begin{eqnarray*}
\sqrt{-\mathsf{g}}R & = & \mid \mathsf{e}\mid \eta ^{ce}\, ^{\star }\! e_{e}^{\beta }\, ^{\star }\! e_{f}^{\gamma }R_{\beta \gamma c}{}^{f}\\
 & = & sgn(\mathsf{e})\mathsf{ee}\, ^{\star }\! e_{c}^{\beta }\, ^{\star }\! e_{f}^{\gamma }\varepsilon ^{cf}\, _{b}(\partial _{[\beta }\varpi _{\gamma ]}{}^{b}+\varpi _{\beta }^{c}\varpi _{\gamma }^{d}\varepsilon _{cd}{}^{b})\\
 & = & sgn(\mathsf{e})\eta _{ab}\epsilon ^{\alpha \beta \gamma }e_{\alpha }^{a}\left(\partial _{[\beta }\varpi _{\gamma ]}^{b}+\varpi _{\beta }^{c}\varpi _{\gamma }^{d}\varepsilon _{cd}{}^{b}\right).
\end{eqnarray*}

The factor $sgn(\mathsf{e})$ will be generally forgotten if we choose
the non-degenerate soldering form to have a positive determinant.
An important remark has to be done. The inverse of the soldering form
has disappeared from the formula (\ref{hilbertpalatiniaction}), hence
we could change the theory by forgetting the assumption that the soldering
form has to be invertible. A priori, extensions of gravity incorporating
degenerate metrics have been considered for a long time in order to
allow change of the spatial topology in a lorentzian space-time. This
extension will be called Chern-Simons gravity, and it is a current
belief that this extension is not a big price to pay in order to quantize
gravity \cite{key-28}. The consequences of this choice will be studied
further in the next subsection. 

The extremization of this action under variations of basic fields
is given by

\begin{eqnarray*}
\delta S_{HP}[e,\varpi ] & = & \frac{1}{\vartheta }\int \! \! \! \! \! \int \! \! \! \! \! \int _{\mathcal{M}}d^{3}x\, \eta _{ab}\epsilon ^{\alpha \beta \gamma }\left(\delta e_{\alpha }^{a}C_{\beta \gamma }^{b}[e,\varpi ]+\delta \varpi _{\alpha }^{a}T_{\beta \gamma }^{b}[e,\varpi ]\right)\\
 &  & -\frac{2}{\vartheta }\int \! \! \! \! \! \int _{\Sigma _{t_{2}}-\Sigma _{t_{1}}}d^{2}x\, \eta _{ab}\epsilon ^{ij}e_{i}^{a}\delta \varpi _{j}^{b}.
\end{eqnarray*}

It appears that this action is differentiable only under variations
of basic fields compatible with fixed values of the spatial part of
the spin-connection on boundaries $\Sigma _{t_{1}},\Sigma _{t_{2}}$.
This class of paths appears to be far from what we are considering
in the ADM action principle, the careful study of additional boundary
terms necessary to obtain a satisfactory variational principle and
gauge symmetry properties is postponed to the next subsection. The
canonical description is trivial, due to the specific form of this
first-order action. Indeed, the Dirac Poisson bracket is given by\begin{eqnarray*}
\left\{ \Omega _{i}^{a}(x),E_{j}^{b}(y)\right\}  & = & -\frac{\vartheta }{2}\eta ^{ab}\epsilon _{ij}\delta ^{(2)}(x-y)\\
\left\{ e_{x^{0}}^{a}(x),\chi _{b}(y)\right\}  & = & 
\left\{ \varpi _{x^{0}}^{a}(x),\psi _{b}(y)\right\} =\delta _{b}^{a}\delta ^{(2)}(x-y).
\end{eqnarray*}

The first-class constraints are $\chi _{a}\approx \psi _{a}\approx \epsilon ^{ij}C_{ij}^{b}(e,\varpi )\approx \epsilon ^{ij}T_{ij}^{b}(e,\varpi )\approx 0,$
and the total hamiltonian, defined using arbitrary functions $\eta ^{a},\sigma ^{a}$
smearing primary constraints, is\begin{eqnarray*}
H_{HP}^{tot}[E,\Omega ,e_{x^{0}},\varpi _{x^{0}},\chi ,\psi ,\eta ,\sigma ] & = & \frac{-1}{\vartheta }\int \! \! \! \! \! \int _{\Sigma }d^{2}x\, \eta _{ab}\epsilon ^{ij}\left(e_{x^{0}}^{a}C_{ij}^{b}(E,\Omega )+\varpi _{x^{0}}^{a}T_{ij}^{b}(E,\Omega )\right)\\
 &  & +\frac{-1}{\vartheta }\int \! \! \! \! \! \int _{\Sigma }d^{2}x\, \eta ^{a}\chi _{a}+\sigma ^{a}\psi _{a}.
\end{eqnarray*}

And the first order problem is then given by the total action\begin{eqnarray}
S_{HP}^{tot}[E,\Omega ,e_{x^{0}},\varpi _{x^{0}},\chi ,\psi ,\eta ,\sigma ] & = & \int \! \! dx^{0}\! \! \int \! \! \! \! \! \int _{\Sigma _{x^{0}}}\! \! \! d^{2}x\, \frac{-2}{\vartheta }\eta _{ab}\epsilon ^{ij}E_{i}^{a}\dot{\Omega }\, _{j}^{b}+\chi _{a}\dot{e}\, _{x^{0}}^{a}+\psi _{a}\dot{\varpi }\, _{x^{0}}^{a}\nonumber \\
 &  & -\int dx^{0}H_{tot}[E,\Omega ,e_{x^{0}},\varpi _{x^{0}},\chi ,\psi ,\eta ,\sigma ].\label{einsteinhilbertfirstorder}
\end{eqnarray}

At least to obtain more compact expressions, it will be nice to use
Chern-Simons notations. If we introduce a $so(3,1)-$connection $A_{\alpha }=A_{\alpha }^{I}\xi _{I}$
with $(A_{\alpha }^{I})_{\alpha \in I_{\mathcal{M}},I\in I_{\mathfrak{g}}}$
defined to be $A_{\alpha }^{L,a}=\varpi _{\alpha }^{a},\: A_{\alpha }^{T,a}=\frac{1}{l_{c}}e_{\alpha }^{a}$
(notations and properties relative to the Lie algebra $so(3,1)$ are
recalled in the appendix \ref{sub:Basic-algebraic-results}), the
equations (\ref{eqmove1})(\ref{eqmove2}) can be rewritten as a zero
curvature equation\begin{equation}
F_{\alpha \beta }[A]^{I}=\partial _{[\alpha }A_{\beta ]}^{I}+\varepsilon ^{I}{}_{JK}A_{[\alpha }^{J}A_{\beta ]}^{K}=0.\label{zerocurvature}\end{equation}

The Chern-Simons action is defined by\begin{eqnarray*}
S_{CS}^{bulk}[A] & = & \frac{l_{c}}{\vartheta }\int \! \! \! \! \! \int \! \! \! \! \! \int _{\mathcal{M}}d^{3}x\, \epsilon ^{\alpha \beta \gamma }\left(<A_{\alpha },\partial _{\beta }A_{\gamma }>+\frac{1}{3}<A_{\alpha },[A_{\beta },A_{\gamma }]>\right)\\
 & = & \frac{l_{c}}{\vartheta }\int dx^{0}\int \! \! \! \! \! \int _{\Sigma }d^{2}x\, \epsilon ^{ij}\left(-<A_{i},\partial _{x^{0}}A_{j}>+<A_{x^{0}},F_{ij}[A]>\right)
\end{eqnarray*}

As a result we have $S_{HP}[A]=S_{CS}^{bulk}[A]+S_{\Sigma }[A],$
with\[
S_{\Sigma }[A]\equiv \frac{l_{c}}{\vartheta }\int \! \! \! \! \! \int _{\Sigma _{t_{2}}-\Sigma _{t_{1}}}d^{2}x\, \epsilon ^{ij}<A_{i}^{(L)},A_{j}^{(T)}>.\]

We will also define $\Psi \equiv \chi _{a}\xi ^{a,L}+\psi _{a}\xi ^{a,T}$
and $\Sigma \equiv \eta ^{a}\xi _{a,L}+\sigma ^{a}\xi _{a,T},$ in
order to simplify the expression of the total action :\begin{eqnarray*}
S_{HP}^{tot}[A,\Psi ,\Sigma ] & = & \frac{l_{c}}{\vartheta }\int dx^{0}\int \! \! \! \! \! \int _{\Sigma }d^{2}x\, \left(-2\epsilon ^{ij}<A_{i}^{(T)},\dot{A}\, _{j}^{(L)}>+\vartheta <\Psi ,\dot{A}\, _{x^{0}}>\right.\\
 &  & \left.+<A_{x^{0}},\epsilon ^{ij}F_{ij}[A]>+<\Psi ,\Sigma >\right).
\end{eqnarray*}

\subsection{The symmetries of general relativity}

\subsubsection{\emph{first digression : the status of general covariance}}

Before studying the basic symmetries of these dynamical problems,
let us discuss the status of one of these, that is reparametrization
symmetry. Although some relevant problems connected to general covariance
can find some interesting solutions in the presence of matter degrees
of freedom, we will begin the discussion in the case of pure gravity
(after all, gravity field has a sort of ontological preeminence because
it says to other fields how to move causally). These problems being
intimately entangled with the identification of deterministically
predictable observables we will try to stay as close as possible from
the hamiltonian approach, which is the better way to attack the initial
value problem and to study observability. 

General relativity in vacuo is a \emph{geometric} theory, its basic
object is a \emph{localized universe}, i.e. a smooth manifold $\mathcal{M}$
and a lorentzian metric tensor $g$ defined on it \cite{key-15,key-2}.
The mappings $\phi $ which preserve the structure of $\mathcal{M}$
induce a mapping on the set of localized universes ($\phi $ acts
by pull-back on $g$). The choice of a particular atlas 
and system of coordinates (among equivalence classes of them defining
$\mathcal{M}$ as a smooth manifold) is necessary to write the kinematical
and dynamical problem of gravity in terms of differential equations
and to solve them explicitely, but $g$ is a tensorial field and Einstein's
equations are generally covariant (they are invariant under a smooth
change of system of coordinates on $\mathcal{M}$). However, in absence
of matter degrees of freedom, there is a priori no dynamical field,
except $g$ itself, allowing an individuation of points on the manifold,
and it seems to be a basic requirement to be able to do that in order
to mention the ``where'' and the ``when'' of an event and then to
speak about local observables. The choice of a particular system of
coordinates, called the \emph{coordinatization}, is a way to individuate
points by brute force, by an explicit mapping $p\in \mathcal{M}\mapsto x(p)$.
It is an additional data attached to the smooth manifold $\mathcal{M}$.
It is then possible to describe a \emph{localized universe} as being
given by the components of the metric tensor $g$ in the system of
coordinates $x.$ Now, a distinction has to be made between two types
of transformations\emph{.} The \emph{passive diffeomorphisms}, i.e.
smooth changes of the chosen coordinatization of $\mathcal{M}$, obviously
does not change the value of the metric tensor at a given geometric
point, but changes components of $g$ in this coordinatization. The
\emph{active diffeomorphisms}, draging geometric points onto $\mathcal{M}$
through a diffeomorphism $\phi $, entails an active redistribution
of the metric over the manifold (the metric tensor is transformed
by the pull-back of $\phi $), and we do not have to require any transformation
of the coordinatization under this active transformation. Obviously,
these actions have to be related, as soon as we choose to use coordinatization
to obtain a description of localized universes in terms of components
of the metric. Indeed, let $\phi $ be an active diffeomorphism which
maps $\mathcal{M}$ onto itself (and its induced action on tensorial
fields). Then, it is obvious that the components, in the coordinatization
$x$ and in a given chart, of the actively transformed metric tensor
are the same as the components of the untransformed metric tensor
in the passively transformed coordinatization $\mathcal{S}_{\phi }x$,
defined such that $\left(\mathcal{S}_{\phi }x\right)\left(\phi (p)\right)=x(p).$
Nevertheless, this point of view disregards a deep difference between
both types of transformations. Indeed, an element of the active transformations
group has to be seen as a regular map associating to a metric tensor
field another one, diffeomorphically related to it by a certain element
$\phi $ of the diffeomorphism group of the manifold. Obviously, this
element of the active transformations group would be different, for
the same $\phi $, if we had considered its action on a different
metric tensor field. Hence, an element of this gauge group, viewed
in terms of the components of the metric in a given system of coordinates
is necessarily written as metric dependent coordinates transformations
$x\mapsto \varphi \left(x,g(x)\right)$. It is not innocent to remark
that, besides passive diffeomorphisms which are lagrangian symmetries
of Einstein's action, the set of equations (\ref{einsteineq}) has
its own symmetries (called \emph{dynamical symmetries}) defined on
its set of solutions. In fact, the biggest group of passive dynamical
symmetries of (\ref{einsteineq}) is the group $\mathcal{Q}$ of metric
dependent coordinate transformations $x\mapsto \varphi \left(x,g(x)\right)$
(the groups $Diff_{P}\mathcal{M}$ and $Diff_{A}\mathcal{M}$ of passive
and active diffeomorphisms are two disjoint non-normal subgroups of
$\mathcal{Q}$, and the space of Einstein's universes can be recovered
as the quotient of the space of solutions of (\ref{einsteineq}) by
any one of these three groups \cite{key-12}). 

This confusion between these two conceptually different transformations
is at the origin of a central problem in general relativity. Indeed,
smooth changes of coordinates, acting through Lie derivative on the
components of the metric field, are local Noether symmetries of the
action (\ref{einsteinhilbertaction}). The equations (\ref{einsteineq})
are then covariant under these symmetries, and it is then generally
considered that we have to identify solutions of (\ref{einsteineq})
which components are related by such transformations. Although it
seems natural that this procedure will eliminate the dependence of
the interesting observables with respect to the freely chosen coordinatization,
it will also erase any reference to individuated geometric points.
Doing this, we have to make a critic choice. If we consider diffeomorphisms
as a gauge symmetry, we abandon the hope to find simple local observables.
This catastrophic consequence has led many authors to reject this
point of view, on the grounds that a particular choice of coordinates
just reflects the physical properties of the reference systems with
respect to which measurements are done. If we do not consider diffeomorphism
as a gauge symmetry, we have to abandon the hope to build a deterministic
theory, due to the hole argument \cite{key-3}. Indeed, although we
can fix a set of coordinates in order to individuate points of $\mathcal{M}$,
the active diffeomorphisms continue to be a continuous symmetry of
the equations of motion. Let us consider two metrics which differ by an active
diffeomorphism only in a region $\mathcal{R}$ of $\mathcal{M}$ such
that we are able to choose a complete Cauchy surface $\mathcal{S}$
for the gravitational field, $\mathcal{R}$ being in the past of $\mathcal{S}$.
If the description an observer can make of these universes has to
be deterministic, we have to conclude that these spacetimes can not
be distinguished by any measure or prediction made by this observer.
This schizophrenic situation has its origin in our desire to localize geometric
points using non-dynamical degrees of freedom as reference system
\cite{key-2}. 

As remarked by C. Rovelli, it is not at all natural to consider the
reference system as an external object in general relativity (like
our fixed coordinate system). If we consider, as a reference, a material
body, we can neglect the impact of its presence on the gravitational
field in a certain approximation where it is sufficiently light, however
we can not disregard the impact of gravitational field on it, because
even its inertial motion needs the metric to be defined! In order
to produce local observables invariant under active diffeomorphisms,
it is sufficient to localize geometric points using dynamical degrees
of freedom which transformations under active diffeomorphisms can
be combined with that of the metric tensor in order to build invariant
objects. We could imagine to use gauge invariant quantities made of
dynamical degrees of freedom contained in the metric itself, however
we will reject this point of view for some reasons: this program is
not technically easy and it gives a very non-local description; this
localization is impossible if the universe is too symmetric and will
then never survive to the special relativity limit; this attempt is
hopeless in three spacetime dimensions where only global aspects exist; 
 endly matter degrees of freedom is always the ground on which
real-life observables are defined and measured. The most simple object
we can use to this aim is a free falling observer with an internal
clock. In this description \cite{key-2,key-4}, we then choose to
fix a ``platform'' of ``spectator-observers'' that is locally proper-time
Einstein-synchronizable, and choose a zero of time for their synchronized
clocks, this constitutes a ``system of observers''. They are free-falling
test bodies with internal clocks which are assumed to be able to gather
information from the universe without disturbing it and without disturbing
each other. It does not forbid obviously to add other matter degrees
of freedom interacting in the usual way with the gravitational field.

These observers are used to ``freeze'' the action of a part of the
group of diffeomorphisms \cite{key-4}. Indeed, a first partial gauge
fixing can be made by choosing coordinates compatible with the flow
of world lines of the observers and their internal clocks. Hence,
the residual gauge symmetry amounts to infinitesimal coordinate transformations
preserving the previous conditions. These transformations preserve
also the system of observers if and only if they are equivalent to
a relabelling of spatial coordinates and a rigid time translation.
The choice of the system of observers is then associated to a procedure
where we rigidify a part of the gauge symmetry, and treat it as a
noether symmetry (this procedure is very different from a gauge fixing
and we will call it a ``gauge freezing''). The observables are then
defined to be the invariant objects under these diffeomorphisms preserving
the system of observers (it is important to remark that this notion
does not forbid an evolution of the value of these observables with
respect to the time given by their clocks). This point of view will
be called \emph{``observer-dependent''}.

It seems that we have lost the general covariance of general relativity.
In fact, the question is really to decide which interpretation can
be given to the set of equations (\ref{einsteineq}) written relatively
to a given system of coordinates, concerning things which can be observed.
The problems of determinism and initial values can not be unambiguously
adressed at this level only, and the problem of gauge symmetry has
to be studied more intrinsically. The previous meaningful gauge freezing
disregards only invariance under passive transformations as a basic
fact, but emphasizes a certain part of dynamical symmetries which
has to be considered as a real gauge symmetry. And the question of
predictability can then be adressed only because this gauge freezing
is attached to a given observer's theory. This theory is intimately
related to a certain off-shell extension of the diffeomorphism symmetry
we will study in the next subsection.

The degrees of freedom associated to these observers can be added
to the pure gravity dynamical problem in order to study quantization
\cite{key-2}. The observer-dependent approach is particularly fruitful
if we want to study the special relativity limit of general relativity\cite{key-4}
or to develop the canonical quantization program in the usual way
\cite{key-5}. Moreover, it is nothing more than an attempt to clarify
the status of the observer in Einstein's equation. Nevertheless, they
are different sort of problems with this point of view. Firstly, due
to gravitational redshift it is not always possible to define globally
such a system of comoving-observers. Then, due to the requirements
made about our set of observers, and if our reference system is made
from real matter, this description can only be viewed as an approximation
of general relativity in which we forbid back-reaction of the gravitational
field on matter degrees of freedom. Hence, it seems not to be a very nice
departure point to explore a quantum theory of gravity. Endly, it
is important to analyze the modification of the status of gravitational
degrees of freedom in the gauge freezing procedure, which could have
dramatic consequences for the quantum theory \cite{key-17}. 

In order to solve part of these problems, some authors have chosen
to restrict these spectator-observers to live only in an asymptotic
region of spacetime \cite{key-16}. We will prefer to develop an approach
where our observer is really enclosed in the universe and interacts
with the gravitational field. Then, our ``participant-observer'' will
be described by matter degrees of freedom (point masses), coupled
to the gravitational field, in a closed universe. We want to insist
that choosing the spectator-observer or participant-observer perspective
does not forbid to fix coordinates externally (after all, it is the
better way to write our equations...) as soon as the physical observables
we consider do not refer to them but are intrisically build from variables
defining the observer degrees of freedom and the metric field. In
the same perspective, trying to develop a participant-observer description,
we do not want to abandon the observer-dependent framework which is
nothing more than a clarification of Einstein's framework which sheds
light on its possible off-shell extensions. We will just disregard
observables made of any spectator-observer degrees of freedom. Before
that, we have first to discuss off-shell extensions of the diffeomorphism
symmetry and then to describe the coupling of gravity with matter
degrees of freedom.

\subsubsection{\emph{off-shell symmetries in hamiltonian gravity}}

Only a small part of the symmetries in $\mathcal{Q}$ can be extended
off-shell \cite{key-10,key-19} . 

Indeed, in ADM formulation, the action of diffeomorphisms generated
by Lie derivative along the vector field $\zeta $ on the components
of the metric is given by\[
\delta _{D}[\zeta ]g_{\alpha \beta }=\zeta ^{\gamma }\partial _{\gamma }g_{\alpha \beta }+g_{\alpha \gamma }\partial _{\beta }\zeta ^{\gamma }+g_{\gamma \beta }\partial _{\alpha }\zeta ^{\gamma }.\]

If we restrict $\zeta $ to be a function of coordinates only (passive
diffeomorphisms), it induces a transformation of lapse function and
shift vector which explicitely depends on their velocities $\dot{N}\, ^{\alpha }\equiv \partial _{x^{0}}N^{\alpha }.$
However, the Legendre map $\mathcal{FL}$ (defined from Einstein's
configuration space to ADM phase space) is such that its pull-back
$\mathcal{FL}^{\star }$(defined from the set of functions on Einstein's
configuration space to the set of functions on ADM phase space) is
such that $\frac{\partial }{\partial \left(\dot{N}\, ^{\alpha }\right)}\circ \mathcal{FL}^{\star }=0$.
Hence this type of transformation is not \emph{projectable}, i.e.
cannot be a projection onto the Einstein's configuration space of
a canonical transformation defined on the ADM phase space. Nevertheless,
if we allow $\zeta $ to depend freely on the lapse function, shift
vector and the spatial metric we can identify a subgroup of projectable
gauge symmetries. More precisely, we have to choose $\zeta ^{a}=\delta _{a}^{a}\kappa ^{a}+n^{\alpha }\kappa ^{x^{0}}$,
i.e. $\zeta ^{x^{0}}=\frac{\kappa ^{x^{0}}}{N^{x^{0}}}$ and $\zeta ^{a}=\kappa ^{a}-\frac{N^{a}}{N^{x^{0}}}\kappa ^{x^{0}},$
where the vector $\kappa $ is function of the coordinates and the
spatial metric only. It is important to remark that, in order for
these transformations to form a group, we cannot avoid the explicit
dependence of $\zeta $ in terms of the spatial metric. Notice that
we are not restricting the set of infinitesimal diffeomorphisms acting
on a specific fixed metric because, for $N\neq 0,$ we can choose
$\kappa ^{x^{0}}=N\zeta ^{x^{0}},\kappa ^{a}=\zeta ^{a}+N^{a}\zeta ^{x^{0}}.$
Moreover, this group of projectable transformations denoted $\mathcal{Q}_{can},$
is another non-normal subgroup of the group $\mathcal{Q}$ of metric
dependent coordinate transformations, and the space of Einstein's
universes can as well be recovered as the quotient of the space of
solutions of (\ref{einsteineq}) by $\mathcal{Q}_{can}$ (see \cite{key-12}
for details).The program can be achieved by identifying the generators
of the corresponding canonical transformations on ADM phase space,
in terms of first class constraints, as well as their Dirac Poisson
algebra \cite{key-10,key-12,key-19}. The previous canonical transformations
correspond to transformations leaving the canonical action off-shell
invariant \emph{up to boundary terms}. However, the relevant gauge
fixings are intimately related to the form of these boundary terms.
Every constraints except $H_{t}$ are linear in the canonical momenta,
and the ADM action is fully invariant under the corresponding symmetries,
they can be canonically gauge fixed. The gauge fixing 
associated to hamiltonian constraint cannot be a canonical one, in 
the action principle we have chosen, 
without restrictions of the phase space. However, it can obviously be fixed by
a derivative gauge\cite{key-6}.

Let us now study the case of first-order gravity. The configuration
space is larger than the previous one, due to the redundant parametrization
(\ref{metric}) of the metric in terms of the soldering form. However,
we have to identify solutions of (\ref{eqmove1})(\ref{eqmove2})
related by Noether Lagrangian symmetries formed by local Lorentz transformations.
This infinitesimal Lagrangian gauge symmetry is described by a lorentz
vector $(\lambda ^{a})_{a\in I_{\mathbf{M}}}$\begin{eqnarray}
\delta _{L}[\lambda ]e_{\alpha }^{a} & = & \varepsilon _{bc}{}^{a}\, e_{\alpha }^{b}\lambda ^{c}\label{lorloce}\\
\delta _{L}[\lambda ]\varpi _{\alpha }^{a} & = & \partial _{\alpha }\lambda ^{a}+\varepsilon _{bc}{}^{a}\, \varpi _{\alpha }^{b}\lambda ^{c}.\label{lorlocom}
\end{eqnarray}

In order to consider the action of this symmetry on the action, we
can use Chern-Simons notations. We have $\delta _{L}[\lambda ]A_{\alpha }=\partial _{\alpha }\underline{\lambda }+[A_{\alpha },\underline{\lambda }],$
where $\underline{\lambda }=\lambda ^{a}\xi _{a,L}$. And, defining
the $so(3,1)-$gauge transformations\begin{equation}
\delta _{G}[\Gamma ]A_{\alpha }\equiv D_{\alpha }^{A}\Gamma \equiv \partial _{\alpha }\Gamma +[A_{\alpha },\Gamma ]\label{gaugetransfoinfinitesimal}\end{equation}
 for any mapping $\Gamma :\mathcal{M}\rightarrow \mathfrak{g},$ we
can verify the following off-shell transformation properties

\begin{eqnarray*}
\delta _{G}[\Gamma ]S_{CS}^{bulk}[A] & = & \frac{l_{c}}{\vartheta }\int \! \! \! \! \! \int _{\Sigma _{t_{2}}-\Sigma _{t_{1}}}\! \! \! d^{2}x\, \epsilon ^{ij}\left(<\partial _{i}A_{j}^{(L)},\Gamma ^{(T)}>+<\partial _{i}A_{j}^{(T)},\Gamma ^{(L)}>\right)\\
\delta _{G}[\Gamma ]S_{\Sigma }[A] & = & \frac{l_{c}}{\vartheta }\int \! \! \! \! \! \int _{\Sigma _{t_{2}}-\Sigma _{t_{1}}}\! \! \! d^{2}x\, \epsilon ^{ij}\left(<\partial _{i}A_{j}^{(L)}+[A_{i}^{(L)},A_{j}^{(L)}]-[A_{i}^{(T)},A_{j}^{(T)}],\Gamma ^{(T)}>\right.\\
 &  & \left.-<\partial _{i}A_{j}^{(T)},\Gamma ^{(L)}>\right)
\end{eqnarray*}

It is a trivial consequence that $S_{HP}[e,\varpi ]$ is fully gauge
invariant under $so(2,1)-$gauge symmetries $\delta _{G}[\underline{\lambda }]=\delta _{L}[\lambda ]$
(because $\underline{\lambda }^{(T)}=0$).

The absence of the velocities $\dot{e}\, _{x^{0}}^{a},\dot{\varpi }\, _{x^{0}}^{a}$in
these formulas ensures that this transformation is projectable to
phase space, without changes, in an off-shell symmetry of the total
action. The associated canonical generator and the explicit transformations
of the lagrange multipliers are found in the standard way\cite{key-8,key-20,key-21}.
More precisely, for any infinitesimal generator $\Gamma \in \mathfrak{l}$,
any function $f$ of the set of canonical variables transforms as
$\delta _{G}[\Gamma ]f=\left\{ f,\textrm{G}(\Gamma )\right\} $, with
$\textrm{G}(\Gamma )\equiv <\Gamma ,\epsilon ^{ij}F_{ij}[A]>+<\Psi ,D_{x^{0}}^{A}\Gamma >,$
and the lagrange multipliers have to transform accordingly to $\delta _{G}[\Gamma ]\Sigma =[\Sigma ,\Gamma ]-D_{x^{0}}^{A}\dot{\Gamma }$
. The total action is then fully off-shell invariant under this symmetry
(the previous formulas can be used as defining the off-shell extension
of the complete $so(3,1)-$gauge symmetry although it does not let
the action fully invariant in itself).

Let us now consider the diffeomorphism symmetries. The infinitesimal
passive transformations associated to components of a vector field
$(\zeta ^{\alpha })_{\alpha \in I_{\mathcal{M}}}$ are given by the
Lie derivative\begin{eqnarray}
\delta _{D}[\zeta ]e_{\alpha }^{a} & = & e_{\beta }^{a}\partial _{\alpha }\zeta ^{\beta }+\zeta ^{\beta }\partial _{\beta }e_{\alpha }^{a}\label{diffeoe}\\
\delta _{D}[\zeta ]\varpi _{\alpha }^{a} & = & \varpi _{\beta }^{a}\partial _{\alpha }\zeta ^{\beta }+\zeta ^{\beta }\partial _{\beta }\varpi _{\alpha }^{a}.\label{diffeoom}
\end{eqnarray}

For the same reasons as in the ADM description, such a transformation
is not projectable, because of the presence of the velocities $\dot{e}\, _{x^{0}}^{a},\dot{\varpi }\, _{x^{0}}^{a}$
in the formulas describing reparametrization along time direction.
We have then to consider metric dependent coordinates transformations.
For example, if $\, ^{\star }\! e_{0}^{x^{0}}\neq 0,$ we can, once
again, choose $\zeta ^{x^{0}}\equiv \, ^{\star }\! e_{0}^{x^{0}}\kappa ^{x^{0}}$and
$\zeta ^{i}\equiv \kappa ^{i}+\, ^{\star }\! e_{0}^{i}\kappa ^{x^{0}},$
and we can check that we obtain a projectable symmetry. However, another
choice is usually made, using extensively the invertibility of the
soldering form. To any field of lorentz vectors $(\tau ^{a})_{a\in I_{\mathbf{M}}},$
eventually depending of spatial dynamical variables (this dependence is probably unavoidable if
we want these symmetries to form a group, but this point will not be studied here), we will associate
the vector field $\zeta _{(\tau )}^{\alpha }\equiv \, ^{\star }\! e_{a}^{\alpha }\tau ^{a}$
and the field of lorentz vectors $\lambda _{(\tau )}^{a}\equiv -\varpi _{\alpha }^{a}\zeta _{(\tau )}^{\alpha }.$
A basic computation gives the following projectable diffeomorphisms
\begin{eqnarray}
\left(\delta _{D}[\zeta _{(\tau )}]+\delta _{L}[\lambda _{(\tau )}]\right)
\varpi _{\alpha }^{a} & = & \tau ^{d}\, ^{\star }\! e_{d}^{\beta }(\partial _{\beta }
\varpi _{\alpha }^{a}-\partial _{\alpha }\varpi _{\beta }^{a}+\varepsilon _{bc}\, ^{a}
\varpi _{\beta }^{b}\varpi _{\alpha }^{c})\label{projdiffeoexp}\\
 & = & \frac{1}{l_{c}^{2}}\tau ^{b}e_{\alpha }^{c}
\varepsilon _{bc}\, ^{a}+\tau ^{d}\, ^{\star }\! e_{d}^{\beta }C_{\beta \alpha }^{a}(e,\varpi )\\
\left(\delta _{D}[\zeta _{(\tau )}]+\delta _{L}[\lambda _{(\tau )}]\right)e_{\alpha }^{a} & = & 
\partial _{\alpha }\tau ^{a}+\tau ^{d}\, ^{\star }\! 
e_{d}^{\beta }(\partial _{\beta }e_{\alpha }^{a}-\partial _{\alpha }
e_{\beta }^{a}+\varepsilon _{bc}\, ^{a}\varpi _{\beta }^{b}e_{\alpha }^{c})\nonumber \\
 & = & \partial _{\alpha }\tau ^{a}+\tau ^{b}\varpi _{\alpha }^{c}\varepsilon _{bc}\, ^{a}+\tau ^{d}\, ^{\star }\! 
e_{d}^{\beta }T_{\beta \alpha }^{a}(e,\varpi ),
\end{eqnarray}
 which do not contain the velocities $\dot{e}\, _{x^{0}}^{a},\dot{\varpi }\, _{x^{0}}^{a}$
and are then projectable. 

We will denote the previous lagrangian transformations \begin{equation}
\delta _{L,PD}[\lambda ,\tau ]\equiv \delta _{D}[\zeta _{(\tau )}]+\delta _{L}[\lambda _{(\tau )}+\lambda ],\label{projdiffeodef}\end{equation}
 and we have to notice that, reciprocally, we have $\delta _{D}[\zeta ]+\delta _{L}[\lambda ]=\delta _{L,PD}[\lambda +\varpi _{\alpha }\zeta ^{\alpha },e_{\alpha }\zeta ^{\alpha }]$.
We will denote $\mathbb{g}_{L,PD}$ the algebra of these gauge symmetries,
and $\mathbb{G}_{L,PD}$ the group of associated large transformations.
As soon as this mapping is invertible, i.e. the soldering form is
non-degenerate, $\mathbb{G}_{L,PD}$ is equal to the group of diffeomorphisms
$+$ local lorentz transformations.

The previous projectable symmetries can be related to another projectable
symmetry, i.e. $so(3,1)$ gauge transformations $\delta _{G}$, at
least infinitesimally and on-shell. Indeed, defining $\underline{\tau }=\tau ^{a}\xi _{a,T}$
and $\Gamma _{(\lambda ,\tau )}=\underline{\lambda }+\underline{\tau }$
we can check that \begin{equation}
\delta _{G}[\Gamma _{(\lambda ,\tau )}]A_{\alpha }^{I}=\delta _{L,PD}[\lambda ,\tau ]A_{\alpha }^{I}+\delta _{TS}[\zeta _{(\tau )}]A_{\alpha }^{I}.\label{diffgauge}\end{equation}

with the trivial gauge symmetry part given by $\delta _{TS}[\zeta ]A_{\alpha }^{I}\equiv \zeta ^{\beta }F_{\alpha \beta }^{I}(A)$.
It is trivial in the sense that it is null if the zero curvature equations,
i.e. the constraints and equations of motion, are verified. It then
seems appealing to replace $\delta _{L,PD}$ by the more simple transformations
$\delta _{G}$ in the study of the reduced phase space of gravity.
Generally, this remark is taken as the basic fact allowing for a reinterpretation
of gravity as a Chern-Simons theory. However, we have to take it with
caution because, even when infinitesimal symmetries coincide on-shell,
they can differ dramatically in two different aspects: concerning
their action on boundary conditions defining the variational problem
and at the level of their group structure. 

Let us, firstly, study the problem of boundary conditions. Plugging
$\delta A_{\alpha }^{I}\equiv \delta _{TS}[\zeta ]A_{\alpha }^{I}$
into the following off shell-formulas\begin{eqnarray*}
\delta S_{CS}^{bulk}[A] & = & \frac{l_{c}}{\vartheta }\int \! \! \! \! \! \int \! \! \! \! \! \int _{\mathcal{M}}d^{3}x\, \epsilon ^{\alpha \beta \gamma }<\delta A_{\alpha },F_{\beta \gamma }(A)>+\frac{l_{c}}{\vartheta }\int \! \! \! \! \! \int _{\Sigma _{t_{2}}-\Sigma _{t_{1}}}d^{2}x\, \epsilon ^{ij}<\delta A_{i},A_{j}>\\
\delta S_{\Sigma }[A] & = & \frac{l_{c}}{\vartheta }\int \! \! \! \! \! \int _{\Sigma _{t_{2}}-\Sigma _{t_{1}}}d^{2}x\, \epsilon ^{ij}\left(<\delta A_{i}^{(L)},A_{j}^{(T)}>+<A_{i}^{(L)},\delta A_{j}^{(T)}>\right)
\end{eqnarray*}
 gives, as soon as the vector field $\zeta $ is tangent to the boundary,
the following boundary contributions\begin{eqnarray*}
\delta _{TS}[\zeta ]S_{CS}^{bulk}[A] & = & \frac{l_{c}}{\vartheta }\int \! \! \! \! \! \int _{\Sigma _{t_{2}}-\Sigma _{t_{1}}}\! \! \! d^{2}x\, \epsilon ^{ij}<\zeta ^{\alpha }A_{\alpha },\partial _{i}A_{j}>\\
\delta _{TS}[\zeta ]S_{\Sigma }[A] & = & \frac{l_{c}}{\vartheta }\int \! \! \! \! \! \int _{\Sigma _{t_{2}}-\Sigma _{t_{1}}}\! \! \! d^{2}x\, \epsilon ^{ij}\left(<\partial _{i}A_{j}^{(L)}+[A_{i}^{(L)},A_{j}^{(L)}]-[A_{i}^{(T)},A_{j}^{(T)}],\zeta ^{\alpha }A_{\alpha }^{(T)}>\right.\\
 &  & \left.-<\partial _{i}A_{j}^{(T)},\zeta ^{\alpha }A_{\alpha }^{(L)}>\right).
\end{eqnarray*}

As a result, we then conclude that $S_{HP}[A]$ is fully off-shell
invariant under transformations $\delta _{L,PD}[\lambda ,\tau ]$
as soon as $\, ^{\star }\! e_{a}^{x^{0}}\tau ^{a}\mid _{\Sigma _{t_{1,2}}}=0,$
this restriction being due to the presence of a gauge symmetry
associated to a first class constraint non-linear in the canonical
momenta. As a result of previous discussions, the variational principle
of $S_{HP}[A]$ describes an action associated to the class of paths
$(\Omega _{i},E_{i})_{i\in I_{\Sigma }}$ (we do not mention \emph{$\varpi _{x^{0}},e_{x^{0}}$}
which are not subject to dynamical equations) \emph{up to symmetries}
$\delta _{L,PD}[\lambda ,\tau ],\: s.t.\: \mathcal{N}_{a}(E)\tau ^{a}\mid _{\Sigma _{t_{1,2}}}=0$,
the variations of \emph{$(\Omega _{i},E_{i})_{i\in I_{\Sigma }}$}
being \emph{restricted at the boundaries} by $\delta \Omega _{i}\mid _{\Sigma _{t_{1,2}}}=0,\: \forall i\in I_{\Sigma }$
( \emph{$(E_{i})_{i\in I_{\Sigma }}$} are freely varied and in particular
no constraint is required to be fullfilled at the boundaries \emph{}).
This physical problem is different from that described by ADM action.
It would be more natural to have a variational principle associated
to a class of paths $(\Omega _{i},E_{i})_{i\in I_{\Sigma }}$ \emph{up
to symmetries} $\delta _{L,PD}[\lambda ,\tau ],\: s.t.\: \mathcal{N}_{a}(E)\tau ^{a}\mid _{\Sigma _{t_{1,2}}}=0$,
the variations of \emph{$(\Omega _{i},E_{i})_{i\in I_{\Sigma }}$}
being taken eventually \emph{on the constraint surface} $\varepsilon ^{ij}T_{ij}^{a}(E,\Omega )\mid _{\Sigma _{t_{1,2}}}=0,\: \forall a\in I_{\mathbf{M}}$
\emph{}and \emph{restricted at the boundaries} by $\delta \left(q_{ij}(E)\right)\mid _{\Sigma _{t_{1,2}}}=0,\: \forall i,j\in I_{\Sigma }$
. However, boundary terms corresponding to such a variational principle
seem to be unknown in the litterature. 

Then, it seems difficult to implement the symmetries $\delta _{G}$
to replace $\delta _{L,PD}$ in the study of the reduced phase space,
because $S_{HP}[A]$ is not invariant, not even on-shell, by general
transformations of the type $\delta _{G}[\Gamma _{(\lambda ,\tau )}],$
due to the boundary terms. Obviously, boundary terms can always be
added to an action in order to change its invariance properties \cite{key-20,key-22}.
This procedure has been developped especially to deal with the problems
associated to gauge symmetries coming from first class constraints
non-linear in the momenta (our case is of this type). We have to take
care that this procedure induces at the same time a change in the
class of paths defining the variational problem, or necessitates the
addition of new degrees of freedom. In fact, these changes, although
apparently irrelevant, generally affect the physics described by the
resulting action principle especially in the type of observables we
will consider. This procedure deals with the problems concerning gauge
symmetries, but disregards invariance properties under other rigid
symmetries of the action principle and also external physical informations
about observables concerning measurement process. It appears sometimes
that some restrictions of the symmetries at boundaries have to be
considered as physically significant facts for the observer theory.
It is clear, for example, that for the free relativistic particle,
as soon as we are interested in measuring positions, we have to take
care of boundary restrictions on time reparametrization symmetry,
and the proper-time gauge is, at least for this purpose, as physically meaningful
as any fully gauge invariant extension of this theory. However, if
our observer's theory disregards observables affected by such a gauge
symmetry in the bulk, it is unsense to take care of boundary restrictions
on the gauge. Again in the case of the free relativistic particle,
if we are only interested in measuring moments, which are gauge invariant
under this symmetry, we do not have to take care of the boundary restrictions
imposed by our action on this gauge freedom. 

If we are really interested in the invariance of the action under
$\delta _{G}[\Gamma _{(\lambda ,\tau )}],$ a conceivable method
is to consider $\tau \mid _{\Sigma _{t_{1,2}}}=0,$ this choice (
less restrictive but related choices are made in the case of asymptotic
boundaries) is often implicit in the litterature. Obviously, it is
also possible to complete Chern-Simons action in order to obtain a
fully $\delta _{G}[\Gamma _{(\lambda ,\tau )}]-$gauge invariant extension
of gravity. This program has been achieved in \cite{key-23}, but
despite of attractive algebraic features, the resulting theory is
far from the physical problem we want to describe in canonical gravity
for different reasons. Firstly, both alternatives are problematic
in the sense that they force us to deal with the boundary values of
the metric, or other geometric fields, in a given spatial coordinate system. Although perfectly
compatible with a spectator-observer description of gravity these
approaches seem to be completely off the scope of a participant-observer
description, in the sense that their purpose is to give an answer
to a question which has no sense in this perspective. Indeed, as soon
as, in a participant observer perspective, the observables we will
consider, based on matter degrees of freedom, will be invariant under
$\delta _{G}$ in the bulk, it is totally useless to take care of
boundary restrictions on this symmetry. As a conclusion, it will be
correct, for the participant observer description, to disregard boundary
restrictions on the gauge symmetry $\delta _{G}.$

Let us now study the other source of problems with the identification of these two symmetries, i.e.
 the difference between the group structures of
the symmetries $\delta _{G},$ $\delta _{L,PD}.$ This difference will have dramatic consequences on
 the global aspects of the physical phase space.
 We will denote $\Delta _{G}[g],$ for $g$ in the gauge
group $\mathbb{G}_{G}$ associated to $SL(2,\mathbb{C})_{\mathbb{R}}$
(the universal covering of $SO(3,1)$), the large gauge transformations, and $\Delta _{L,PD}[u],$ for $u$ in the gauge
group $\mathbb{G}_{L,PD}$ the large projectable diffeomorphisms $+$ local Lorentz transformations.
We do not want to enter into the details of the structure of the group
$\mathbb{G}_{L,PD}$ of projectable diffeomorphisms and local lorentz
transformations. However, a key point has to be mentioned. The large
gauge transformations, integrated from (\ref{gaugetransfoinfinitesimal}),
are given by\begin{equation}
\Delta _{G}[g](A_{\alpha })=g^{-1}A_{\alpha }g+g^{-1}\partial _{\alpha }g.\label{largegaugetransfo}\end{equation}

Let us consider a connection $A$ solution of the zero-curvature equation,
with $A^{(T)}$ corresponding to a nowhere-degenerate soldering form.
It is clear that, at least locally, a gauge transformation can always
be found such that $(\Delta _{G}[g](A_{\alpha }))=0$, which is incompatible
with the requirement that $(\Delta _{G}[g](A)){}^{(T)}$ is associated
to a non-degenerate soldering form. This is a crucial difference between
these symmetry groups. Indeed, it is never possible to find any projectable
diffeomorphism and local lorentz transformation such that $(\Delta _{L,PD}[g](A_{\alpha }))^{(T)}=0.$ 

We have already mentioned how Chern-Simons extension of gravity was
obtained by incorporating degenerate soldering forms. The extension
of the allowed range for the dynamical variables is simultaneous to
an extension of the gauge group. Indeed, the gauge symmetry is then
associated to $\Delta _{G}$ and not to $\Delta _{L,PD}$ anymore.
In fact, the space of classical solutions of the Hilbert-Palatini
theory is obtained as the space of $so(3,1)$connections $(A_{\alpha })_{\alpha \in I_{\mathcal{M}}},$
with a non degenerate translation part, i.e. \begin{equation}
\epsilon ^{\alpha \beta \gamma }<A_{\alpha }^{(T)},[A_{\beta }^{(T)},A_{\gamma }^{(T)}]>\neq 0,\label{non_degenerate}\end{equation}
 obeying the zero curvature equations (\ref{zerocurvature}), equipped
with the equal-time Poisson bracket\begin{eqnarray}
\{A_{i}^{I}(x),A_{j}^{J}(y)\}_{2} & = & \frac{2l_{c}}{\vartheta }\epsilon _{ij}t^{IJ}\delta ^{(2)}(x-y)\; ,
\end{eqnarray}

and moded out by the action of the gauge group $\mathbb{G}_{L,PD}$
given by (\ref{projdiffeoexp})(\ref{projdiffeodef}). 

The space of classical solutions of Chern-Simons gravity is obtained
as the space of $so(3,1)$connections $(A_{\alpha })_{\alpha \in I_{\mathcal{M}}},$
without restriction of its translation part, obeying the zero curvature
equations (\ref{zerocurvature}), equipped with the same equal-time
Poisson bracket, and moded out by the action of the gauge group $\mathbb{G}_{G}$
given by (\ref{largegaugetransfo}). We have moreover to decide, in
each of these cases, if these spaces must be moded out by the group
of orientation preserving diffeomorphisms of the base space not connected to the identity,
i.e. the mapping class group.

\emph{A contrario} from the common opinion exposed in current litterature on the subject, the reduced
phase space of Chern-Simons gravity is in fact smaller than that of
gravity. This point is very subtle, and generally disregarded by most
of the authors (see however the fundamental H.J.Matschull's paper
\cite{key-32}, see also \cite{key-34}), we prefer to develop these points after having completed
our perspective on the place of the observers and having emphasized
some fundamental observables allowing us to emphasize precisely the
discrepancy between the two phase spaces.

\subsubsection{\emph{Overview on the reduced phase space of Chern-Simons deSitter
gravity in vacuo}}

Before to study the coupling to particles, we want to explain an interesting
description of the space of classical solutions of Chern-Simons gravity,
the so-called \emph{combinatorial approach}, through the study of
a simple example. 

Let us take $\Sigma $ to be a $2-$torus parametrized by two angles
$(\varphi ,\psi )\in [0,2\pi [^{\times 2}.$ The set $\mathcal{O}=\left\{ (t,\varphi ,\psi )\in ]t_{1},t_{2}[\times ]0,2\pi [^{\times 2}\right\} $
is a connected and simply connected open subset of $]t_{1},t_{2}[\times \Sigma ,$
on which the zero curvature equations can be solved up to the action
of the gauge group $\mathbb{G}_{G}$ by $A_{\alpha }=k^{-1}\partial _{\alpha }k,\: \alpha \in \left\{ t,\varphi ,\psi \right\} $
where $k$ is a smooth mapping from $\mathcal{O}$ to $G.$ We will
not consider restrictions of $k$ coming from the boundaries $\Sigma _{t_{1}},\Sigma _{t_{2}}.$
The ambiguities on $k$ are linked to the action of the gauge group,
given by a smooth mapping $g$ from $]t_{1},t_{2}[\times \Sigma $
to $G,$ as $k\mapsto kg,$ and the Noether rigid symmetry given by
a fixed element $s$ of the group $G$ as $k\mapsto sk.$ If we introduce
the notations $v_{1}(t,\varphi )=\overrightarrow{Pexp}\int _{0}^{2\pi }d\psi \, A_{\psi }(t,\varphi ,\psi )$
and $v_{2}(t,\psi )=\overrightarrow{Pexp}\int _{0}^{2\pi }d\varphi \, A_{\varphi }(t,\varphi ,\psi ),$
we can verify that $v_{1}(t,\varphi )\equiv k^{-1}(t,\varphi ,0^{+})k(t,\varphi ,2\pi ^{-})$
and $v_{2}(t,\psi )\equiv k^{-1}(t,0^{+},\psi )k(t,2\pi ^{-},\psi ).$
Separately, we can introduce $u_{1}(t,\varphi )\equiv k(t,\varphi ,0^{+})k^{-1}(t,\varphi ,2\pi ^{-})$
and $u_{2}(t,\psi )\equiv k(t,0^{+},\psi )k^{-1}(t,2\pi ^{-},\psi ).$
We can check, using the action of the gauge, that $u_{1},u_{2}$ are
gauge invariant and, using the smoothness of the connection, that
they are independent of their parameters, i.e. $u_{1}(t,\varphi )\equiv u_{1}$
and $u_{2}(t,\psi )\equiv u_{2}$. The zero curvature equation implies
moreover that $u_{1}u_{2}u_{1}^{-1}u_{2}^{-1}=I,$ so $u_{1},u_{2}$
can be reduced in a common basis. Endly, the Noether symmetry acts
on them as $u_{1,2}\mapsto su_{1,2}s^{-1}$ then, only their conjugacy
classes $c_{1},c_{2}$ are physically relevant. Reciprocally, let be given two
connections $A,A'$ solutions of the zero curvature equation and associated
to the same couple $c_{1},c_{2}$ defined along previous procedure.
If we choose smooth gauges $k,k'$ respectively associated to $A,A',$
the mapping $g=k^{-1}k'$ can be shown to be smooth on the whole $]t_{1},t_{2}[\times \Sigma ,$
using the fact that $k,k'$ are associated to the same $c_{1},c_{2}.$
Then $A$ and $A'$are related by a gauge transformation.

Generically, $u_{1},u_{2}$ can be diagonalized simultaneously, using
the noether symmetry, their conjugacy class being respectively given
by $c_{1,2}=e^{2\pi (\lambda _{1,2}\xi _{0,T}+\rho _{1,2}\xi _{0,L})}.$
These datas define completely an element of the space of classical
solutions of Chern-Simons theory, and a connection $A_{\alpha }$
solution of the zero-curvature equation compatible with them is, for
example, $A=(\lambda _{1}\xi _{0,T}+\rho _{1}\xi _{0,L})d\psi +(\lambda _{2}\xi _{0,T}+\rho _{2}\xi _{0,L})d\varphi .$
Any observable can be written in terms of $c_{1},c_{2}$ however the
Poisson algebra appears to be not so easy to describe in these terms.
Nevertheless it can be done and lead to beautiful algebraic structures,
see \cite{key-36} for the original paper, \cite{key-29}for the classical
description, and \cite{key-31} for the quantization and the study
of the quantum phase space of the vacuum Chern-Simons deSitter gravity.

\section{Point masses in deSitter gravity}

Before to study the coupling of Chern-Simons deSitter gravity to point
particles in three dimensions we have to recall some basic facts about
free particles on a fixed deSitter background.

\subsection{\emph{Free relativistic particle on deSitter spacetime\label{sub:Free-relativistic}}}

Notations and basic properties of $SL(2,\mathbb{C})$ used in this
section can be found in the appendix \ref{sub:Basic-algebraic-results}.

\subsubsection{\emph{Basic notions on deSitter spacetime}}

The deSitter spacetime $dS_{3}$ is defined as the maximally symmetric
solution to 2+1 dimensional Einstein's equations (\ref{einsteineq}).
Its metric can be described, for example, in cylindrical coordinates
as\begin{eqnarray*}
ds^{2} & = & -(1-\frac{r^{2}}{l_{c}^{2}})dt^{2}+\frac{dr^{2}}{(1-\frac{r^{2}}{l_{c}^{2}})}+r^{2}d\phi ^{2}.
\end{eqnarray*}
We have to notice that this metric describes only a part of $dS_{3}.$

It can be embedded into the 4-dimensional Minkowski space $\mathbf{M}_{4}$
as follows. $\mathbf{M}_{4}$ will be identified with the set of $2\times 2$
hermitian matrices $\mathcal{H}$ by the isomorphism of vector space: 

\begin{eqnarray}
\mathcal{M}_{4}\; \longrightarrow \; \mathcal{H}\; ,\; \; \; \; x\; \longmapsto \; X=x^{\mu }\sigma _{\mu }=
\left(\begin{array}{cc}
 x^{0}+x^{1} & x^{2}-ix^{3}\\
 x^{2}+ix^{3} & x^{0}-x^{1}\end{array}
\right). &  & \label{isoM4H}
\end{eqnarray}
 This map is an isometry as soon as we equip $\mathcal{H}$ with the
pseudo-norm $|X|^{2}=-\text {det}(X)$. $dS_{3}$ can be mapped to
the subset $\mathcal{D}$ of $\mathcal{H}$ defined by: 

\begin{equation}
\mathcal{D}=\{Q\in \mathcal{H}|\det Q=-l_{c}^{2}\}.\label{defD}\end{equation}

A distance is induced on $dS_{3}$ from that on $\mathbf{M}_{4}$,
$d(Q,Q')^{2}=l_{c}^{2}tr(Q^{-1}Q'-I)$ for any couple $Q,Q'\in \mathcal{D}.$ 

The action of the (universal covering of the) Lorentz group $G=SL(2,\mathbb{C})$
on $\mathbf{M}_{4}$ is translated into the following action on $\mathcal{H}$:
\begin{eqnarray}
G\times \mathcal{H}\longrightarrow \mathcal{H}\; , &  & (\Lambda ,X)\longmapsto {}^{\Lambda }X=\Lambda X\Lambda ^{\dagger }.\label{actionofthegroup}
\end{eqnarray}
 This action descends to the subset $\mathcal{D}$ and promotes $G=SL(2,\mathbb{C})$
as the group of moves on $dS_{3}$; a basic subgroup of $G,$ i.e.
$\mathcal{L}_{o},$ forms the $2+1$ Lorentz group.

As a consequence, if we consider an arbitrary element $o\in \mathcal{D}$,
we can define an application from the configuration space $\mathcal{C}\equiv SL(2,\mathbb{C})$
to $\mathcal{D}$ as follows: \begin{eqnarray}
\mathcal{C}\longrightarrow \mathcal{D}\; , &  & M\longmapsto Q_{M}=MoM^{\dagger }\; .\label{applicationCtoD}
\end{eqnarray}

As a remark, we have the following interesting formula \begin{equation}
d(Q_{M},Q'_{M})^{2}=l_{c}^{2}tr((M^{-1}M')(M^{-1}M')^{\star }-I).\label{distance}\end{equation}

Obviously this map is multivalued. In fact, we can decompose $\mathcal{D}$
as the union $\mathcal{D}_{o}^{+}\cup \mathcal{D}_{o}^{-}$ with $\mathcal{D}_{o}^{\pm }=\{Q\in \mathcal{D}|\text {Tr}(I\pm Qo^{-1})>0\}$
such that on each of these parts the application \begin{equation}
\Upsilon _{\pm }:\mathcal{D}_{o}^{\pm }\longrightarrow \mathcal{T}_{o}\; \; \; ,\; \; \; Q\longmapsto \Upsilon _{\pm }(Q)=\frac{I\pm Qo^{-1}}{\sqrt{\text {Tr}(I\pm Qo^{-1})}}\label{applicationDtoT}\end{equation}

is invertible with inverse defined by $\Upsilon _{\pm }^{-1}(\tau )=\pm \tau ^{2}o.$ 

Moreover, if $Q$ and $Q'$ are two elements of $\mathcal{D}$, it
is always possible to find an element $o$ such that these two elements
belong to $\mathcal{D}_{o}^{+}$. 

We will denote by $(\theta ^{T,i})_{i=0,1,2}$ a system of local coordinates
on $dS_{3},$ we will denote $x^{i}:=\theta ^{T,i}$ as well. Using
the mappings $\Upsilon _{\pm }$ we define an atlas of $\mathcal{T}_{o}$ given
by a set of mappings from $\mathbb{R}^{3}$ to $T_{o}$ we will denote
abusively $x\mapsto \tau (x).$ We will introduce also $(\theta ^{L,i})_{i=0,1,2}$
a system of local coordinates on $\mathcal{L}_{o},$ we will denote
$r^{i}:=\theta ^{L,i}$ as well and will denote abusively $r\mapsto \lambda (r)$
the corresponding mappings from $\mathbb{R}^{3}$ to $\mathcal{L}_{o}$. Due
to factorization theorems recalled in the appendix \ref{sub:Basic-algebraic-results},
these datas allow us to parametrize the configuration space $\mathcal{C}\equiv SL(2,\mathbb{C})$
as a whole and we will denote $(\theta ^{I})_{I\in \{T,L\}\times \{0,1,2\}}\mapsto M(\theta )=\tau (\theta ^{T})\lambda (\theta ^{L})$
the corresponding matrix of coordinate functions. 

The metric on $dS_{3}$ is given by $dx^{i}g_{x^{i}x^{j}}(x)dx^{j}=l_{c}^{2}\; \text {det}(d(\tau (x)^{2}))$
and $e_{\alpha }^{a}$ and $\omega _{\alpha }^{a}$ defined such that
$\tau ^{-1}\partial _{x^{i}}\tau =(\frac{1}{l_{c}}e_{x^{i}}^{a}\xi _{T,a}+\omega _{x^{i}}^{a}\xi _{L,a})$
is a solution of (\ref{eqmove1})(\ref{eqmove2}) giving a first order
description of $dS_{3}$. Let us notice how the first order datas
transform under their basic local symmetry associated to Lorentz group.
If $\lambda $ is a given mapping from an open subset of $\mathbb{R}^{3}$
to $\mathcal{L}_{o},$ then the local Lorentz transformations associated to
$\lambda $ are defined by $(\frac{1}{l_{c}}e_{x^{i}}^{a}\xi _{T,a}+\omega _{x^{i}}^{a}\xi _{L,a})\mapsto \lambda ^{-1}(\frac{1}{l_{c}}e_{x^{i}}^{a}\xi _{T,a}+\omega _{x^{i}}^{a}\xi _{L,a})\lambda +\lambda ^{-1}\partial _{x^{i}}\lambda \equiv (\frac{1}{l_{c}}e'{}_{x^{i}}^{a}\xi _{T,a}+\omega '{}_{x^{i}}^{a}\xi _{L,a}).$
Then it is obvious that any mapping $k$ from an open subset of $\mathbb{R}^{3}$
to $G$ defines a first order description of $dS_{3}$ by $k^{-1}\partial _{x^{i}}k=(\frac{1}{l_{c}}e'{}_{x^{i}}^{a}\xi _{T,a}+\omega '{}_{x^{i}}^{a}\xi _{L,a}),$
which associated metric is $dx^{i}g_{x^{i}x^{j}}(x)dx^{j}=l_{c}^{2}\text {det}(d(kk^{\star })).$

The usual expression of timelike geodesic equations on $dS_{3}$ in
terms of the proper time $\sigma $ can be simplified in this framework,
i.e. if we express them in terms of $Q$, into the following equation
$\frac{d^{2}Q}{d\sigma ^{2}}=Q.$ In fact, these trajectories are
parametrized by $Q(\sigma )=M_{0}e^{2\sigma \xi _{T,0}}M_{0}^{\star }o.$

\subsubsection{\emph{The relativistic particle on} $dS_{3}$}

The basic action principle used to describe a relativistic particle
of mass $m$ and spin $s$ on $dS_{3}$ is \begin{equation}
S_{p}[M]=\int _{t_{1}}^{t_{2}}dt<\chi ^{b_{+},b_{-}}|M^{-1}\frac{dM}{dt}>\; ,\label{actionoftheparticle}\end{equation}

where$\: \chi ^{b_{+},b_{-}}\equiv b_{+}(m,s)\xi _{L,0}+b_{-}(m,s)\xi _{T,0}$with
$b_{+}(m,s)\equiv ml_{c},\: b_{-}(m,s)\equiv s.$ The dynamical variable
$M$ belongs to $\mathcal{C}\equiv SL(2,\mathbb{C})$. In order to
define the variational problem, half of the dynamical variables are
kept fixed at the boundaries such that in particular $<\chi ^{b_{+},b_{-}}|M^{-1}\delta M>(t_{1,2})=0$,
i.e. the position on deSitter spacetime $Q_{M}$ in the case of a
purely massive particle, the problem of boundary conditions will be
discussed soon. The canonical analysis of this system has been already
done in the litterature (see for example \cite{key-27}\cite{DSG}). Better than
working with coordinates $(\theta ^{I})_{I\in I_{\mathfrak{g}}}$
and their momenta $(\pi _{I})_{I\in I_{\mathfrak{g}}}$, we will work
with the matrix of coordinate functions $M$ and $P=p_{I}\xi ^{I}$,
where $p_{I}$ are certain functions of coordinates and momenta chosen
such that the Poisson brackets are given by: \begin{equation}
\{M_{1},M_{2}\}=0\; \; ,\; \; \{P_{1},M_{2}\}=t_{12}M_{2}\; \; ,\; \; \{P_{1},P_{2}\}=-[t_{12},P_{1}]\; \; \label{Poissonbracket}\end{equation}
 (here we used the tensorial notation $A_{1}=A_{n}^{m}E_{m}^{n}\otimes 1$
and $A_{2}=1\otimes A_{n}^{m}E_{m}^{n}$, for any $2\times 2$ matrix
$A$ and $(E_{m}^{n})_{m,n}$ is the canonical basis of the space
of $2\times 2$ matrices).

The set of primary constraints is given by \begin{equation}
C[M,P]=M^{-1}PM+\chi ^{b_{+},b_{-}}\thickapprox 0.\label{contrainte}\end{equation}

The canonical Hamiltonian is equal to zero. The action (\ref{actionoftheparticle})
is equivalent to the first order action:\begin{equation}
S_{p}^{tot}[M,P,\mu ]=-\int dt(<P|\frac{dM}{dt}M^{-1}>+H_{tot}[X,P,\mu ])\label{firstorderparticle}\end{equation}
 where $\mu \in \mathfrak{g}=sl(2,\mathbb{C})$ and the total Hamiltonian\begin{equation}
H_{p}^{tot}[M,P,\mu ]=<\mu |C[M,P]>.\label{canonhamilt}\end{equation}
 Conservation of the constraints under time evolution imposes no secondary
constraints and fix $\mu $ to belong to $\mathfrak{h}$, i.e. \begin{equation}
\mu =\mu _{\mathfrak{h}}=\rho _{L}\xi _{L,0}+\rho _{T}\xi _{T,0}.\label{valmu}\end{equation}

The Dirac bracket can be easily computed (see Appendix \ref{sub:Dirac-free}).
As $P$ commutes with the second class constraints, the Poisson brackets
(\ref{Poissonbracket}) involving $P$ are not modified by the Dirac
reduction:\[
\{P_{1},M_{2}\}_{D}=t_{12}M_{2}\; \; ,\; \; \{P_{1},P_{2}\}_{D}=-[t_{12},P_{1}]\; \; \]

However, the Poisson bracket (\ref{Poissonbracket}) between matrix
elements of $M$ is modified into the following bracket: \[
\{M_{1},M_{2}\}_{D}=M_{1}M_{2}\mathfrak{R}_{12}^{(0)}(M,P)\]
 where $\mathfrak{R}_{12}^{(0)}(M,P)=r_{12}^{(0)}(\tilde{b}_{+},\tilde{b}_{-})$
with $\tilde{b}_{+}\equiv b_{+}+<C[M,P],\xi _{0,T}>,$ $\tilde{b}_{-}\equiv b_{-}+<C[M,P],\xi _{0,L}>$
and $r^{(0)}$ is defined in the appendix \ref{sub:Dirac-gravit}. 

In the following we will restrict ourself to the purely massive case,
i.e. $b_{-}=0.$

The dynamical equations are given by \begin{equation}
\frac{dM}{dt}=\{M,H_{p}^{tot}\}_{D}=-M\mu _{\mathfrak{h}}\; \; ,\; \; \frac{dP}{dt}=\{P,H_{p}^{tot}\}_{D}=0.\label{eqmoveparticle}\end{equation}

Solutions of equations of motion and constraints are parametrized
by a fixed $M_{0}\in G$ as 

\begin{equation}
M(t)=M_{0}exp\left(-\int _{t_{1}}^{t}dv\, \mu _{\mathfrak{h}}(v)\right)\; \; ,\; \; P(t)=-b_{+}M_{0}\xi _{L,0}M_{0}^{-1}.\label{soleqmoveparticle}\end{equation}

The corresponding motion of the particle on deSitter spacetime is
given by $Q_{M}(t)=M(t)oM(t)^{\dagger }=M_{0}exp(-2\int _{t_{1}}^{t}dv\, \rho _{T}(v)\xi _{T,0})oM_{0}^{\dagger }$.
It is clear, from the expression of the length element on the trajectory
$dl^{2}=-det(dQ)=-l_{c}^{2}\rho _{T}(t)^{2}dt^{2},$ that $l_{c}\left|\rho _{T}(t)\right|$
is the flow rate of proper-time. The proper time is then given by
\begin{equation}
\vartheta (t)=l_{c}\int _{t_{1}}^{t}dv\left|\rho _{T}(v)\right|.\label{propertime}\end{equation}
 From (\ref{soleqmoveparticle}), it is easy to see that, if $v\mapsto \rho _{T}(v)$
is a continuous function such that $\rho _{T}(v)\neq 0,\forall v$,
$Q(\sigma )=Q_{M}(\vartheta ^{-1}(\sigma ))$ obeys the timelike geodesic
equation $\frac{d^{2}Q}{d\sigma ^{2}}=Q$. In order to avoid multiple
counting of histories and to check if a gauge choice allows us to select
causal histories according to previous remark, we have to study carefully
the gauge symmetries of our action. 

Infinitesimal gauge symmetries generated by the first class constraints
are: \begin{equation}
\delta _{H}[\gamma ]M=-M\gamma \; \; ,\; \; \delta _{H}[\gamma ]P=0\; \; ,\; \; \mathrm{where}\; \gamma \in \mathfrak{h}.\label{symmetryofMandP}\end{equation}
 If we allow also the Lagrange multipliers $\mu $ to transform as:
\begin{equation}
\delta _{H}[\gamma ]\mu =-\dot{\gamma }+[\gamma ,\mu ]\label{symmetryofmu}\end{equation}
we obtain $\delta _{H}[\gamma ]S_{p}^{tot}[M,P,\mu ]=\left[-b_{+}<\gamma \mid \xi _{L,0}>\right]_{t_{1}}^{t_{2}}$.
If we do not change our action principle we have to require $<\gamma \mid \xi _{L,0}>(t_{1,2})=0$,
and the natural partial gauge fixing compatible with this gauge freedom
is obtained by the usual derivative gauge $<\dot{\mu }\mid \xi _{L,0}>=0.$
The residual gauge freedom $\gamma =\kappa \xi _{L,0}$, which corresponds
to an internal gauge symmetry, can then be fixed by a canonical gauge.
However, we will not pay attention to these gauge degrees of freedom.
\emph{Up to trivial gauge symmetries}, the gauge symmetry corresponding
to the reparametrization of the worldline of the particle $t\mapsto t+\varsigma (t)$
can be recovered from the previous gauge transformations by taking
$\gamma =\varsigma (t)\mu _{\mathfrak{h}}$. It appears that our derivative
gauge fixes also the reparametrization freedom according to proper-time
gauge which selects causal histories, this fact justifies in itself
the interest of this gauge-fixing. The identification of boundary
terms changing our action principle in order to ensure its complete
invariance seems then to be of no particular interest.

Endly, it is important to notice that our action is also invariant
under some rigid Noether symmetries. The first one, corresponding
to the change of origin and reference frame corresponds to the isometry
group of deSitter spacetime. It is simply given by\begin{eqnarray}
M\mapsto g^{-1}M\quad  & P\mapsto g^{-1}Pg\quad  & \mu \mapsto \mu \label{noether}
\end{eqnarray}

$g$ being a fixed element of the group $G.$

\subsection{\emph{Coupling gravity to point masses}}

\subsubsection{\emph{the Kerr-deSitter solution}}

The Kerr-deSitter metric describing a gravity field coupled to a particle
at rest, of mass $m$ and spin $s$, in presence of a positive cosmological
constant was studied in \cite{Jac}, is given, in cylindrical coordinates, by\begin{eqnarray*}
ds^{2} & = & -(C_{(m)}-\frac{r^{2}}{l_{c}^{2}}+\frac{l_{c}^{2}D_{(s)}}{4r^{2}})dt^{2}+\frac{dr^{2}}{(C_{(m)}-\frac{r^{2}}{l_{c}^{2}}+\frac{l_{c}^{2}D_{(s)}}{4r^{2}})}+r^{2}(d\phi -\frac{l_{c}D_{(s)}}{2r^{2}}dt)^{2}
\end{eqnarray*}

(we will choose $C_{(m)}=1-\frac{\vartheta m}{2\pi },\: D_{(s)}=\frac{\vartheta s}{2\pi l_{c}}$).
This metric is obtained as the rotationally symmetric and stationary
solution of Einstein's equation with a positive cosmological constant
which tends to deSitter spacetime asymptotically ($C_{(m)},D_{(s)}$are
the constants appearing in the integration of this dynamical problem)\cite{key-24}.
We have to notice that the presence of the pure point mass ($m\neq 0,s=0$)
causes the metric to have a conical singularity at the origin, indeed\[
ds^{2}\propto _{r\propto 0}-(1-\frac{r^{2}}{l_{c}^{2}})dt^{2}+\frac{dr^{2}}{(1-\frac{r^{2}}{l_{c}^{2}})}+C_{(m)}r^{2}d\phi ^{2}.\]

We will have to relax the requirement of smoothness of the dynamical
fields in the vicinity of particles, this point will be studied in
the next subsection.

In order to simplify the expressions of the first-order description,
we notice that this metric can be reparametrized into the following
nice formula\begin{eqnarray*}
ds^{2} & = & -l_{c}^{2}cos^{2}(\rho )(a_{+}d\tau -a_{-}d\phi )^{2}+l_{c}^{2}d\rho ^{2}+l_{c}^{2}sin^{2}(\rho )(a_{-}d\tau +a_{+}d\phi )^{2}
\end{eqnarray*}

where $a_{\pm }(m,s)=\sqrt{\frac{\pm C_{(m)}+\sqrt{C_{(m)}^{2}+D_{(s)}^{2}}}{2}}$,
and the adimensional real variables $\rho ,\tau $ are $t=l_{c}\tau ,\; r^{2}=l_{c}^{2}(a_{+}^{2}sin^{2}(\rho )-a_{-}^{2}cos^{2}(\rho ))$,
$\rho \in [\rho _{min}\equiv Arctan(\frac{a_{-}}{a_{+}}),\frac{\pi }{2}[$
.

A soldering form can be chosen according to this metric:\begin{eqnarray*}
e_{\tau }^{2}=l_{c}a_{-}sin(\rho )\:  & e_{\tau }^{1}=0\:  & e_{\tau }^{0}=l_{c}a_{+}cos(\rho )\\
e_{\phi }^{2}=l_{c}a_{+}sin(\rho )\:  & e_{\phi }^{1}=0\:  & e_{\phi }^{0}=-l_{c}a_{-}cos(\rho )\\
e_{\rho }^{0}=0\:  & e_{\rho }^{1}=-l_{c}\:  & e_{\rho }^{0}=0\: 
\end{eqnarray*}

and the spin-connection, associated to it, is given by:\begin{eqnarray*}
\varpi _{\tau }^{2}=a_{+}sin(\rho )\:  & \varpi _{\tau }^{1}=0\:  & \varpi _{\tau }^{0}=-a_{-}cos(\rho )\\
\varpi _{\phi }^{2}=-a_{-}sin(\rho )\:  & \varpi _{\phi }^{1}=0\:  & \varpi _{\phi }^{0}=-a_{+}cos(\rho )\\
\varpi _{\rho }^{0}=0\:  & \varpi _{\rho }^{1}=0\:  & \varpi _{\rho }^{0}=0\: 
\end{eqnarray*}

The $so(3,1)-$connection built from these datas is simply given by\begin{eqnarray*}
A_{\tau } & = & e^{\rho \xi _{1,T}}(a_{+}\xi _{0,T}-a_{-}\xi _{0,L})e^{-\rho \xi _{1,T}}\\
A_{\phi } & = & -e^{\rho \xi _{1,T}}(a_{-}\xi _{0,T}+a_{+}\xi _{0,L})e^{-\rho \xi _{1,T}}\\
A_{\rho } & = & -\xi _{1,T}
\end{eqnarray*}

It is interesting to remark that the previous formula can be rewritten
as

\begin{eqnarray*}
A=V^{-1}BV+V^{-1}dV,\;  & B\equiv -(a_{-}\xi _{0,T}+a_{+}\xi _{0,L})d\phi ,\;  & V\equiv e^{-\tau (a_{+}\xi _{0,T}-a_{-}\xi _{0,L})}e^{-\rho \xi _{1,T}}.
\end{eqnarray*}

\subsubsection{\emph{Coupling point masses to gravity: the regularization problem} \textmd{}}

Till now we have not mentioned what are the constraints and equations
of motion giving previous expressions as solutions. The problem is
that the stress-energy tensor of the particle has no canonically defined
meaning in the distributional sense, and we have to regularize carefully
the kinematical and dynamical equations to produce previous solutions.  
In order to incorporate conical singularities, we have to slightly change our
assumptions about the base manifold. Hence, we will consider the worldlines
of $N$ particles as mappings $(x_{(n)})_{n=1,\cdots ,N}$ from $[t_{1},t_{2}]$
to $\Sigma $. For this purpose, we replace $\mathcal{M}$ by the
three dimensional compact manifold with boundaries, denoted $\mathcal{M}^{*},$
obtained from $\mathcal{M}$ by removing $N$ disjoint open tubular
neighbourhoods $(c_{(n)})_{n=1...N}$ of the worldlines. $\mathcal{M}^{*}$
inherits a foliation from $\mathcal{M}$ and we will denote by $\Sigma _{t}^{*}$
the slice at label $t$ of $\mathcal{M}^{*}$. The boundary of $\Sigma _{t}^{*}$
consists in the union of $N$ disjoint circles $(l_{(n)}(t))_{n=1,\cdots ,N}$.
The spatial boundary will be denoted $\mathcal{B}\equiv \cup _{t\in [t_{1},t_{2}]}s_{t},\: s_{t}\equiv \cup _{i=1}^{n}l_{(n)}(t)$.
Any smooth function $f$ defined on $\mathcal{M}^{*}$can be restricted
to $\mathcal{B}$ and we will denote $\bar{f}$ this restriction.
Points $x_{(n)}$ of $l_{(n)}(t)$ are parametrized by $(t,\phi )$,
where $t\in [t_{1},t_{2}]$ is the time coordinate and $\phi \in [0,2\pi [$
is an angular coordinate, $\int _{l_{(n)}(t)}f$ will denote $\int _{0}^{2\pi }d\phi \: f(x_{(n)}(t,\phi )),$
$\int _{s_{t}}\equiv \sum _{n=1}^{N}\int _{l_{(n)}(t)}$ and $\int \! \! \! \! \! \int _{\mathcal{B}}\equiv \int _{t_{1}}^{t_{2}}dt\int _{s_{t}}$.
Note that the orientation of each circle $l_{(n)}(t)$ is chosen such
that $\int _{\Sigma _{t}^{*}}d^{2}x\epsilon ^{ij}\partial _{i}u_{j}=-\sum _{n=1}^{N}\int _{l_{(n)}(t)}d\phi \; u_{\phi }$
for any one-form $u=\sum _{l=1,2}u_{x^{l}}dx^{l}$. Here and in the
sequel, we will currently denote $\bar{u}_{\phi }(t,\phi )\equiv \sum _{l=1,2}u_{x^{l}}(x_{(n)}(t,\phi ))\frac{\partial x_{(n)}^{l}}{\partial \phi }$
and $\partial _{\phi }\bar{f}(t,\phi )\equiv \sum _{l=1,2}(\partial _{x^{l}}f)(x_{(n)}(t,\phi ))\, \frac{\partial x_{(n)}^{l}}{\partial \phi }$,
for any one-form $u$ and function $f$ defined on $\Sigma ^{*}$
with value in $\mathfrak{g}$ . For any $f\in C^{\infty }(\mathcal{B},\mathfrak{g}),$
we define $f_{(n)}^{av.}\in C^{\infty }([t_{1},t_{2}],\mathfrak{g})$
by $f_{(n)}^{av.}(t)\equiv \frac{1}{2\pi }\int _{0}^{2\pi }d\phi \; f\left(x_{(n)}(t,\phi )\right)\; ,$
and $\breve{f}\in C^{\infty }(\mathcal{B},\mathfrak{g})$ by $\breve{f}(x_{(n)}(t,\phi ))\equiv f_{(n)}^{av.}(t).$
According to our spectator-observer perspective, we will disregard
information coming from the choice of coordinates, and because there
is no global topological obstruction to do that, we will suppose,
in the following, that $\frac{\partial x_{(n)}^{\mu }}{\partial t}=\delta _{0}^{\mu }$.

We will require the soldering form, the spin connection and also the
mappings defining the gauge transformations to be smooth on $\mathcal{M}^{*}$
and its boundaries.

These assumptions are too general for our purpose of describing the
worldline of a point particle, and we have to impose some boundary
conditions on the connection in order to ensure, at least, that the distance, computed from
the metric associated to the translational part of the connection, from the points 
$x_{(n)}(t,\phi )$ and $x_{(n)}(t,\phi '))=0$ is null $\forall n\in {1,\cdots ,N},\forall t\in [t_{1},t_{2}],
\forall \phi ,\phi '\in [0,2\pi [$
(it means that a worldline is not a cylinder), and also to fix kinematical
properties of the particle. 

There are two relatively different points of view in the litterature
about these boundary conditions. 

The first construction introduced by E.Witten \cite{key-25,key-26}
is an attempt to restore a complete gauge symmetry in presence of
matter. To this aim, we associate to each particle a dynamical variable
$M_{(n)}$ constant along each circle $l_{(n)}$. The action of a
free particle (\ref{actionoftheparticle}) being, as a fundamental
fact, invariant under the rigid left action of the group $G$ defined
in (\ref{noether}). We can gauge this symmetry thanks to a $so(3,1)-$
gauge field $A$ which dynamics will be dictated by the action studied
in the previous section. The gauge algebra $\mathbb{g}$ is chosen
to be 

\[
\mathbb{g}\equiv \left\{ \Gamma \in \mathcal{C}^{\infty }(\mathcal{M}^{*},\mathfrak{g})/\bar{\Gamma }=\breve{\Gamma }\right\} \]
 and the action of $\xi \in \mathbb{g}$ on the gauge field $A$ and
on dynamical variables $M_{(n)}$, for $n\in \{1,\cdots ,N\}$ will
be respectively given by: \begin{eqnarray}
\delta _{G}[\Gamma ]A_{\mu }=D_{\mu }^{A}\Gamma \; \; \; , & \; \; \; \delta _{G}[\Gamma ]M_{(n)}=\Gamma _{(n)}^{av.}M_{(n)}\; \; \; ,\; \; \;  & \forall \Gamma \in \mathbb{g}.\label{symmetry}
\end{eqnarray}

As a result, the minimal coupling action for a particle given by:
\begin{eqnarray}
S_{c}[M_{(n)},A] & = & \int _{t_{1}}^{t_{2}}\! \! \! dt\; <\chi ^{b_{+}^{(n)},b_{-}^{(n)}}|M_{(n)}^{-1}\frac{dM_{(n)}}{dt}\! +\! M_{(n)}^{-1}(A_{t})_{(n)}^{av.}M_{(n)}>\; ,\label{couplingaction1}
\end{eqnarray}
 is let invariant by the previous local symmetry, because of the formula
$\delta _{G}[\Gamma ]\left((A_{t})_{(n)}^{av.}\right)=\partial _{t}\Gamma _{(n)}^{av.}+[(A_{t})_{(n)}^{av.},\Gamma _{(n)}^{av.}]$. 

The dynamic of the gauge field $A$ is, a priori, described in terms
of the Hilbert-Palatini action, but the action is now defined on the
manifold with boundary $\mathcal{M}^{*}$, i.e. $S_{HP}[A]=S_{CS}^{bulk}[A]+S_{\Sigma }[A]$
with\begin{eqnarray*}
S_{CS}^{bulk}[A] & = & \frac{l_{c}}{\vartheta }\int \! \! \! \! \! \int \! \! \! \! \! \int _{\mathcal{M}^{*}}d^{3}x\, \epsilon ^{\alpha \beta \gamma }\left(<A_{\alpha },\partial _{\beta }A_{\gamma }>+\frac{1}{3}<A_{\alpha },[A_{\beta },A_{\gamma }]>\right)\\
 & = & \frac{l_{c}}{\vartheta }\int dt\int \! \! \! \! \! \int _{\Sigma ^{*}}d^{2}x\, \epsilon ^{ij}\left(-<A_{i},\partial _{t}A_{j}>+<A_{t},F_{ij}[A]>\right)+\frac{l_{c}}{\vartheta }\int dt\int _{s_{t}}<\bar{A}_{t}|\bar{A}_{\phi }>\\
S_{\Sigma }[A] & = & \frac{l_{c}}{\vartheta }\int \! \! \! \! \! \int _{\Sigma _{t_{2}}^{*}-\Sigma _{t_{1}}^{*}}d^{2}x\, \epsilon ^{ij}<A_{i}^{(L)},A_{j}^{(T)}>.
\end{eqnarray*}

For the moment, we will not be interested in the possible supplementary
terms associated to boundaries $\Sigma _{t_{1}}$ and $\Sigma _{t_{2}}$
in order to change the variational principle. It is manifest that
the presence of the tubes corresponding to particles breaks gauge
invariance of the Chern-Simons action and we have:\begin{eqnarray}
\delta _{G}[\Gamma ]S_{CS}^{bulk}[A] & = & -\frac{l_{c}}{\vartheta }\int \! \! \! \! \! \int _{\mathcal{B}}<\partial _{t}\bar{\Gamma }|\bar{A}_{\phi }>+\frac{l_{c}}{\vartheta }\int \! \! \! \! \! \int _{\Sigma _{t_{2}}^{*}-\Sigma _{t_{1}}^{*}}\! \! \! d^{2}x\, \epsilon ^{ij}<A_{i},\partial _{j}\Gamma >\label{gaugetransformationofCSaction}\\
\delta _{G}[\Gamma ]S_{\Sigma }[A] & = & \frac{-l_{c}}{\vartheta }\int \! \! \! \! \! \int _{\Sigma _{t_{2}}^{*}-\Sigma _{t_{1}}^{*}}\! \! \! d^{2}x\, \epsilon ^{ij}<A_{i},\partial _{j}\Gamma >+\frac{2l_{c}}{\vartheta }\int _{s_{t_{2}}-s_{t_{1}}}\! \! <\Gamma ^{(T)},A_{\phi }^{(L)}>\label{gaugetransformationofsigmaaction}\\
 &  & +\frac{l_{c}}{\vartheta }\int \! \! \! \! \! \int _{\Sigma _{t_{2}}-\Sigma _{t_{1}}}\! \! \! d^{2}x\, \epsilon ^{ij}<2\partial _{i}A_{j}^{(L)}+[A_{i}^{(L)},A_{j}^{(L)}]-[A_{i}^{(T)},A_{j}^{(T)}],\Gamma ^{(T)}>\nonumber 
\end{eqnarray}

In order to recover the gauge invariance, we choose to add the following
term to the action: \begin{eqnarray}
S_{B}[A] & = & \frac{l_{c}}{\vartheta }\int \! \! \! \! \! \int _{\mathcal{B}}<\bar{A}_{t}|\bar{A}_{\phi }>\; .\label{actioncylinders}
\end{eqnarray}
 It is immediate to show that $S_{B}[A]$ transforms as follows under
a gauge transformation: \begin{eqnarray*}
\delta _{G}[\Gamma ]S_{B}[A] & = & \frac{l_{c}}{\vartheta }\int \! \! \! \! \! \int _{\mathcal{B}}<\partial _{t}\bar{\Gamma }|\bar{A}_{\phi }>
\end{eqnarray*}
This expression is exactly the opposite of the problematic term which
appears in (\ref{gaugetransformationofCSaction}); as a consequence,
the resulting action $S_{CS}[A]+S_{B}[A]$ is invariant by gauge transformations.

\medskip

In order to describe the coupling between particles and gravity, we
will then study the action defined by: \begin{eqnarray}
S_{W}[A,M] & = & S_{HP}[A]+S_{B}[A]+\sum _{n=1}^{N}S_{c}[M_{(n)},A]\; .\label{couplingaction}
\end{eqnarray}

The second construction introduced by H.J.Matschull in \cite{key-35}
follows a different perspective to regularize particles, trying to obtain an action 
of gauge symmetries, in the neighbourhood of the particle, as close as possible 
from diffeomorphisms and local Lorentz invariance. 
This regularization appears
to be a bit more difficult to exploit for a canonical analysis. Our point of view, is 
that both approaches must have a common gauge fixed description, and that for the 
purpose of capturing the whole geometrical aspects of the whole universe, the choice of 
regularization in the vicinity of particles does not have any importance. However, in order 
to develop the canonical analysis of our problem in presence of boundaries, we have to take care
of some subtleties in the computations of Poisson brackets \cite{Ber}.

\subsubsection{\emph{Dirac reduction of Witten's type action}}

It will be convenient to express Witten's type action as follows:
\begin{eqnarray}
S_{W}[A,M] & = & \frac{2l_{c}}{\vartheta }\int \! dt\! \! \left(\int \! \! \! \! \! \int _{\Sigma _{t}^{*}}d^{2}x\, \epsilon ^{ij}\! \! <A_{j}^{(T)}\mid \partial _{t}A{}_{i}^{(L)}>+\frac{1}{2}<A_{t}|\epsilon ^{ij}F_{ij}(A)>\right.\label{Lagrangian}\nonumber \\
 &  & \! \! \! \! \! \! \left.+\sum _{n=1}^{N}<\chi _{n}|M_{(n)}^{-1}\frac{dM_{(n)}}{dt}>+\int _{s_{t}}\! \! <\bar{A}_{t}|\bar{A}_{\phi }-\breve{X}>\right)\label{actionwitten}
\end{eqnarray}
 where $\breve{X}$ is defined to be the mapping from the spatial
boundary $\mathcal{B}$ to $\mathfrak{g}$ defined by \begin{eqnarray}
\breve{X}(x_{(n)}(t,\phi )) & \equiv  & \frac{-M_{(n)}(t)\chi _{n}M_{(n)}^{-1}(t)}{2\pi },\label{definitionX}\\
\chi _{n} & = & \frac{\vartheta }{2l_{c}}\chi ^{b_{+}^{(n)},b_{-}^{(n)}},
\end{eqnarray}

and we will also introduce $\kappa _{n}=\frac{2l_{c}}{\vartheta }\kappa ^{b_{+}^{(n)},b_{-}^{(n)}}.$

We denote by $(\theta _{(n)}^{I})_{I\in I_{\mathfrak{g}}}$ the local
coordinates parametrizing $M_{(n)}$, according to the previous section,
and $(\pi _{I}{}_{(n)})_{I\in I_{\mathfrak{g}}}$ their canonical
momenta. As in the free particle case, it will be convenient to introduce
the functions $P_{(n)}\in \mathfrak{g}$ of the previous variables,
defined such that: \begin{eqnarray}
\{M_{(n)1},M_{(l)2}\}=0, & \{P_{(n)1},M_{(l)2}\}=t_{12}M_{(n)2}\delta _{n,l}, & \{P_{(n)1},P_{(l)2}\}=-[t_{12},P_{(n)1}]\delta _{n,l}.\nonumber \\
 &  & \label{PoissonbracketMP}
\end{eqnarray}
 We will denote by $(B_{I}^{\mu })_{I\in I_{\mathfrak{g}},\mu \in I_{\mathcal{M}}}$,
the canonical momenta associated to the gauge fields in order that:
\begin{eqnarray}
\{A_{\mu }^{I}(x),B_{J}^{\nu }(y)\}=\delta _{I}^{J}\; \delta _{\nu }^{\mu }\; \delta ^{(2)}(x-y)\; , &  & \; \; \forall x,y\in \Sigma ^{*}.\label{PoissonbracketAB}
\end{eqnarray}
 For any one-form $u=\sum _{r\in I_{\mathbf{M}}}u_{x^{r}}dx^{r}$
and vector field $v=\sum _{r\in I_{\mathbf{M}}}v^{x^{r}}\partial _{x^{r}}$
with value in the Lie algebra $\mathfrak{g},$ we will define $B(u)\equiv \int \! \! \! \! \! \int _{\Sigma ^{*}}d^{2}x\; <B^{\mu }|u_{\mu }>$and
$A(v)=\int \! \! \! \! \! \int _{\Sigma ^{*}}d^{2}x\; <A_{\mu }|v^{\mu }>\; .$ 

Let us examine the constraints derived from the action (\ref{actionwitten}).
It is immediate to show that the set of primary constraints is given
by: \begin{eqnarray}
C_{(n)} & \equiv  & M_{(n)}^{-1}P_{(n)}M_{(n)}+\frac{2l_{c}}{\vartheta }\chi _{n}\thickapprox 0,\; \; \; \forall n\in \{1,\cdots N\}.\label{constraintsCn}\\
\Xi (u) & \equiv  & B(u)-\frac{2l_{c}}{\vartheta }A(\epsilon ^{t\mu \nu }u_{\mu }^{(L)}\partial _{\nu })\thickapprox 0,\; \; \; \forall u=\sum _{r\in I_{\mathbf{M}}}u_{x^{r}}dx^{r}.\label{ConstraintsGamma}
\end{eqnarray}
 We will call respectively $\mathcal{C}_{P}$ and $\mathcal{C}_{\Xi }$
these sets of constraints. We deduce, from the action, the total Hamiltonian
: \begin{eqnarray}
H_{W}^{tot}[A,B;M,P;\mu ,\rho ] & = & \frac{-2l_{c}}{\vartheta }\left(\frac{1}{2}\int \! \! \! \! \! \int _{\Sigma ^{*}}d^{2}x\, \epsilon ^{ij}<A_{t}|F_{ij}(A)>+\int _{s_{t}}\! \! <\bar{A}_{t}|\bar{A}_{\phi }-\breve{X}>\right)\nonumber \\
 &  & +\sum _{n=1}^{N}<\mu _{(n)}|C_{(n)}>+\Xi (\rho )\; .\label{hamiltoniantotalwitten}
\end{eqnarray}
where we have introduced the Lagrange multipliers $(\mu _{(n)})_{n=1,\cdots ,N}\in \mathfrak{g}^{\otimes N}$,
and $\rho $ a one-form with value in $\mathfrak{g}$. Conservation
of the constraints (\ref{constraintsCn}) under time evolution imposes
the following conditions: \begin{eqnarray}
0\thickapprox \frac{dC_{(n)}}{dt} & =
\left\{ C_{(n)},H_{W}^{tot}\right\}  & \thickapprox \frac{2l_{c}}{\vartheta }[\mu _{(n)}-M_{(n)}^{-1}(A_{t})_{(n)}^{av.}M_{(n)},\chi _{n}]\; .\label{fixationofmu}
\end{eqnarray}
 As a result, the equations (\ref{fixationofmu}) do not impose any
secondary constraint. Indeed, using the projector introduced in (\ref{projector}),
we can show that Lagrange multipliers $\mu _{(n)}$ are fixed by:
\begin{eqnarray}
\mu _{(n)} & = & \mu _{(n)\mathfrak{h}}-\left[\kappa _{n},[M_{(n)}^{-1}(A_{t})_{(n)}^{av.}M_{(n)},\chi _{n}]\right]
\; ,\label{restrictionmu}
\end{eqnarray}
 where $(\mu _{(n)\mathfrak{h}})$ is a freely chosen element of the
Cartan subalgebra $\mathfrak{h}$ . We will prefer to introduce a
family $\upsilon _{(n)}$ such that \[
\mu _{(n)}=\upsilon _{(n)}+M_{(n)}^{-1}(A_{t})_{(n)}^{av.}M_{(n)}.\]
 The restriction (\ref{restrictionmu}) is then easily rewritten as
\begin{equation}
\upsilon _{(n)}\in \mathfrak{h}.\label{restrictionv}\end{equation}

Let us now examine the requirement that constraints (\ref{ConstraintsGamma})
must be preserved in time. In that case, we obtain that: \begin{eqnarray}
0\thickapprox \frac{d\Xi (u)}{dt} & = & \frac{-2l_{c}}{\vartheta }\int \! \! \! \! \! \int _{\Sigma ^{*}}d^{2}x\; \epsilon ^{ij}<u_{i}|\rho _{j}-D_{j}A_{t}>\label{fixationofrhoi0}\\
 & + & \frac{2l_{c}}{\vartheta }\left(\frac{1}{2}\int \! \! \! \! \! \int _{\Sigma ^{*}}d^{2}x\: \epsilon ^{ij}<u_{t}|F_{ij}(A)>+\int _{s_{t}}<\bar{u}_{t}|\bar{A}_{\phi }-\breve{X}>\right).\label{fixationofrhoi0}\nonumber 
\end{eqnarray}
It is straightforward to see that equation (\ref{fixationofrhoi0})
imposes a restriction on the Lagrange multiplier $\rho $: \[
\rho _{i}=D_{i}^{A}A_{t},\: i\in I_{\Sigma }.\]
 However, we obtain no restriction on the component $\rho _{t}$ of
the Lagrange multiplier. Furthermore, (\ref{fixationofrhoi0}) also
imposes the following secondary constraint: \begin{eqnarray}
\Omega (v) & \equiv  & \frac{1}{2}\int \! \! \! \! \! \int _{\Sigma ^{*}}d^{2}x\; \epsilon ^{ij}<v|F_{ij}(A)>+\int _{s_{t}}<\bar{v}|\bar{A}_{\phi }-\breve{X}>\thickapprox 0\; ,\label{constraintOmega}\\
 &  & \; v\in C^{\infty }(\Sigma ^{*},\mathfrak{g}).\nonumber 
\end{eqnarray}
 We will denote by $\mathcal{C}_{\Omega }$ this set of constraints.
The previous constraints can be written as bulk constraints and boundary
constraints:\begin{eqnarray}
\bar{A}_{\phi }=\breve{X} &  & \label{fixationontheboundary}\\
F_{ij}(x)=0 &  & \forall x\in \Sigma _{t}^{*}.\label{fixationinthebulk}
\end{eqnarray}

The requirement that the constraint $\Omega (v)$ must be preserved
in time implies tertiary constraints. Indeed, using the restrictions
of the Lagrange multipliers $\mu $ (\ref{restrictionmu}), we can
show that time evolution of $\Omega (v)$ is given by: \begin{eqnarray}
\frac{d\Omega (v)}{dt} & = & \Omega ([A_{t},v])+\int _{s_{t}}<\bar{v}_{t}|\partial _{\phi }\bar{A}_{t}+[\breve{X},\bar{A}_{t}-\breve{A}_{t}]>\label{timeevolutionOmega}
\end{eqnarray}

The set of tertiary constraints, denoted $\mathcal{C}_{\Upsilon }$,
is\begin{eqnarray}
\Upsilon (\bar{w})\equiv \int _{s_{t}}<\bar{w}|\partial _{\phi }\bar{A}_{t}+[\breve{X},\bar{A}_{t}-\breve{A}_{t}]>
\thickapprox 0\; . &  & \bar{w}\in C^{\infty }(s_{t},\mathfrak{g})\label{Upsilon}
\end{eqnarray}
The study of the operator $K_{\phi }^{X}:C^{\infty }(\mathcal{B},\mathfrak{g})\longrightarrow C^{\infty }(\mathcal{B},\mathfrak{g}),\: \bar{u}\mapsto \partial _{\phi }\bar{u}+[\breve{X},\bar{u}-\breve{u}]$
is detailed in the appendix \ref{sub:Operator}. As a result, these
constraints imply that \begin{eqnarray}
\bar{A}_{t} & = & \breve{A}_{t}\label{restrictionAt}
\end{eqnarray}

Conservation of the constraint (\ref{Upsilon}) under time evolution produces
no more constraint, but constraints the Lagrange multiplier $\rho _{t}$.
Using the previous constraint, we can simplify this fixation to obtain\begin{eqnarray}
\bar{\rho }_{t} & = & \breve{\rho }_{t}\label{restrictionrhot}
\end{eqnarray}
The Dirac process ends and we are left with the set of constraints
$\mathcal{C}_{0}=\mathcal{C}_{P}\cup \mathcal{C}_{\Xi }\cup \mathcal{C}_{\Omega }\cup \mathcal{C}_{\Upsilon }$.
In order to study the constrained surface, it will be useful to replace
the constraints $\Omega (v)$and $\Upsilon (\bar{w})$ by an equivalent
set of constraints. Hence, denoting \begin{eqnarray*}
\tilde{X}(x_{(n)}(t,\phi )) & = & \breve{X}(x_{(n)}(t,\phi ))+\frac{\vartheta }{4\pi l_{c}}M_{(n)}(t)(C_{n})_{\mathfrak{h}}M_{(n)}^{-1}(t)\\
 & = & \frac{\vartheta }{4\pi l_{c}}M_{(n)}(M_{(n)}^{-1}P_{(n)}M_{(n)})_{\mathfrak{h}}M_{(n)}^{-1}
\end{eqnarray*}
 we define

\begin{eqnarray*}
\tilde{\Omega }(v) & \equiv  & \frac{1}{2}\int \! \! \! \! \! \int _{\Sigma ^{*}}d^{2}x\; \epsilon ^{ij}<v|F_{ij}(A)>+\int _{s_{t}}<\bar{v}|\bar{A}_{\phi }-\tilde{X}>\\
\tilde{\Upsilon }(\bar{w}) & \equiv  & \int _{s_{t}}<\bar{w}|K_{\phi }^{\tilde{X}}\bar{A}_{t}>
\end{eqnarray*}
 and we will call $\tilde{\mathcal{C}}_{\Omega }$ and $\tilde{\mathcal{C}}_{\Upsilon }$
these new sets of constraints. 

The Poisson bracket between constraints of the subset $\mathcal{C}_{P}$
is given by: \begin{eqnarray}
\{C_{(n)1},C_{(l)2}\} & = & -[\frac{2l_{c}}{\vartheta }(\chi _{n})_{2}-C_{(n)2},t_{12}]\; \delta _{n,l}\; .
\end{eqnarray}

and the computation of the Dirac bracket has already been done in
the free particle case. As a result, Poisson brackets involving $P_{(n)}$
are unchanged and the Poisson bracket between the elements $M_{(n)}$ is modified
into the following quadratic bracket: \begin{eqnarray}
\{M_{(n)1},M_{(l)2}\}_{1} & = & M_{(n)1}M_{(n)2}r_{12}^{(0)}(\tilde{b}{}_{+}^{(n)},\tilde{b}{}_{-}^{(n)})\; \delta _{n,l}\; \label{DiracbracketM}
\end{eqnarray}
 where $r_{12}$ is the classical dynamical r-matrix introduced in
the previous section and $\chi ^{\tilde{b}{}_{+}^{(n)},\tilde{b}{}_{-}^{(n)}}=\chi ^{b{}_{+}^{(n)},b{}_{-}^{(n)}}-(C_{(n)})_{\mathfrak{h}}=-(M_{(n)}^{-1}P_{(n)}M_{(n)})_{\mathfrak{h}}$
. The subset of first class constraints $\mathcal{C}_{P}^{f}$ corresponds 
to the Cartan part of the constraints $C_{(n)}$.

After this partial reduction, we are left with a Poisson bracket $\{\cdot ,\cdot \}_{1}$
and the following set of constraints $\mathcal{C}_{1}=\mathcal{C}_{P}^{f}\cup \mathcal{C}_{\Xi }\cup \tilde{\mathcal{C}}_{\Omega }\cup \tilde{\mathcal{C}}_{\Upsilon }$.
$\mathcal{C}_{P}^{f}$ are also first class in this complete set. 

The subset $\mathcal{C}_{\Gamma }$ is treated as in vacuo. The subset
of first class (resp. second class) constraints is given by $\mathcal{C}_{\Xi }^{0}=\{\Xi (u)\mid u=u_{t}dt\}$(resp.
$\mathcal{C}_{\Xi }^{s}=\{\Xi (u)\mid u=\sum _{l=1,2}u_{x^{l}}dx^{l}\}$).
The Poisson brackets involving particles degrees of freedom are not
modified by this reduction because $P_{(n)}$ and $M_{(n)}$, for
$n\in \{1,\cdots ,N\}$, commute with the second class constraints
belonging to $\mathcal{C}_{\Xi }$. The Dirac bracket between the
gauge field components is modified into the bracket: \begin{eqnarray}
\{A(u),A(v)\}_{2} & = & \frac{-\vartheta }{2l_{c}}\int \! \! \! \! \! \int _{\Sigma ^{*}}d^{2}x\; \epsilon _{ij}<u^{i},v^{j}>\; .
\end{eqnarray}

and the momenta corresponding to the spatial part of the connection
can be explicitely eliminated by the constraints $\mathcal{C}_{\Xi }^{s}$.
Finally, we have to consider the set of constraints $\mathcal{C}_{2}=\mathcal{C}_{P}^{f}\cup \mathcal{C}_{\Gamma }^{0}\cup \tilde{\mathcal{C}}_{\Omega }\cup \tilde{\mathcal{C}}_{\Upsilon }$
and the Poisson bracket is given by $\{\cdot ,\cdot \}_{2}$. 

The non-vanishing Poisson brackets between elements of $\mathcal{C}_{2}$
are given by: \begin{eqnarray}
\{\tilde{\Omega }(u),\tilde{\Omega }(v)\}_{2}=\frac{\vartheta }{2l_{c}}(\tilde{\Omega }([u,v])-\int _{s_{t}}<\bar{u}|K_{\phi }^{\tilde{X}}\bar{v}>) &  & \; u,v\in C^{\infty }(\Sigma ^{*},\mathfrak{g})\label{PoissonbracketOmegaOmega}\\
\{\Xi (udt),\tilde{\Upsilon }(\bar{w})\}_{2}=-\int _{s_{t}}<\bar{w}|K_{\phi }^{\tilde{X}}\bar{u}> &  & \; ,\; \bar{w}\in C^{\infty }(\mathcal{B},\mathfrak{g})\; .\label{Poissonbracketgammaupsilon}
\end{eqnarray}
 We have to distinguish first class from second class constraints.
It is immediate to see that, given $\bar{w}\in C^{\infty }(\mathcal{B},\mathfrak{g})$,
$\tilde{\Upsilon }(\bar{w})$ would be first class if and only if
$\bar{w}=\breve{w}$, but, in this case $\tilde{\Upsilon }(\bar{w})=0$,
hence, $\tilde{\mathcal{C}}_{\Upsilon }$ is a set of second class
constraints. From the Poisson brackets (\ref{Poissonbracketgammaupsilon}),
we see that, the constraint $\Xi (u_{t}dt)$ is first class if and
only if $u_{t}\in \mathbb{g},$ and we will denote $\Xi _{0}(u)\equiv \Xi (udt),\forall u\in \mathbb{G}.$
There is no canonical way to characterize the set of second class
constraints. However, the Dirac bracket does not depend on this choice.
This procedure is achieved by a choice of a space of functions $\check{\mathbb{g}}$
such that $C^{\infty }(\Sigma ^{*},\mathfrak{g})=\mathbb{g}\oplus \check{\mathbb{g}},$
and $\forall u\in \check{\mathbb{g}},\: \breve{u}=0$. The detail
of this space in the bulk is not important and we can always choose
$\check{\mathbb{g}}$ such that the support on $\Sigma ^{*}$ of its
elements is a neighbourhood of the boundary chosen as small as necessary, this point will be detailed in
the sequel. We will denote $\tilde{\mathcal{C}}\, _{\Xi }^{f}$ (resp.$\tilde{\mathcal{C}}\, _{\Xi }^{s}$)
the set of first (resp. second) class constraints $\Xi (u_{t}dt)$
with $u_{t}\in \mathbb{g}$ (resp. $u_{t}\in \check{\mathbb{g}}$).
All the way, the implementation of $\tilde{\mathcal{C}}\, _{\Xi }^{s}$
imposes \begin{equation}
\bar{B}{}^{t}=\breve{B}{}^{t}\label{restrictionBt}\end{equation}
From Poisson brackets (\ref{PoissonbracketOmegaOmega}), we see that,
given $u\in C^{\infty }(\Sigma ^{*},\mathfrak{g})$, the constraint
$\tilde{\Omega }(u)$ is first class if and only if $\bar{u}$=$\breve{u}$.
Then, as before, we will denote $\tilde{\mathcal{C}}\, _{\Omega }^{f}$
(resp.$\tilde{\mathcal{C}}\, _{\Omega }^{s}$) the set of first (resp.
second) class constraints $\tilde{\Omega }(u)$ with $u\in \mathbb{g}$
(resp. $u\in \check{\mathbb{g}}$).

Finally, the set of first class constraints $\mathcal{C}_{2}^{f}$
is given by $\mathcal{C}_{2}^{f}=\mathcal{C}_{P}^{f}\cup \mathcal{C}_{\Xi }^{f}
\cup \tilde{\mathcal{C}}{}_{\Omega }^{f}$.
In order to compute the Dirac bracket, it will be useful to introduce
the antisymmetric bilinear form $K$ as follows: \begin{eqnarray}
\check{\mathbb{g}}\, ^{\times 2} & \longrightarrow  & \mathbb{C}\nonumber \\
(u,v) & \longmapsto  & \tilde{K}(u,v)=\int _{s_{t}}<\bar{u}|K_{\phi }^{\tilde{X}}\bar{v}>.
\end{eqnarray}
 Given an element $u\in \check{\mathbb{g}}$, $K(u,\cdot ):\check{\mathbb{g}}\longrightarrow \mathbb{C}$
is invertible and we will denote by $K^{-1}(\cdot ,u)$ its inverse,
i.e. \begin{eqnarray}
\int \! \! \! \! \! \int _{\check{\mathbb{g}}\times \check{\mathbb{g}}}[\mathcal{D}w]\tilde{K}(u,w)\tilde{K}{}^{-1}(w,v)=\delta _{\check{\mathbb{g}}}(u-v)\; , &  & \forall \; u,v\in \check{\mathbb{g}}\; .
\end{eqnarray}

We have to notice that the strong expressions of Poisson brackets
between constraints of the sets $\tilde{\mathcal{C}}\, _{\Xi }^{s}$
and $\tilde{\mathcal{C}}_{\Upsilon }$ are exactly given by the operator
$K,$ however the expressions of Poisson brackets between constraints
of the set $\tilde{\mathcal{C}}\, _{\Omega }^{s}$ are only weakly
given by $K,$ the problematic term being a combination of constraints
of the sets $\tilde{\mathcal{C}}\, _{\Omega }^{f}$ and $\tilde{\mathcal{C}}\, _{\Omega }^{s}.$
It is however a fundamental property of Dirac bracket that we can
compute it using only the weak expression of the Dirac matrix as soon
as we restrict ourself to functions of the phase space which are Poisson commuting
to elements of $\tilde{\mathcal{C}}\, _{\Omega }^{f}.$

As a result, the final Dirac bracket between two functions $f$ and
$g$ of dynamical variables, which are Poisson commuting to elements
of $\tilde{\mathcal{C}}\, _{\Omega }^{f}$, is formally given by:
\begin{eqnarray}
\{f,g\}_{D} & = & \{f,g\}_{2}+\frac{2l_{c}}{\vartheta }\int \! \! \! \! \! \int _{\check{\mathbb{g}}\times \check{\mathbb{g}}}[\mathcal{D}u][\mathcal{D}v]\{f,\tilde{\Omega }(u)\}_{2}\tilde{K}{}^{-1}(u,v)\{\tilde{\Omega }(v),g\}_{2}\label{definitiveDiracbracket}\\
 &  & +\int \! \! \! \! \! \int _{\check{\mathbb{g}}\times \check{\mathbb{g}}}[\mathcal{D}u][\mathcal{D}w]\left(\{f,\tilde{\Upsilon }(\bar{w})\}_{2}\tilde{K}{}^{-1}(w,u)\{\Xi _{0}(u),g\}_{2}-f\leftrightarrow g\right)\; .\nonumber 
\end{eqnarray}

These results have to be compared with results of \cite{P2}. In the following we will go beyond 
this formal result for particular functionals of dynamical variables. 

Let us now study the gauge symmetries of our dynamical problem. Now,
the total action of our problem is given by \begin{eqnarray}
S_{W}^{tot} & = & \int _{t_{1}}^{t_{2}}dt\: \left(\int \! \! \! \! \! \int _{\Sigma *}d^{2}x\; \frac{-2l_{c}}{\vartheta }\epsilon ^{ij}<A_{i}^{(T)}\mid \partial _{t}A_{j}^{(L)}>+<B^{t}|\partial _{t}A_{t}>\right.\nonumber \\
 &  & \left.-\sum _{n=1}^{N}<P_{(n)}|\frac{dM_{(n)}}{dt}M_{(n)}^{-1}>-H_{W}^{tot}\right)\label{actiontotalewitten}\\
H_{W}^{tot} & = & \left(\sum _{n=1}^{N}<P_{(n)}|\bar{A}_{t(n)}>+\sum _{n=1}^{N}<\upsilon _{(n)}|C_{(n)}>+\int \! \! \! \! \! \int _{\Sigma ^{*}}d^{2}x\; <B^{t}|\rho _{t}>\right.\nonumber \\
 &  & \left.-\frac{l_{c}}{\vartheta }\int \! \! \! \! \! \int _{\Sigma ^{*}}d^{2}x\, \epsilon ^{ij}<A_{t}|F_{ij}(A)>-\frac{2l_{c}}{\vartheta }\int _{s_{t}}\! \! <\bar{A}_{t}|\bar{A}_{\phi }>\right)\; .\label{hamiltontotalwitten}
\end{eqnarray}

where, the dynamical variables $(A_{i})_{i\in I_{\Sigma }},A_{t},B^{t};M,P;\upsilon ,\rho _{t}$
are restricted by (\ref{restrictionv})(\ref{restrictionAt})(\ref{restrictionrhot})(\ref{restrictionBt}).
Given a family $(\varsigma _{(n)})_{n=1...N}\in \mathfrak{h}^{\times N}$,
and a function $\Gamma \in \mathbb{g}$, the following gauge transformations
of the dynamical variables: \begin{eqnarray}
\delta _{W}[\varsigma ,\Gamma ]M_{(n)}=-M_{(n)}\varsigma _{(n)}-\overline{\Gamma }_{(n)}M_{(n)} & \: ,\:  & \delta _{W}[\varsigma ,\Gamma ]P_{(n)}=[P_{(n)},\overline{\Gamma }_{(n)}]\label{gaugewitten}\\
\delta _{W}[\varsigma ,\Gamma ]A_{\mu }=D_{\mu }\Gamma  & \: ,\:  & \delta _{W}[\varsigma ,\Gamma ]B^{t}=0\nonumber \\
\delta _{W}[\varsigma ,\Gamma ]\rho _{t}=-\partial _{t}(D_{t}\Gamma ) & \: ,\:  & \delta _{W}[\varsigma ,\Gamma ]\upsilon _{(n)}=\partial _{t}\varsigma _{(n)}\nonumber 
\end{eqnarray}

transform the total action as follows \begin{eqnarray*}
\delta _{W}[\varsigma ,\Gamma ]S_{W}^{tot} & = & \frac{l_{c}}{\vartheta }\int \! \! \! \! \! \int _{\Sigma _{t_{2}}-\Sigma _{t_{1}}}\! \! \! d^{2}x\, \epsilon ^{ij}<2\partial _{i}A_{j}^{(L)}+[A_{i}^{(L)},A_{j}^{(L)}]-[A_{i}^{(T)},A_{j}^{(T)}],\Gamma ^{(T)}>\\
 &  & +\frac{2l_{c}}{\vartheta }\int _{s_{t_{2}}-s_{t_{1}}}\! \! <\Gamma ^{(T)},A_{\phi }^{(L)}>+\sum _{n=1}^{N}[<\varsigma _{(n)}|\chi ^{b_{+}^{(n)},b_{-}^{(n)}}>]_{t_{1}}^{t_{2}}.
\end{eqnarray*}

According to the conclusions obtained for the same problem in the
pure gravity case and the free particle case, we will not try to add
boundary terms in order to achieve the complete covariance under the
gauge symmetries. Our choice, in the purely massive case $b_{-}^{(n)}=0$,
will be to consider observables invariant under the symmetries $\delta _{W}[\varsigma _{(n)}^{(L)},\Gamma ]$,
and to fix a proper-time gauge for the symmetries of the form $\delta _{W}[\varsigma _{(n)}^{(T)},0].$ 

As a summary, the phase space of Chern-Simons deSitter gravity is
given by the space of dynamical variables $A_{\mu },B_{t},M_{(n)},P_{(n)}$
constrained by (\ref{fixationontheboundary})(\ref{fixationinthebulk})(\ref{restrictionAt})(\ref{constraintsCn})
and the Lagrange multipliers of first class constraints $\rho _{t},\upsilon _{(n)}$
restricted by (\ref{restrictionrhot})(\ref{restrictionv}), this
space having to be moded out by large gauge transformations denoted
$\Delta _{W}[\lambda ,g]$ obtained by exponentiation from (\ref{gaugewitten}),
and the Poisson algebra of these dynamical variables being given by
the Dirac bracket (\ref{definitiveDiracbracket}). The hamiltonian
is given by (\ref{hamiltoniantotalwitten}), and the dynamical equations
are then \begin{eqnarray*}
\dot{M}_{(n)}=-\bar{A}_{t(n)}M_{(n)}-M_{(n)}\upsilon _{(n)} &  & \dot{P}_{(n)}=[P_{(n)},\bar{A}_{t(n)}]\\
\dot{A}_{i}=D_{i}A_{t} &  & \dot{A}_{t}=\rho _{t}
\end{eqnarray*}

\subsubsection{\emph{Overview on the reduced phase space} }

Following the participant observer point of view, we do not want to
enter into details of any gauge fixing concerning the symmetry $\delta _{G}.$
The only gauge fixing, required for our purpose, concerns the reparametrization
of the worldlines. As in the case of the free particle we will restrict
ourself to the purely massive case, $b_{-(n)}=0$ , i.e. $\chi _{n}=\chi _{n}^{(L)}.$
And we will require the gauge condition\[
<\dot{\upsilon }_{(n)}\mid \xi _{L,0}>=0\]

which, combined with the restrictions $\varsigma _{(n)}^{(T)}\mid _{\Sigma _{t_{1,2}}}=0$
imposes $\varsigma _{(n)}^{(T)}(t)=0,\forall t.$ 

The aim of this subsection is to analyse the content of the reduced
phase space of Chern-Simons deSitter gravity. Let us for convenience
take $\Sigma $ to be a $2-$sphere, and we will consider that
the particles are indexed by an element of $\mathbb{Z}/N\mathbb{Z}$
rather than an integer from $1$ to $N.$  Let us distinguish
$N$ continuous families ($t$ and $\eta $ are the continuous parameters)
of curves $S_{n}^{(\eta )}(t),\, n\in \left\{ 1,...,N\right\} ,0\leq \eta \leq \pi ,\, t\in [t_{1},t_{2}]$
chosen to be such that \begin{eqnarray*}
\left(S_{n}^{(\eta )}(t)\in \Sigma _{t}^{*}\right), &  & \left(S_{n}^{(\eta )}(t)\cap S_{n'}^{(\eta ')}(t')\neq \textrm{Ø}\: \Rightarrow \: \eta =\eta ',n=n',t=t'\right)\\
d(S_{n}^{(\eta )}(t))=x_{(n)}(t,\eta ), &  & e(S_{n}^{(\eta )}(t))=x_{(n+1)}(t,2\pi -\eta ),
\end{eqnarray*}

and, if we denote \begin{eqnarray*}
E_{n}^{(+)(\eta ,\eta ')}(t) & \mathrm{the}\: \mathrm{curve} & \theta \in [0,\left|\eta -\eta '\right|]\mapsto x_{(n)}(t,\mathrm{min}(\eta ,\eta ')+\theta )\\
E_{n}^{(-)(\eta ,\eta ')}(t) &  & \theta \in [0,\left|\eta -\eta '\right|]\mapsto x_{(n)}(t,2\pi -\mathrm{max}(\eta ,\eta ')+\theta )\\
F_{n}^{(+)(\eta ,\eta ')}(t) &  & \theta \in [0,2\pi -2\mathrm{max}(\eta ,\eta ')]\mapsto x_{(n)}(t,\mathrm{max}(\eta ,\eta ')+\theta )\\
F_{n}^{(-)(\eta ,\eta ')}(t) &  & \theta \in [0,2\mathrm{min}(\eta ,\eta ')]\mapsto x_{(n)}(t,2\pi -\mathrm{min}(\eta ,\eta ')+\theta )
\end{eqnarray*}
 we require also that, if $\eta <\eta '$ the curve $S_{n}^{(\eta )}(t)\circ E_{n+1}^{(-)(\eta ,\eta ')}(t)^{-1}\circ S_{n+1}^{(\eta ')}(t)^{-1}\circ E_{n}^{(+)(\eta ,\eta ')}(t)$
be a contractible curve in $\Sigma _{t}^{*}$ surrounding a connected
simply connected open subset of $\Sigma _{t}^{*}$ which will be denoted
$\mathcal{Z}_{n,n+1}^{(\eta )}(t).$ 
\begin{figure}
\centering
\includegraphics[scale=0.5]{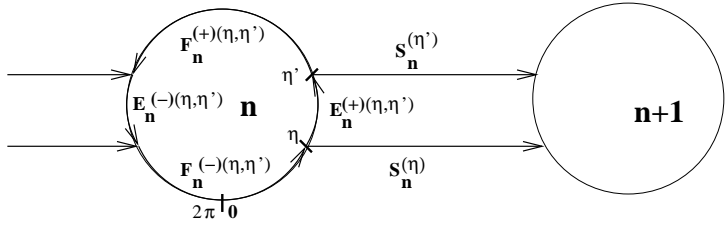}
\caption{notations}
\end{figure}

In the following, $\eta $ is fixed and we impose $0<\eta <\frac{\pi }{2}.$
We will denote $N_{t}^{(+,\eta )}$ the connected, simply connected,
open subset of $\Sigma _{t}^{*}$ which boundary, followed in the counterclockwise
sense, is the curve \[
\bigcirc _{i=1}^{n}\left(S_{i}^{(\eta )}(t)E_{i+1}^{(-)(\pi -\eta ,\eta )}(t)^{-1}F_{i+1}^{(+)(\pi -\eta ,\eta )}(t)^{-1}E_{i+1}^{(+)(\pi -\eta ,\eta )}(t)^{-1}\right).\]
 We will denote $N_{t}^{(-,\eta )}$ the connected, simply connected,
open subset of $\Sigma _{t}^{*}$which boundary, followed in the clockwise
sense, is the curve \[
\bigcirc _{i=1}^{n}\left(S_{i}^{(\pi -\eta )}(t)E_{i+1}^{(-)(\pi -\eta ,\eta )}(t)F_{i+1}^{(-)(\pi -\eta ,\eta )}(t)E_{i+1}^{(+)(\pi -\eta ,\eta )}(t)\right).\]
 The intersection of $N_{t}^{(+,\eta )}$and $N_{t}^{(-,\eta )}$
is the disjoint union $\bigcup _{i=1}^{N}\mathcal{Z}_{i,i+1}^{(\eta )}(t),$
and $\Sigma _{t}^{*}=N_{t}^{(+,\eta )}\cup N_{t}^{(-,\eta )}.$ The previous notations are summarized 
in the figures $1$ and $2$.
\begin{figure}
\centering
\includegraphics[scale=0.5]{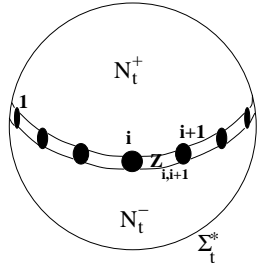}
\caption{decomposition of $\Sigma _{t}^{*}$ in two connected simply connected closed subsets}
\end{figure}

Endly, we will denote $\mathcal{N}^{(\pm ,\eta )}=\bigcup _{t\in [t_{1},t_{2}]}N_{t}^{(\pm ,\eta )}$
and $Z_{i,i+1}^{(\eta )}=\bigcup _{t\in [t_{1},t_{2}]}\mathcal{Z}_{i,i+1}^{(\eta )}(t),$
and $\mathcal{S}_{n}^{(\eta )}=\cup _{t\in [t_{1},t_{2}]}S_{n}^{(\eta )}(t).$
The zero curvature equation (\ref{zerocurvature}) can be solved on
each subset $\mathcal{N}^{(\pm )}$in terms of smooth mappings \begin{eqnarray}
k_{(\pm )}:\mathcal{N}^{(\pm ,\eta )}\rightarrow G &  & A=k_{(\pm )}^{-1}dk_{(\pm )}.\label{solutiongauge}
\end{eqnarray}
 If we denote $u_{i,i+1}$ the mappings defined on $Z_{i,i+1}^{(\eta )}$
by $k_{(+)}k_{(-)}^{-1},$ the univaluation of the connection imposes
$du_{i,i+1}=0$ and then $u_{i,i+1}$ is a constant. The boundary
conditions imposes restrictions on the gauge. Indeed, (\ref{fixationontheboundary})
imposes $\bar{k}_{(\pm )}(x_{(n)}(t,\varphi +\frac{3\pi }{4}\pm \frac{\pi }{2}))=K_{(\pm ),n}(t)e^{\chi _{n}\varphi }M_{(n)}^{-1}(t),$
for $\varphi \in [0,\frac{3\pi }{2}]$ and (\ref{restrictionAt})
imposes $(K_{(\pm ),n}^{-1}{}\partial _{t}K_{(\pm ),n})=w_{(\pm ),n}(t)\in \mathfrak{h},$
and then $K_{(\pm ),n}(t)=K_{(\pm ),n}(t_{1})e^{\int _{t_{1}}^{t}d\tau \, w_{(\pm ),n}(\tau )}.$
The dynamical equations impose the relations $w_{(\pm ),n}=\upsilon _{(n)}.$ 

The gauge group $\mathbb{G}$ acts on $A$ as $\Delta _{W}[\lambda ,g]A=g^{-1}Ag+g^{-1}dg$,
hence, it acts on $k_{(\pm )}$ as $\Delta _{W}[\lambda ,g]k_{(\pm )}=k_{(\pm )}g.$
$M_{(n)}$transforms as $\Delta _{W}[\lambda ,g]M_{(n)}=\bar{g}{}_{(n)}^{-1}M_{(n)}\lambda _{(n)},$
and on $e^{\int _{t_{1}}^{t}\upsilon _{(n)}}$ as $\Delta _{W}[\lambda ,g]e^{\int _{t_{1}}^{t}\upsilon _{(n)}}=e^{\int _{t_{1}}^{t}\upsilon _{(n)}}\lambda _{(n)}(t)\lambda _{(n)}^{-1}(t_{1}),$
however we have already fixed partially the gauge symmetry associated
to the reparametrization of the worldline, then $\lambda _{(n)}=\lambda _{(n)}^{(L)}\in H^{(L)}.$
Endly, $\mathbb{G}$ acts on $K_{(\pm ),n}(t)$ as $\Delta _{W}[\lambda ,g]K_{(\pm ),n}(t)=K_{(\pm ),n}(t)\lambda _{(n)}^{(L)}(t).$
We are then led to introduce \begin{eqnarray*}
X_{(\pm ),n}(t) & \equiv  & \bar{k}_{(\pm )}(x_{(n)}(t,\frac{\pi }{2}))M_{(n)}(t)e^{-\int _{t_{1}}^{t}d\tau \, \upsilon _{(n)}(\tau )},
\end{eqnarray*}

which owns the properties\begin{eqnarray*}
\partial _{t}X_{(\pm ),n}(t)=0 &  & \Delta _{W}[\lambda ,g]X_{(\pm ),n}(t)=X_{(\pm ),n}(t)\lambda _{(n)}^{(L)}(t_{1}).
\end{eqnarray*}

It has to be noticed that the redundant parametrization of the connection
induces a supplementary rigid symmetry $k_{(\pm )}\rightarrow a_{(\pm )}k_{(\pm )}$
associated to a pair of fixed elements of the Lie group, and this
freedom will be used to impose \begin{equation}
X_{(+),1}(t_{1})=X_{(-),1}(t_{1})\in H^{(L)}.\label{1atrest}\end{equation}

Now, we want to identify functions on the phase space encoding the
gauge invariant part of the phase space and then parametrizing the
reduced phase space, in a way which will allow us to describe the
Poisson brackets in these terms. To this aim, we will introduce some
quantities which will be associated to a given point in the phase
space. Let us define \begin{eqnarray*}
D_{(\pm ),n}(t,t') & \equiv  & e^{(t_{1}-t)\upsilon _{(1)}^{(T)}}D_{(\pm ),n}e^{(t'-t_{1})\upsilon _{(n)}^{(T)}},\\
D_{(\pm ),n} & \equiv  & X_{(\pm ),1}^{-1}(t_{1})X_{(\pm ),n}(t_{1})
\end{eqnarray*}

Fundamental properties of these datas are\begin{eqnarray*}
\partial _{t'}D_{(\pm ),n}(t,t')=D_{(\pm ),n}(t,t')\upsilon _{(n)}^{(T)} &  & \Delta _{W}[\lambda ,g]D_{(\pm ),n}(t,t')=\lambda _{(1)}^{(L)}(t_{1})^{-1}D_{(\pm ),n}(t,t')\lambda _{(n)}^{(L)}(t_{1})
\end{eqnarray*}

Let us introduce an equivalence relation on the extended phase space.
Let be given two points $(A,M,\upsilon ),(A',M',\upsilon ')$ in this
space, we choose mappings $k_{(\pm )},k'_{(\pm )}$ respectively associated
to $A,A',$ and build the corresponding quantities $D_{(\pm ),n},D'_{(\pm ),n}.$ 

We will say $(A,M,\upsilon )\sim (A',M',\upsilon ')$ if and only
if \begin{eqnarray}
\forall n=1...N, &  & \upsilon _{(n)}^{(T)}=\upsilon '{}_{(n)}^{(T)}\label{requirement1}\\
\exists (\omega _{n}^{(L)})_{n=1...N}\in (H^{(L)})^{\times N}, & s.t. & D'_{(\pm ),n}=\omega _{1}^{(L)}D_{(\pm ),n}\omega _{n}^{(L)-1}.\label{requirement2}
\end{eqnarray}

The first point to notice is that, \begin{eqnarray*}
\exists (\lambda ,g)\in \mathbb{G}/\qquad \qquad  & \Rightarrow  & \qquad (A,M,\upsilon )\sim (A',M',\upsilon ').\\
\Delta _{W}[\lambda ,g](A',M',\upsilon ')=(A,M,\upsilon )\qquad  &  & 
\end{eqnarray*}
 Reciprocally, \begin{eqnarray*}
(A,M,\upsilon )\sim (A',M',\upsilon ')\qquad  & \Rightarrow  & \qquad \exists (\lambda ,g)\in \mathbb{G},\exists (\omega _{n})_{n=1...N}\in (H^{(L)})^{\times N}/\\
 &  & \qquad \Delta _{W}[\lambda ,g](A',M',\upsilon ')=(A,M\omega ,\upsilon ).
\end{eqnarray*}
 Indeed, we choose mappings $k_{(\pm )},k'_{(\pm )}$ respectively
associated to $A,A',$ and we denote $a_{(\pm )}\equiv X'_{(\pm ),1}(t_{1})\omega _{1}X_{(\pm ),1}(t_{1})^{-1}$
and $g\equiv k'{}_{(\pm )}^{-1}a_{(\pm )}k_{(\pm )}\: \mathrm{on}\: \mathcal{N}^{(\pm ,\eta )}.$
Relation (\ref{requirement2}), through $D_{(+),n}D_{(-),n}^{-1}=D'_{(+),n}D'{}_{(-),n}^{-1}$
and the univaluation of the connection implies $\forall i=1...N,\: a_{+}u_{i,i+1}a_{-}^{-1}=u'_{i,i+1}$
, and then, the mapping $g$ is univalued and smooth, and we have
$A=g^{-1}A'g+g^{-1}dg$. Relation (\ref{requirement2})(\ref{requirement1})
and dynamical equations, imply $M_{(n)}=\bar{g}{}_{n}^{-1}M'_{(n)}\omega _{(n)}e^{\int _{t_{1}}^{t}d\tau \, (\upsilon {}_{(n)}^{(L)}(\tau )-\upsilon '{}_{(n)}^{(L)}(\tau ))}.$
Then, if we denote $\lambda _{(n)}(t)\equiv e^{\int _{t_{1}}^{t}d\tau \, (\upsilon {}_{(n)}^{(L)}(\tau )-\upsilon '{}_{(n)}^{(L)}(\tau ))},$
we obtain the announced result. The ambiguity associated to $\omega $
will be completely irrelevant for the observables we will consider.
As a conclusion, $(\upsilon _{(n)}^{(T)})_{n=1...N},(D_{(\pm ),n})_{n=2...N}$
appear to be interesting variables for our purpose. However, these
variables are not independent. Indeed, we have easily \begin{eqnarray*}
u_{n,n+1} & = & \bar{k}_{(+)}(x_{(n)}(t,\frac{\pi }{2}))\bar{k}_{(-)}(x_{(n)}(t,\frac{\pi }{2}))^{-1}=X_{(+),n}X_{(-),n}^{-1}.\\
 & = & \bar{k}_{(+)}(x_{(n+1)}(t,\frac{3\pi }{2}))\bar{k}_{(-)}(x_{(n+1)}(t,\frac{3\pi }{2}))^{-1}=X_{(+),n+1}e^{2\pi \chi _{n+1}}X_{(-),n+1}^{-1}.
\end{eqnarray*}
 where we have used the formula \[
\forall n=1...N,\: k_{(\pm )}(x_{(n)}(t,\frac{3\pi }{2}))=k_{(\pm )}(x_{(n)}(t,\frac{\pi }{2}))M_{(n)}(t)e^{\pm \chi _{n}\pi }M_{(n)}(t)^{-1}.\]
 These relations allow us to express the family $\left\{ (D_{(-),n})_{n=1...N}\mid D_{(-),1}=1\right\} $
in terms of the other datas, more precisely \begin{equation}
D_{(-),i}=\left(\prod _{k=2}^{i-1}(D_{(+),k}^{-1}e^{2\pi \chi _{k}}D_{(+),k})\right)D_{(+),i}e^{2\pi \chi _{i}}.\label{D-D+}\end{equation}

Endly, the combinatorial datas defining a point in the reduced phase
space are the class of elements \begin{eqnarray}
(\upsilon _{(n)}^{(T)})_{n=1...N},(D_{(+),n})_{n=2...N} &  & \mathrm{fixed}\: \mathrm{up}\: \mathrm{to}\: \mathrm{symmetries}\nonumber \\
D_{(+),n}\mapsto \omega _{1}^{(L)}D_{(+),n}(t_{1})\omega _{n}^{(L)-1}, &  & \; (\omega _{n}^{(L)})_{n=1...N}\in (H^{(L)})^{\times N}\label{combinatorialdatas}
\end{eqnarray}

\section{The combinatorial description of $3$ dimensional gravity}

This section is devoted to the study of the participant-observer approach
and the technical tool adapted to the development of this approach,
i.e. the combinatorial description. This approach is radically different
from the usual ones abundantly developed in the litterature (see for example \cite{Sch}), closer
from a local approach or a spectator-observer approach, which are,
to our point of view, failing conceptually and technically to capture
the physical phase space of gravity in a really covariant way.

\subsection{Towards a participant-observer description\label{sub:participant-observer}}

\subsubsection{\emph{second digression: clocks and measurements of a participant
observer}}

Till now we have taken a local point of view, in the sense that, the
coordinate system is fixed, a part of the gauge freedom is freezed.
We do not want to develop a spectator-observer approach but rather
use previous observables associated to matter degrees of freedom to
build a participant-observer description of the dynamic of self-gravitating
particles in a closed universe. Two questions could be adressed to
this program: which physical information can be recovered from such
observables, and what is new in this approach which could solve the
well-known problem of observables in gravity? It is obvious that most
natural observables of gravity are based on causal processes between
different material bodies in the universe. Let us discuss some examples.
The distance between a celestial body and the earth can be measured
by measuring the time a light signal takes to run to it and to run back
to the earth. The angle between the two light signals arriving at
the same time, on the earth, is also observable. What sort of local
physical measure are we thinking about when we say ``at the time''
or ``during the time''? Two different properties of our ``clock''
seem to be required in previous experiments: capability to distinguish
moments of time and to measure intervals of time. We have to carefully
distinguish between the properties we require from the physical process
at work in our clock and the properties of the measurement process
itself. The observer can choose, as a clock, any process interacting
with him\cite{key-33}. For example, the observer on the earth is
able to measure time using processes which are not directly related
to astrophysical considerations, like oscillations in an atomic clock,
but it is also possible to use astrophysical measures by using, for
example, the varying angle between two celestial bodies to measure
the flow of time. In any case, it is necessary, in order to discuss
simultaneity of the reception/emission of the signals, that our measurement
process allocates at most one moment of time to any event. However,
requiring this property to the measurement process is not sufficient
if the sub-system constituting the clock has a periodic behaviour.
Hence, we require moreover for the measurement process to have the
capability to order moments of time. These properties being fullfiled,
the measurement process has no capability in itself to have a trace
of a priviledge flow of time. These properties of the measurement
process are directly encoded in the parametrization of the worldlines
by classes of smooth and bijective maps modulo diffeomorphisms. It
is a fundamental fact that, in the description of free-particles as
well as of self-gravitating particles, the boundary restrictions on
the gauge freedom degrees $(\upsilon _{(n)}^{(T)})_{n=1...N}$ appeal
gauge fixings compatible with the gauge fixing of the reparametrization
invariance of the worldlines of particles. Any process interacting
with the $n-$th particle and giving a varying index (a varying angle
between stars, an internal process...) can be used to gauge fix $\upsilon _{(n)}^{(T)}$
as the value of this index. Obviously, this choice of clock will have
no consequence on relative measurements as described before. Our proper-time
gauge is one of these choices.

\subsubsection{\emph{reconstruction of the geometry from the participant-observer
measurements\label{sub:reconstruction}}}

Now, let us study how to recover the natural observables (distances,
angles,velocities,...) we have just described in terms of the datas
parametrizing the reduced phase space. 

Let be given a set of datas (\ref{combinatorialdatas}) defining a
point in the reduced phase space and a representant $(A,M,\upsilon )$
of this state, chosen such that the condition (\ref{1atrest}) as
well as the partial gauge fixing\begin{equation}
M_{(n)}(t)=e^{\int _{t_{1}}^{t}d\tau \, \upsilon _{(n)}^{(L)}(\tau )},\forall n=1...N.\label{partialgaugeM}\end{equation}

be verified.

As noticed in section \ref{sub:Free-relativistic}, a \emph{smooth}
and \emph{injective} mapping $\tau $ from an open subset of $\mathbb{R}^{3}$
to $T_{o}$ being given, its image is isometric to an open subset
of $dS_{3}.$ If we dress $\tau $ by any mapping $\lambda $ from
$\mathbb{R}^{3}$ to $\mathcal{D}_{o}$ to form a mapping $k=\tau \lambda $
from an open subset of $\mathbb{R}^{3}$ to $G,$ then the connection
$A=k^{-1}dk$ gives a first order description of an open subset of
$dS_{3}.$ Reciprocally, any mapping $k$ from an open subset of $\mathbb{R}^{3}$
to $G,$ can be linked to a mapping $\tau $ from an open subset of
$\mathbb{R}^{3}$ to $T_{o}$ using (\ref{applicationCtoD})(\ref{applicationDtoT}),
however this mapping is not necessary smooth and injective and its
image can be very different from a region of $dS_{3}.$ If we ask
this mapping to be smooth and injective, by requiring the property
(\ref{non_degenerate}), we then obtain an image which is isometric
to a part of $dS_{3}.$ Then, the relation (\ref{solutiongauge})
solving the zero curvature equation for $A$ ensures, that the image
of $\mathcal{N}^{(\pm ,0)}$in $\mathcal{D}$ by the mapping $k_{(\pm )}k_{(\pm )}^{\star }o$
is isometric to a closed subset of $dS_{3}$ as soon as this mapping
is injective. The first order datas associated to the constantly curved
geometry of $dS_{3}$ are encoded in $A,$ the geometric properties
(distances, angles, ...) can then be recovered from $k_{(\pm )},$
however the value of $k_{(\pm )}$ has no gauge invariant meaning
except at the location of particles. It is then natural to introduce
the datas $\bar{k}_{(+)}(x_{(1)}(t,\frac{\pi }{2}))^{-1}\bar{k}_{(+)}(x_{(n)}(t',\frac{\pi }{2})).$
But in general, if we do not choose the previous partial gauge fixing,
and if we consider location of particles corresponding to different
time labels, we have to consider $D_{(\pm ),n}(t,t').$ We are then
able to compute distances , or the apparent angle between the $n-$th
and the $p-$th particle seen from the $1-$st particle, ..., from
these datas. For example, the distance between the location of the
$1-$st particle when its time label is $t$ and the location of the
$n-$th particle when its time label is $t'$ is given by $l_{c}^{2}tr(D_{(+),n}(t,t')D_{(+),n}(t,t')^{\star }-I)$
(compare this formula with (\ref{distance})). We could ask if the
choice of $x_{(n)}(t_{1},\frac{\pi }{2})$ rather than any $x_{(n)}(t_{1},\phi )$ in
the definition of our observables has a consequence. However, due
to boundary conditions such a change would affect $D_{(+),n}(t,t')$
only by a right action of an element of $H^{(L)}$ and the geometric
observables are not affected by this change.

A particular representative of an element of the reduced phase space
being chosen, it is possible to have a picture of the entire spacetime
geometry corresponding to this solution. Indeed, let us consider a point
in the reduced phase space by the datas (\ref{combinatorialdatas}).
The condition (\ref{partialgaugeM}) being verified, we then have
\begin{eqnarray*}
X_{(+),i}=D_{(+),i} &  & X_{(-),i}=\left(\prod _{k=2}^{i-1}(D_{(+),k}^{-1}e^{2\pi \chi _{k}}D_{(+),k})\right)D_{(+),i}e^{2\pi \chi _{i}}.\\
u_{i,i+1} & = & \left(\prod _{k=2}^{i}(D_{(+),k}^{-1}e^{-2\pi \chi _{k}}D_{(+),k})\right)
\end{eqnarray*}

and the condition (\ref{1atrest}) being imposed, we have \[
\bar{k}_{(+)}(x_{(n)}(t,\frac{\pi }{2}))=D_{(+),i}e^{(t-t_{1})\upsilon _{(n)}^{(T)}}.\]

We have then fixed $2N$ free falling trajectories on $dS_{3}$, 
$$\mathcal{T}_{n}^{(\pm )}:t\mapsto 
Q_{\bar{k}_{(\pm )}(x_{(n)}(t_{1},\frac{\pi }{2}))e^{(t'-t_{1})\upsilon _{(n)}^{(T)}}}$$ 
(the boundary $l_{n}(t)$ is mapped onto a point by the previous mapping).
The first particle plays the role of the observer, and is considered
at rest (condition (\ref{1atrest})). For any $n=1,...,N,$ the surface
$Q_{k_{(-)}(\mathcal{S}_{n}^{(0)})}$ is mapped onto $Q_{k_{(+)}(\mathcal{S}_{n}^{(0)})}$
by the isometry $Q\mapsto u_{n,n+1}Qu_{n,n+1}^{\dagger }.$ The images
by the smooth and injective mappings $\mathcal{N}^{(\pm ,0)}\rightarrow \mathcal{D},\: (x^{\mu })_{\mu \in I_{M}}\mapsto Q_{k_{(\pm )}(x)}$
are two regions of $dS_{3}$, these regions are separated by a third
one which has to be cutted out from $dS_{3}$ in order to glue them
together along their boundaries by $Q_{k_{(+)}(x)}\sim Q_{k_{(-)}(x)},\forall x\in \cup _{n=1...N}\mathcal{S}_{n}^{(0)}$
in order to obtain the geometry of the spacetime described by our
classical solution (this procedure is the usual way to build multiconical
solutions by removing regions associated to conical defects and then
glue along the cuts). 

There is a nice way, a classical solution being chosen, to compute
the datas necessary to describe the previous geometrical observables.
Indeed, if we define the holonomy $U_{1,n}(A)=\overrightarrow{Pexp}\int _{C}A$
of the connection $A$ along a curve $C$ entirely contained in $N_{t_{1}}^{(+,0)}$
going from $x_{(1)}(t_{1},\frac{\pi }{2})$to $x_{(n)}(t_{1},\frac{\pi }{2}).$
We have to notice that $e^{(t_{1}-t)\upsilon _{(1)}^{(T)}}M_{(1)}^{-1}(t_{1})U_{1,n}(A)M_{(n)}(t_{1})e^{(t'-t_{1})\upsilon _{(n)}^{(T)}}$
does not depend of the representant we have chosen. The zero-curvature
equation obeyed by the connection ensures that we do not change the
value of these elements by replacing the chosen curve $C$ by any
other curve $C'$ being such that $C\circ C'$ is a contractible curve
in $\sum _{t_{1}}^{*}.$ The curve $C$ being entirely contained in
$N_{t_{1}}^{(+,0)},$ the relation (\ref{solutiongauge}) implies
that $e^{(t_{1}-t)\upsilon _{(1)}^{(T)}}M_{(1)}^{-1}(t_{1})U_{1,n}(A)M_{(n)}(t_{1})e^{(t'-t_{1})
\upsilon _{(n)}^{(T)}}=D_{(+),n}(t,t').$ 

We have emphasized a certain homotopy class of curves, which choice
has been guided by the decomposition of our manifold, in order to
compute previous observables. However, what would be the result if
we change this homotopy class ? Let us study this point through an
example. Let us choose a simple curve $\tilde{C}$ going from $x_{(1)}(t_{1},\frac{\pi }{2})$to
$x_{(p)}(t_{1},\frac{\pi }{2})$ through $N_{t_{1}}^{(+,0)}$ and
then going from $x_{(p)}(t_{1},\frac{\pi }{2})$to $x_{(n)}(t_{1},\frac{\pi }{2})$
through $N_{t_{1}}^{(-,0)}.$ The holonomy $U_{\tilde{C}}(A)=\overrightarrow{Pexp}\int _{\tilde{C}}A$
of the connection along this curve can be written as \begin{eqnarray*}
U_{\tilde{C}}(A) & = & \bar{k}_{(+)}(x_{(1)}(t_{1},\frac{\pi }{2}))^{-1}\bar{k}_{(+)}(x_{(p)}(t_{1},\frac{\pi }{2}))\bar{k}_{(-)}(x_{(p)}(t_{1},\frac{\pi }{2}))^{-1}\bar{k}_{(-)}(x_{(n)}(t_{1},\frac{\pi }{2}))\\
 & = & \bar{k}_{(+)}(x_{(1)}(t_{1},\frac{\pi }{2}))^{-1}u_{p,p+1}\bar{k}_{(-)}(x_{(n)}(t_{1},\frac{\pi }{2})).
\end{eqnarray*}

This formula has to be understood in the following way: $u_{p,p+1}$
is an element of the isometry group of moves on $dS_{3}$ such that
$u_{p-1,p}u_{p,p+1}^{-1}$is a move describing the defect angle of
the gravitational lens provocated by the $p-$th particle, to replace
$\bar{k}_{(+)}(x_{(n)}(t_{1},\frac{\pi }{2}))$ by $u_{p,p+1}\bar{k}_{(-)}(x_{(n)}(t_{1},\frac{\pi }{2}))$
consists in replacing the location of the $n-$th particle by its
virtual image through the lens described by the isometry defined by
$u_{p,p+1}.$ As a conclusion, if we replace $U_{1,n}$ by $U_{\tilde{C}}$
in the computation of the distance between particles $1$ and $n$,
it will just consist in replacing the $n-$th particle by this virtual
image and then compute the distance as if the $1-$st particle and
this virtual image were both on $dS_{3}.$ 

We have to notice also that $e^{(t_{1}-t)\upsilon _{(1)}^{(T)}}M_{(1)}^{-1}(t_{1})U_{\tilde{C}}M_{(n)}(t_{1})e^{(t'-t_{1})\upsilon _{(n)}^{(T)}}$
can be written in terms of our basic datas. Indeed, we have easily\begin{eqnarray*}
e^{(t_{1}-t)\upsilon _{(1)}^{(T)}}M_{(1)}^{-1}(t_{1})U_{\tilde{C}}M_{(n)}(t_{1})e^{(t'-t_{1})\upsilon _{(n)}^{(T)}} & = & D_{(+),p}(t,t_{1})D_{(-),p}^{-1}D_{(-),n}(t_{1},t').
\end{eqnarray*}

More generally, for $i,i'\in \left\{ 1,...,N\right\} ,t,t'\in [t_{1},t_{2}],$
we will denote by $\mathcal{C}_{(i,t),(i',t')}$ the set of continuous
curves $s\in [0,1]\mapsto C(s)\in \mathcal{M}^{*}$ such that $C\cap \mathcal{B}=\left\{ C(0),C(1)\right\} $
and $C(0)\in l_{(i)}(t),C(1)\in l_{(i')}(t').$ We introduce an equivalence
relation on this set, denoted $\asymp _{(i,t),(i',t')}$ and defined
such that $C\asymp _{(i,t),(i',t')}C'$ if and only if we can interpolate
between $C$ and $C'$ using a continuous family of elements of $\mathcal{C}_{(i,t),(i',t')},$
and will denote $\bar{\mathcal{C}}_{(i,t),(i',t')}$ the set of classes
under this relation.

In the same way, we can define observables corresponding to distances
for any homotopy class of paths $\bar{C}_{(1,t)(n,t')}$ by the formula
\begin{eqnarray}
d(\bar{C}_{(1,t)(n,t')}) & \equiv  & l_{c}^{2}tr(D_{P,n}(t,t')D_{P,n}(t,t')^{\star }-I)\nonumber \\
D_{P,n}(t,t') & \equiv  & e^{(t_{1}-t)\upsilon _{(1)}^{(T)}}M_{(1)}^{-1}(t_{1})U_{C}(A)
M_{(n)}(t_{1})e^{(t'-t_{1})\upsilon _{(n)}^{(T)}}\label{distancegrav}
\end{eqnarray}

where $U_{C}(A)$ is the holonomy of the connection along any particular
representative $C$ of this class of paths. As before, any of these
observables can be expressed in terms of our basic combinatorial datas.

\subsubsection{\emph{third digression: degenerate metrics from the participant observer's point of view}}

We have emphasized along previous sections that the choice between
Chern-Simons gravity and true gravity consisted in the choice of large
gauge transformations, and especially their capability to transform
a non-degenerate soldering form into a degenerate one. Degenerate
soldering forms have been introduced in Lorentzian gravity as a bill
to pay if we want to incorporate, into the framework, transitions
between topologically different spatial slices along time evolution
of a universe. All the way, spurious degeneracies are always present
in the formalism because of topological obstructions against coordinatization
(for example, the $2-$sphere can not be parallelised by a globally
non degenerate soldering form, although it exists well defined geometry
on it). On another part, the problem of caracterizing true singularities
independently of any coordinatization choice is a difficult task.
In a participant-observer approach of gravity, the problem is posed
independently of any coordinatization and the question of the existence
of true singularities is transfered to the problem of the existence
of physical measurements giving evidence of the existence of such
singularities. We have then to study the consequence of true singularities
on the set of non-local observables based on matter degrees of freedom.
The choice between Chern-Simons gravity and true gravity is however
fundamental in this study because gauge transformations in Chern-Simons
theory can transform a non-degenerate soldering form in a degenerate
one. Reciprocally, in absence of global obstruction, it has no sense
to say that a classical solution of Chern-Simons theory has a degenerate
or non-degenerate translation part, because only the gauge orbit is
physically meaningful. Generically, we can find a representative in
any gauge orbit with no true singularity. Considering naively this
fact, it would seem that, a part a sector of classical solutions owning
true degeneracies which cannot be washed by gauge transformations
because of global obstructions, we could map gauge orbits under $\Delta _{W}$
of Chern-Simons theory onto gauge orbits under $\Delta _{L,PD}$ of
true gravity, however this naive picture is rigourously false. As
we already mentioned, the set of gauge orbits of true gravity is made
of an infinite number of copies of the space of gauge orbits of Chern-Simons
theory (to see this through a simple example, see \cite{key-34}).
Indeed, there are generically an infinite number of diffeomorphically
inequivalent geometries which are however gauge related by $\Delta _{W}$
and then considered as the same object in Chern-Simons gravity. This
result constitutes a fundamental problem for the Chern-Simons approach
and we will try to elucidate how to repare it in a participant observer
perspective. Let us precise the problem in the case of gravitating
particles on a sphere (this discussion is directly inspired by \cite{key-32}). 

The choice of an element of the reduced phase space of Chern-Simons
deSitter gravity consists exactly, the conditions (\ref{partialgaugeM})(\ref{1atrest})
being chosen, in the choice of datas $(\bar{k}_{(+)}(x_{(n)}(t_{1},\frac{\pi }{2})))_{n=1...N},$
up to fixed elements in $H^{(L)},$ and $(\upsilon _{(n)}^{(T)})_{n=1,...,N}.$
The bulk gauge transformations $\Delta _{W}$ act freely on dynamical
variables but let these datas fixed. In the following, we will denote
shortly $Q_{(\pm )}(x)\equiv Q_{k_{(\pm )}(x)}$ because the spin
part of $k_{(\pm )}$ will not be of particular interest for the present
discussion. A particular representative of this state being chosen,
we can build the corresponding spacetime according to previous subsection.
We have to notice that the smooth by parts surface $Q_{(+)}(\mathcal{S}_{n}^{(0)})$
is only constrained along trajectories $\mathcal{T}_{n}^{(+)}$ where
it is eventually not smooth, the rest of the surface being a priori
freely chosen. Once the previous surface is chosen, the surface $Q_{(-)}(\mathcal{S}_{n}^{(0)})$
is uniquely obtained using the fixed isometries associated to $u_{n,n+1}.$
For a given family of free-falling trajectories $\mathcal{T}^{(+)}=(t\mapsto \mathcal{T}_{n}^{(+)}(t))_{n=1,...,N}$in
$dS_{3}$ and a given family $\Im =(s_{n,n+1})_{n=1...N}$ of elements
of the isometry group of moves on $dS_{3}$, we will denote by $\aleph _{(\mathcal{T}^{(+)},\Im )}$the
set formed by families $(\sigma _{n}^{(+)},\sigma _{n}^{(-)})_{n=1,...,N},$
such that $\sigma _{n}^{(\pm )}$ is a mapping from $\mathcal{S}_{n}^{(0)}$to
$dS_{3},$ for any $\phi ,$ $\mathcal{T}_{n}^{(+)}=\left\{ t\mapsto \sigma _{n}^{(\pm )}(x_{n}(t,\phi ))\right\} ,$$\mathcal{T}_{n+1}^{(+)}=\left\{ t\mapsto \sigma _{n}^{(\pm )}(x_{n+1}(t,\phi ))\right\} ,$
$\sigma _{n}^{(-)}(x)=s_{n,n+1}(\sigma _{n}^{(+)}(x)),\forall x\in \mathcal{S}_{n}^{(0)}.$
As a result of the previous discussion, an element of the reduced phase
space of Chern-Simons gravity being chosen we can associate uniquely
a set $(\mathcal{T}^{(+)},\Im )$and for any pair of elements of $\aleph _{(\mathcal{T}^{(+)},\Im )}$we
can obviously find a gauge transformation $\Delta _{W}$ which maps
one element to the other one. If we require the non-degeneracy of $A^{(T)}$ by constraining
the mapping $x\mapsto Q_{(\pm )}(x)$ to be injective the situation
is very different. $\bigcirc _{i=1}^{N}Q_{(+)}(\mathcal{S}_{n}^{(0)})$
and $\bigcirc _{i=1}^{N}Q_{(-)}(\mathcal{S}_{n}^{(0)})$ correspond
to two smooth (except along the lines $\mathcal{T}_{n}^{(\pm )}$
), disjoint, simple surfaces in $dS_{3}$ surrounding the two disjoint
images of $\mathcal{N}^{(\pm ,0)}$ by the smooth and injective mappings
$\mathcal{N}^{(\pm ,0)}\rightarrow \mathcal{D},\: (x^{\mu })_{\mu \in I_{M}}\mapsto Q_{(\pm )}(x),$
which can be both isometrically mapped to a part of $dS_{3}.$ The
entire spacetime being recovered by gluing these two parts along $\bigcirc _{i=1}^{N}Q_{(+)}(\mathcal{S}_{n}^{(0)})$
and $\bigcirc _{i=1}^{N}Q_{(-)}(\mathcal{S}_{n}^{(0)}).$ The action
of bulk gauge transformations $\Delta _{L,PD}$ on the previous datas
is obvious, the local Lorentz transformations act trivially on these
datas, the projectable diffeomorphisms connected to the identity transform
the surfaces $\bigcirc _{i=1}^{N}Q_{(\pm )}(\mathcal{S}_{n}^{(0)})$
by smooth deformations in $dS_{3}$ leaving the lines $\mathcal{T}_{n}^{(\pm )}$
unchanged (as well as the isometries associated to $u_{n,n+1}$).
The projectable diffeomorphisms do not affect the injectivity of the
maps $Q_{(\pm )}$ and then the surfaces $\bigcirc _{i=1}^{N}Q_{(\pm )}(\mathcal{S}_{n}^{(0)})$
stay smooth by parts, simple and disjoint under these transformations.
We will denote $\aleph _{(\mathcal{T}^{(+)},\Im )}^{simple}$ the
subset of $\aleph _{(\mathcal{T}^{(+)},\Im )}$formed by elements
$(\sigma _{n}^{(+)},\sigma _{n}^{(-)})_{n=1,...,N}$ such that $\bigcirc _{i=1}^{N}\sigma _{n}^{(+)}(\mathcal{S}_{n}^{(0)})$
and $\bigcirc _{i=1}^{N}\sigma _{n}^{(-)}(\mathcal{S}_{n}^{(0)})$
are disjoint and simple. We can introduce the equivalence relation
$\approx $ between pairs of elements of $\aleph _{(\mathcal{T}^{(+)},\Im )}^{simple}$
defined such that $(\sigma _{n}^{(+)},\sigma _{n}^{(-)})_{n=1,...,N}\approx (\sigma '{}_{n}^{(+)},\sigma '{}_{n}^{(-)})_{n=1,...,N}$
if and only if we can find a continuous family of elements of $\aleph _{(\mathcal{T}^{(+)},\Im )}^{simple}$interpolating
between $(\sigma _{n}^{(+)},\sigma _{n}^{(-)})_{n=1,...,N}$and $(\sigma '{}_{n}^{(+)},\sigma '{}_{n}^{(-)})_{n=1,...,N}.$
Each equivalence class will be called \emph{a skeleton} of the corresponding
geometry. Endly, $\bar{\aleph }{}_{(\mathcal{T}^{(+)},\Im )}^{simple}$will
denote the set of equivalence classes under this relation. For a given
point in the reduced phase space, and then, for a given family $(\mathcal{T}^{(+)},\Im ),$
the set $\bar{\aleph }{}_{(\mathcal{T}^{(+)},\Im )}^{simple}$is generically
not reduced to a single element. This result proves clearly the announced
result that the phase space of true deSitter gravity is in fact larger
than the phase space of Chern-Simons deSitter gravity, and our combinatorial
datas (\ref{combinatorialdatas}) do not manage to capture the whole
information about a classical solution of deSitter gravity. Moreover,
once is fixed a set of combinatorial datas and an associated skeleton
the non-degenerate geometry is entirely fixed up to a projectable
diffeomorphism. Hence, we obtain a combinatorial description of the
reduced phase space of true deSitter gravity, because a point in the
reduced phase space of true deSitter gravity is fixed once are chosen
a set of combinatorial datas and a skeleton for these datas. 
The previous discussion shows that classical solutions
associated to different skeletons have to be physically distinguished
as soon as we require the non-degeneracy of the metric. It is important to notice 
that the corresponding geometries are really different and these differences have 
fundamental and easily observable consequences on the measurements made by the observer. 
Let us study this point through a simple example. 
We represent, on the following picture, the images on the sphere of a given region $R$ of $\Sigma_t^*$ 
(with light grey background) by two different mappings. This region contains 
particles $i,i+1$ (black disks) and the first particle, 
playing the role of the observer (medium grey disk). 
We will assume that the two mappings are identical 
except in this region, and that the image of $R$ is such that the point masses $p \not= i,i+1$   
are sufficiently far from particle $1$ in order that the minimal distance between particle $1$ 
and particles $i,i+1$  
are given by the geodesic distance along paths entirely contained in this region.  
On these pictures the images of $R\cap N^+_t$ and 
$R\cap N^-_t$ are separated by a region drawn in dark grey which has 
to be cutted out, the two images being
glued together along the cuts, in order to recover the multiconical geometry 
associated to these mappings. 
We have chosen two mappings corresponding to the same combinatorial datas, 
indeed the isometries associated to
the gluings are the same (rotation of angle $\frac{\pi}{36}$ centered on 
the particle $i+1$ for the gluing of 
the segment $\mathcal{S}_{i+1}^{(0)}$, rotation of angle $\frac{\pi}{6}$ centered on the particle $i$ for the gluing of 
the segment $\mathcal{S}_{i-1}^{(0)}$ and identity map for the segments 
$\mathcal{S}_{N}^{(0)},\mathcal{S}_{1}^{(0)},
\mathcal{S}_{i}^{(0)}$), and the positions of particles on the sphere are the same. However, the 
skeletons differ radically, in a way shown on the picture. 

\begin{figure}
\centering
\includegraphics[scale=0.5]{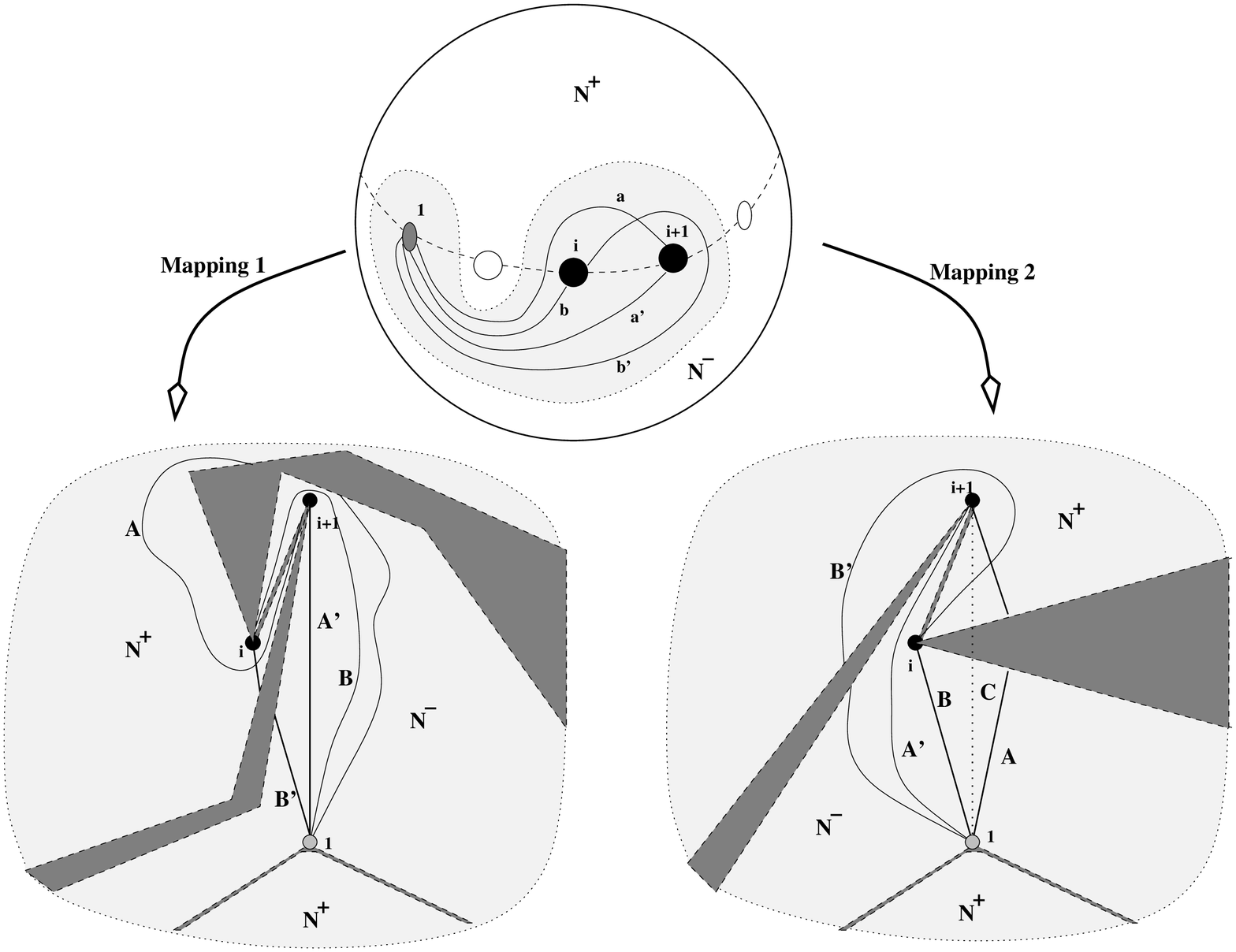}
\label{}
\end{figure}

We have drawn with a thick line the geodesics belonging to homotopy classes of paths 
for which such a minimal distance curve exists and with a thin line paths belonging to homotopy classes of paths 
for which no curve of this class is able to minimize the distance, the corresponding curves on $\Sigma_t^*$ are 
indexed by the 
small letters and their images are indexed by the corresponding capital letters on the two multiconical 
geometries. It is clear, in this example, that these geometries cannot be identified as the same geometrical
object, because distances and angles associated to images of point masses, as measured physically in the sky 
of the observer, are  different. However, the difference is subtle. Indeed, although the ``distance-observable'' 
(\ref{distancegrav}) between 
particles $1$ and $i+1$ can be technically computed (and is a Dirac observable for Chern-Simons theory), for example, 
along the path $a'$ in both cases, 
it gives really a geodesic distance only through the mapping $1$ for which the path $A',$ drawn on 
the multiconical geometry, is actually a geodesic. Through the mapping $2$ there is no geodesic curve 
in the homotopy class of $a',$ the ``distance-observable'' (\ref{distancegrav}) gives the virtual 
distance $C$ which does not correspond to any geodesic curve, the geodesic curve being in fact 
in the homotopy class of the path $a$ and give the geodesic distance $A.$ According to our assumptions, the 
minimal distance at which the particle $i$ (resp. $i+1$) is seen by the observer is the length of 
the segment $B'$ (resp. $A'$) for the mapping $1$ and is the length of 
the segment $B$ (resp. $A$) for the mapping $2.$ And the picture shows clearly that the angle between both 
rays coming from particles $i,i+1$ and received by the observer are in the case of mapping $1$ is more acute 
than that in the case of mapping $2.$ 
As a conclusion, it is important to emphasize 
that the two geometries are neither related by a projectable diffeomorphism connected to the identity 
(because their skeletons are different), nor related by an element of the mapping class group which would just 
exchange homotopy classes $a$ and $a'.$ This point agrees with \cite{key-32} and disagrees with the conclusion of 
\cite{Sch}.

\subsubsection{\emph{the observables of Chern-Simons gravity and true gravity}}

In order for us to develop a combinatorial approach of Chern-Simons
and true deSitter gravity we have to determine a set of observables
which would allow us to distinguish points in the reduced phase space
of each theory and would allow us to compute explicitely their Dirac
algebra. 

We already know how to distinguish points in the reduced phase space
of Chern-Simons deSitter gravity using our combinatorial datas (\ref{combinatorialdatas}).
However, we will introduce, in the next subsection, algebraic objects,
called generalized spin-network observables, built from these datas
and equivalent to them in their capability to capture the whole reduced
phase space of Chern-Simons deSitter gravity. We will see that the Dirac algebra
of these observables can be explicitely computed and emphasizes nice algebraic
structures which seems to be easily quantizable.

It is fundamental to notice that these observables are also observables
of true deSitter gravity because of their larger invariance properties.
Moreover their Dirac algebra do not depend on the theory we are considering,
because the considerations about large gauge transformations do not
concern the definition of the Dirac bracket. However, they are not
sufficient to capture the richness of the reduced phase space of true
deSitter gravity, and we have to identify new observables in order
to complete the description. 

Let us define at least one observable, called ``window-observable'',
allowing us to distinguish between different skeletons. According, to 
the discussion of the previous subsection, it seems that the geometries 
associated to different skeletons differ in the fact that, in the same homotopy
class of paths, a geodesic curve of the multiconical geometry can be found or not. 
If a geodesic curve belongs to a given homotopy class of paths, the distance 
observable associated to this class and computed from combinatorial datas is 
actually the distance at which the observer locate the corresponding point mass
using the physical measurements he can realize. 
Hence, let us
choose a set of acceptable combinatorial datas (according to constructions
of previous subsections), a skeleton associated to them, a class of
paths $\bar{\mathcal{C}}_{(i,t),(i',t')}$ and an element $\bar{C}\in \bar{\mathcal{C}}_{(i,t),(i',t')}$.
We will denote $\mathcal{W}_{\bar{C}}^{light}$ the function defined
as follows: we choose a particular representative $(A,M,\upsilon )$,
obeying as usual the condition (\ref{1atrest}), of the point of the
reduced phase space of true gravity associated to the combinatorial
datas and the skeleton, then we define $\mathcal{W}_{\bar{C}}^{light}$
to be equal to $1$ if there exists a representative $C$ of $\bar{C}$
which is a lightlike-geodesic in the geometry defined by $A^{(T)},$
and to be equal to $0$ elsewhere.

For purely technical reasons it is generally more simple to work with
equal-time geodesics rather than with lightlike geodesics although the latter
are the natural objects to discuss causality. All the
way, the problem of relating these two objects is simplified as soon
as our point masses have inertial trajectories. We are now able to
formulate a conjecture:

\emph{the set of windows and spin-networks observables distinguishes
points in the reduced phase space of true deSitter gravity.}

This point deserves a careful analysis which will be given in a forthcoming
paper.

\subsubsection{\emph{fourth digression: the action of the mapping class group}}

The choice of the decomposition of the manifold as the union $\Sigma _{t}^{*}=N_{t}^{(+,0)}\cup N_{t}^{(-,0)}$
is non-canonical, and then, the distinction we have made between virtual
and true images in subsection \ref{sub:reconstruction} has no intrinsic
sense, and all homotopy classes produce, along previous constructions,
equally acceptable observables for Chern-Simons gravity. Moreover,
our participant-observer has no external tool to distinguish between
homotopy classes by only measuring time intervals associated to exchange
of particles along the different paths, then they have to be considered
on the same footing. However, if we think to these distances as measured
by the exchange of particles, it is natural to distinguish between
different homotopy classes if the observables corresponding to some
of them could not be associated to a physical process. It is clear from previous 
subsection 
that the observables (\ref{distancegrav}) can be considered
as distances between the observer and the point mass if and only if such an exchange
of physical particle can occur along this homotopy class of trajectories. 

A priori, the set of windows observables is a natural complement of
the set of combinatorial datas, indeed it is the necessary and sufficient
information allowing us to distinguish true
images from virtual images in the sky of the observer. Then, a contrario
from the situation of Chern-Simons deSitter gravity where a point
mass was seen from an infinite number of directions and distances
(corresponding to different but indistinguishable homotopy classes
of paths) leading generically to a non-Haussdorf apparent sky for
the observer, none except a finite number of equivalence classes of
paths admits a light-like geodesic and then, even after having moded
out our set of observables by the action of the mapping class group
of the surface, the apparent sky of the observer in true deSitter
gravity, defined to be the respective positions of stars computed
from combinatorial datas by the formula (\ref{distancegrav}) for
classes of paths which window observable has value $1,$ will show
a finite number of stars.

\subsection{Towards a combinatorial description of the N-body problem in deSitter
gravity}

This section is essentially technical and aims to show how to describe
combinatorially the Poisson structure on our algebra of observables.

\subsubsection{Definition and properties of generalized spin-networks}

Let us begin by developing a formalism generalizing spin-network's
formalism for our purposes. We will choose $2N$ new tubular compact
neighbourhoods $(c'_{(n)})_{n=1...N},(c''_{(n)})_{n=1...N}$ of the
worldlines such that $c_{(n)}$ (resp. $c'_{(n)}$) is strictly contained
in the interior of $c'_{(n)}$(resp. $c''_{(n)}$) for any $n$, and
$c''_{(n)}\cap c''_{(p)}=\textrm{Ø}$ if $n\neq p,$ we will denote
$(l'_{(n)})_{n=1...N},(l''_{(n)})_{n=1...N}$ their respective boundaries
and $s'_{t}=\bigcup _{i=1...N}l'_{(i)},s''_{t}=\bigcup _{i=1...N}l''_{(i)}$.
We will denote $\check{\Sigma }{}_{t}^{*}=\Sigma _{t}^{*}\cap \bigcup _{n=1...N}(c''_{(n)}-l''_{(n)}),$
$\: \Sigma '{}_{t}^{*}=\Sigma _{t}-\bigcup _{n=1...N}c'_{(n)},$$\: \Sigma ''{}_{t}^{*}=\Sigma _{t}-\bigcup _{n=1...N}c''_{(n)}.$
Let us define a generalized spin-network. We choose $2M$ distinct
points $(z_{(i)})_{i=1...M},(y_{(i)})_{i=1...M}$ such that $z_{(i)}\in s'_{t},y_{(i)}\in s_{t}''$
and denote $(P_{(i)})_{i=1...M}$ disjoint simple oriented curves,
which interior are included in $\check{\Sigma }{}_{t}^{*},$ $P_{(i)}$
connecting $y_{(i)}$to $z_{(i)}.$ Let be given an open graph drawn
in $\Sigma _{t}^{*},$ $T_{1}$ will denote the ordered set of its
oriented edges and $T_{0}$ the set of its vertices, for any $P\in T_{1}$
$e(P),d(P)$ will denote respectively the end and the departure point
of $P$ and we will as well use the notation $P=[d(P)e(P)]$. We will
impose $y_{(i)},z_{(i)}\in T_{0},\forall i=1...M$ and will denote
$T'_{0}=T_{0}-\left\{ z_{(i)},i=1...M\right\} ,\: T'_{1}=T_{1}-\left\{ P_{(i)},i=1...M\right\} $.
We will also require $T_{1}\cap \Sigma ''{}_{t}^{*}=\bigcup _{i=1...M}P_{(i)}.$
To each $P\in T_{1}$ we associate a representation $\Pi _{P}$ of
the envelopping algebra $\mathcal{U}(\mathfrak{g}),$ and to each
$w\in T'_{0},P\in T_{1}$ we will associate $\Pi _{w,P}=\Pi _{P}\: \mathrm{if}\: w=e(P),\Pi _{P}^{*}\: 
\mathrm{if}\: w=d(P),\: \iota \: \mathrm{elsewhere}\:.$ To
each $w\in T'_{0},$ will be associated an intertwining operator $\Phi _{w}\in \mathrm{Hom}_{\mathcal{U}(\mathfrak{g})}(\bigotimes _{P\in T_{1}}V_{\Pi _{w,P}},\mathbb{C}).$
The set of previous datas (the graph with edges colored by representations
and the set of intertwiners associated to vertices) will be called
generalized spin-network and denoted $\mathcal{S}.$ Two spin networks
will be said to be transverse if the intersection of their graphs
is disjoint of their set of vertices.

We will denote $U_{\mathcal{P}}^{\Pi _{P}}(A)=\overrightarrow{Pexp}\int _{d(P)}^{e(P)}\Pi _{P}(A),$
the element of $\mathrm{End}(V_{\Pi _{P}})=V_{\Pi _{P}}\otimes V_{\Pi _{P}}^{*}$
(which matrix elements are functions of the connection $A$) given
by the path ordered exponential along $P\in T_{1},$ of the connection
$A,$ taken in the representation $\Pi _{P}.$ 

We will associate to the whole set of previous datas the following
functions of the connection 
$\mathcal{O}_{\mathcal{S}}^{bulk}\equiv (\bigotimes _{w\in T'_{0}}\Phi _{w})
(\bigotimes _{P\in T'_{1}}U_{\mathcal{P}}^{\Pi _{P}})\in 
\mathrm{Hom}(\bigotimes _{i=1...M}V_{\Pi _{y_{(i)},P_{(i)}}},\mathbb{C}),$
and $\mathcal{O}_{\mathcal{S}}^{bound}\equiv \bigotimes _{i=1...M}(U_{\mathcal{P}_{(i)}}^{\Pi _{P_{(i)}}}
M^{\Pi _{P_{(i)}}}_{(z_{(i)})})\in \mathrm{End}(\bigotimes _{i=1...M}V_{\Pi _{y_{(i)},P_{(i)}}}).$
We want to emphasize that $\Delta _{G}[g]\mathcal{O}_{\mathcal{S}}^{bulk}=
\mathcal{O}_{\mathcal{S}}^{bulk}(\bigotimes _{i=1...M}\Pi _{P_{(i)}}(g_{y_{(i)}}))$
and $\Delta _{G}[g]\mathcal{O}_{\mathcal{S}}^{bound}(A)=
(\bigotimes _{i=1...M}\Pi _{P_{(i)}}(g_{y_{(i)}}^{-1}))\mathcal{O}_{\mathcal{S}}^{bound}(A).$ 

The reader used to current litterature on Chern-Simons gravity could
misunderstand the sense of previous results. The previous combinatorial
datas contain more informations than the usual set of observables
used in the combinatorial approach, i.e. the observables associated
to spin-networks without ``legs'' ($M=0$). This smaller set of observables
will be a Poisson subalgebra of our Poisson algebra of generalized
spin-networks observables. In fact, these observables correspond to
the set of constants of motion of Chern-Simons deSitter gravity. This
Poisson subalgebra has already been quantized and an irreductible
star representation of the corresponding quantum operator algebra
has been found in \cite{key-31}.

\subsubsection{\emph{Explicit computation of the Dirac algebra of observables}}

Let us recall that the detail in the bulk of the second class
functions $\check{\mathbb{g}}$ defining our Dirac bracket is irrelevant
and we can always choose $\check{\mathbb{g}}$ such that the support
on $\Sigma ^{*}$of its elements is as small as necessary. In the
following, we will impose that the support of the elements of $\check{\mathbb{g}}$
is strictly contained in $\check{\Sigma }{}_{t}^{*}.$ Hence, if we
consider a function $v$ of the connection which depends only on the
connection in the bulk $\Sigma ''{}_{t}^{*}$ and any other function
$w$, from the definition of the Dirac bracket (\ref{definitiveDiracbracket}),
their Dirac brackets are given by: $\{v,w\}_{D}=\{v,w\}_{2}.$ Indeed,
due to the property \begin{eqnarray*}
\left\{ A_{i}(x),\tilde{\Omega }(u)\right\} =\frac{-\vartheta }{2l_{c}}D_{i}^{(A)}u &  & \forall x\in \Sigma {}_{t}^{*},\forall u\in \mathbb{g}
\end{eqnarray*}

we deduce that $\left\{ A_{i}(x),\tilde{\Omega }(u)\right\} =0,\forall x\in \Sigma ''{}_{t}^{*},\forall u\in \check{\mathbb{g}}$
and then the second term in the expression (\ref{definitiveDiracbracket})
is null in this situation. 

As a result, Dirac brackets between two functions of the connection
defined in the bulk is obtained thanks to the usual Chern-Simons bracket.
In particular, the Poisson bracket $\left\{ \mathcal{O}_{\mathcal{S}}^{bulk},\mathcal{O}_{\mathcal{S}'}^{bulk}\right\} _{D}$
for two transverse generalized spin-networks $S,S'$ is given by the
usual formula \cite{key-29,key-30}:\begin{eqnarray*}
\left\{ \mathcal{O}_{\mathcal{S}}^{bulk},\mathcal{O}_{\mathcal{S}'}^{bulk}\right\} _{D} & = & \frac{\vartheta }{2l_{c}}\sum _{x\in S\cap S'}\iota _{x}(S,S')\mathcal{O}_{\mathcal{S}\odot _{x}S'}^{bulk}
\end{eqnarray*}

where the sum extends to all intersection points between graphs associated
to $S$ and $S',$$\iota _{x}(S,S')$ is the index of the intersection
at $x$ between $S$ and $S',$ and $\mathcal{S}\odot _{x}S'$ is
the spin network obtained from $S$ and $S'$ by fusing the colored
graphs and by adding at $x$ the intertwiner $\frac{2l_{c}}{\vartheta }t$
(see \cite{key-31} for a more pedagogical introduction). From previous
considerations, we also have trivially \begin{eqnarray*}
\left\{ \mathcal{O}_{\mathcal{S}}^{bulk},\mathcal{O}_{\mathcal{S}'}^{bound}\right\} _{D} & = & 0
\end{eqnarray*}

Then, in order to obtain a complete description of the Dirac algebra
of generalized spin-networks we have to compute the Poisson bracket
$\left\{ \mathcal{O}_{\mathcal{S}}^{bound},\mathcal{O}_{\mathcal{S}'}^{bound}\right\} _{D}=0$
which obviously decomposes using leibniz rule in terms of Poisson
brackets $\left\{ U_{\mathcal{P}_{(i)}}^{\Pi _{P_{(i)}}}M_{(z_{(i)})},U_{\mathcal{P}_{(j)}}^{\Pi _{P_{(j)}}}M_{(z_{(j)})}\right\} _{D}$.
These Poisson brackets are obviously null if $z_{(i)},z_{(j)}$ belong
to different components of the boundary. Its explicit computation
requires a careful analysis in the other case. This computation is
done in the appendix (\ref{sub:Dirac-gravit}) and we obtain\begin{eqnarray*}
\left\{ (U_{\mathcal{P}_{(i)}}^{\Pi _{P_{(i)}}}M_{(z_{(i)})})_{1},(U_{\mathcal{P}_{(j)}}^{\Pi _{P_{(j)}}}M_{(z_{(j)})})_{2}\right\} _{D} & = & (U_{\mathcal{P}_{(i)}}^{\Pi _{P_{(i)}}}M_{(k)})_{1}(U_{\mathcal{P}_{(j)}}^{\Pi _{P_{(j)}}}M_{(k)})_{2}\times \\
 &  & \times \mathfrak{R}_{12}^{(1)}(M_{(k)},P_{(k)};\phi _{i}-\phi _{j})
\end{eqnarray*}

where $z_{(i)}=x_{k_{i}}(t,\phi _{i}),\: z_{(j)}=x_{k_{j}}(t,\phi _{j})$and
$k_{i}=k_{j}\equiv k.$ This Dirac algebra is rather close from that
found in \cite{key-29} and studied extensively in \cite{key-31}.

Its complete study as well as the study of the Poisson bracket of
windows observables is devoted to a forthcoming paper.

\section{Conclusion}

This paper can be considered as a strong support in favour of the
combinatorial approach to $3$ dimensional gravity, and as a generalization
of this program necessary to capture dynamics of gravity. The nice
structure of the Dirac algebra between observables motivates a study
of its quantization and of the representation of the corresponding
quantum algebra of operators, completing the work already done in
\cite{key-31}. It is, for the first time, possible to hope a \emph{well-posed}
discussion of the spectrum of operators associated to distances between
particles in a completely generally covariant formalism. Having managed
to capture the whole generally invariant information on the phase
space of a set of gravitating particles in a closed universe, we may
hope to pursue the program adressed for a long time by G. 'tHooft
of building from this departure point a second-quantized theory of
the self-gravitating matter field. The present paper is only a modest
contribution in this direction.

\subsubsection*{Aknowledgements:} we thank M.Henneaux for stimulating discussions.

\appendix

\section{Appendix}

\subsection{Basic algebraic results\label{sub:Basic-algebraic-results}}

We will recall in this appendix basic notions on the Lorentz group
$G=SL(2,\mathbb{C})$ and its Lie algebra $\mathfrak{g}=sl(2,\mathbb{C})$.
The Lorentz group is the real Lie group of $2\times 2$ complex matrices
of determinant one and $\mathfrak{g}$ is the real Lie algebra of
$2\times 2$ complex matrices of zero trace. We recall that the set
of $2\times 2$ hermitian complex matrices is denoted by $\mathcal{H}$
and we denote $\mathcal{D}$ the subset of $\mathcal{H}$ defined
by $\mathcal{D}=\{Q\in \mathcal{H}|\det Q=-l_{c}^{2}\}.$ To any element
$o\in \mathcal{D}$, we can associate an antimorphic $\star $-map
on $SL(2,\mathbb{C})$ defined by 

\begin{eqnarray}
\star :{SL(2,\mathbb{C})}\longrightarrow {SL(2,\mathbb{C})}\; \; \; ,\; \; \; \Lambda \longmapsto \Lambda ^{\star }=o\Lambda ^{\dagger }o^{-1}\; , &  & \label{star}
\end{eqnarray}

and define the following two subsets of $SL(2,\mathbb{C})$: \begin{eqnarray}
\mathcal{L}_{o} & = & \{\lambda \in SL(2,\mathbb{C})|\lambda ^{\star }\lambda =\lambda \lambda ^{\star }=\pm 1\}\label{subspaceL}\\
\mathcal{T}_{o} & = & \{\tau \in SL(2,\mathbb{C})|\tau ^{\star }=\tau \; \text {and}\; \text {Tr}(\tau )>0\}\; .\label{subspaceT}
\end{eqnarray}

Any element of the Lorentz group decomposes as follows: \begin{eqnarray}
\forall \Lambda \in SL(2,\mathbb{C}),\; \; \exists \lambda _{\Lambda }\in \mathcal{L}_{o}\; \text {and}\; \tau _{\Lambda }\in \mathcal{T}_{o}\; \; \mathrm{such}\: \mathrm{that}\; \; \Lambda =\tau _{_{\Lambda }}\lambda _{_{\Lambda }}\; \; . &  & \label{polardecomposition}
\end{eqnarray}
This decomposition is not unique and is explicitely given by the following
formulae: 

\begin{enumerate}
\item if $\text {Tr}(1+\Lambda \Lambda ^{\star })>0$, $\exists !(\lambda _{_{\Lambda }},\tau _{_{\Lambda }})\in \mathcal{L}\times \mathcal{T}$
such that $\lambda _{_{\Lambda }}\lambda _{_{\Lambda }}^{\star }=+1$
given by: \begin{eqnarray}
\tau _{_{\Lambda }}=\frac{1+\Lambda \Lambda ^{\star }}{\sqrt{\text {Tr}(1+\Lambda \Lambda ^{\star })}}\; \; \; ,\; \; \; \lambda _{_{\Lambda }}=\frac{\Lambda +\Lambda ^{\star -1}}{\sqrt{\text {Tr}(1+\Lambda \Lambda ^{\star })}}\; \; . &  & 
\end{eqnarray}

\item if $\text {Tr}(1-\Lambda \Lambda ^{\star })>0$, $\exists !(\lambda _{_{\Lambda }},\tau _{_{\Lambda }})\in \mathcal{L}\times \mathcal{T}$
such that $\lambda _{_{\Lambda }}\lambda _{_{\Lambda }}^{\star }=-1$
given by: \begin{eqnarray}
\tau _{_{\Lambda }}=\frac{1-\Lambda \Lambda ^{\star }}{\sqrt{\text {Tr}(1-\Lambda \Lambda ^{\star })}}\; \; \; ,\; \; \; \lambda _{_{\Lambda }}=\frac{\Lambda -\Lambda ^{\star -1}}{\sqrt{\text {Tr}(1-\Lambda \Lambda ^{\star })}}\; \; . &  & 
\end{eqnarray}

\end{enumerate}
This $\star $-map induces a $\star $-map on the Lie algebra $\mathfrak{g}$.
Let us introduce the following subspaces of $\mathfrak{g}$ : $\mathfrak{l}=\{\xi \in \mathfrak{g}|\xi ^{\star }=-\xi \}\; \; ,\; \; \mathfrak{t}=\{\xi \in \mathfrak{g}|\xi ^{\star }=\xi \}.$
The sets $\mathfrak{l}$ and $\mathfrak{t}$ depend on the choice
of the element $o\in \mathcal{D}$ which defines the $\star $-map.
Indeed, given another element $o'\in \mathcal{D}$, we get another
$\star -$map on $SL(2,\mathbb{C})$ denoted by $\star '$. Let us
denote by $\mathfrak{l}'$ and $\mathfrak{t}'$ the corresponding
subspaces. We can relate $\mathfrak{l}$ to $\mathfrak{l}'$ and $\mathfrak{t}$
to $\mathfrak{t}'$. Indeed, $\forall \; o'\in \mathcal{D}\; ,\; \exists \; M\in SL(2,\mathbb{C})\; \mathrm{such}\: \mathrm{that}\; o=Mo'M^{\dagger }\; .$
Then $\forall \xi \in \mathfrak{g}$,$\xi ^{\star }=M(M^{-1}\xi M)^{\star '}M^{-1}$
and $\mathfrak{l}'=\{M^{-1}\xi M|\xi \in \mathfrak{l}\}\; \; \; ,\; \; \; \mathfrak{t}'=\{M^{-1}\xi M|\xi \in \mathfrak{t}\}\; .$
In the sequel, we will choose $o=l_{c}\left(\begin{array}{cc}
 1 & 0\\
 0 & -1\end{array}
\right)$ and the results we will obtain could be generalized using the previous
relations. In that case, we can find an explicit basis of the vector
spaces $\mathfrak{l}$ and $\mathfrak{t}$. To this purpose, let us
introduce the usual Pauli matrices $(\sigma _{a})_{a=0,1,2}$ defined
by: \begin{eqnarray}
\sigma _{0}=\frac{1}{2}\left(\begin{array}{cc}
 1 & 0\\
 0 & -1\end{array}
\right)\; \; ,\; \; \sigma _{1}=\frac{1}{2}\left(\begin{array}{cc}
 0 & -i\\
 -i & 0\end{array}
\right)\; \; ,\; \; \sigma _{2}=\frac{1}{2}\left(\begin{array}{cc}
 0 & -1\\
 1 & 0\end{array}
\right). &  & 
\end{eqnarray}
 Let us introduce the notations $P_{a}\equiv \sigma _{a}$ and $J_{a}\equiv -i\sigma _{a}$,
$\forall a=0,1,2$ and therefore $(P_{a})_{a}$ is a basis of $\mathfrak{t}$
and $(J_{a})_{a}$ is a basis of $\mathfrak{l}$. $\forall a\in \{0,1,2\}$,
we will denote by $\xi _{T,a}=MP_{a}M^{-1}$ and $\xi _{L,a}=MJ_{a}M^{-1}$.
It is immediate to see that $(\xi _{T,a})_{a}$ is a basis of $\mathfrak{t}$
and $(\xi _{L,a})_{a}$ is a basis of $\mathfrak{l}$. Moreover, $(\xi _{T,a},\xi _{J,a})_{a}$
satisfy the following Lie algebra: \begin{eqnarray}
[\xi _{L,a},\xi _{L,b}]=\epsilon _{ab}{}^{c}\xi _{L,c}\; \; ,\; \; [\xi _{T,a},\xi _{L,b}]=\epsilon _{ab}{}^{c}\xi _{T,c}\; \; ,\; \; [\xi _{T,a},\xi _{T,b}]=-\epsilon _{ab}{}^{c}\xi _{L,c}\; , &  & 
\end{eqnarray}
 where $\epsilon _{abc}$ is the antisymmetric tensor defined by $\epsilon _{012}=1$
and the indices are lowered and raised by the metric $\eta =\text {diag}(-1,+1,+1)$.
We will denote $I_{\mathfrak{g}}=\left\{ L,T\right\} \times \left\{ 0,1,2\right\} $
and $(\epsilon _{IJ}{}^{K})_{I,J,K\in I_{\mathfrak{g}}}$ the structure
constants of the Lie algebra defined by \begin{eqnarray}
[\xi _{I},\xi _{J}]=\epsilon _{IJ}{}^{K}\xi _{K}\; . &  & 
\end{eqnarray}
 We can endow $\mathfrak{g}$ with a non degenerate invariant bilinear
form $<|>$ defined by: \begin{eqnarray}
<\xi _{T,a}|\xi _{L,b}>=\eta _{ab}\; \; ,\; \; <\xi _{T,a}|\xi _{T,b}>=0\; \; ,\; \; <\xi _{L,a}|\xi _{L,b}>=0\; . &  & 
\end{eqnarray}

We introduce the Casimir tensor $t=t^{IJ}\xi _{I}\otimes \xi _{J}$
such that $t^{IJ}<\xi _{J}|\xi _{K}>=\delta _{K}^{I}$ and the dual
basis $\xi ^{I}=t^{IJ}\xi _{J}$.

We will denote by $\mathfrak{h}$ the Cartan sub-algebra generated
by the two elements $(\xi _{T,0},\xi _{L,0})$, and by $\mathfrak{h}^{(L)}$
(resp. $\mathfrak{h}^{(T)}$) the Cartan sub-algebra generated by
$\xi _{L,0}$ (resp.$\xi _{T,0}$). We will denote by $H,H^{(L)},H^{(T)}$
the corresponding sub-groups. 

Given an element $\xi \in \mathfrak{g}$, we will use the notation
$\xi _{\mathfrak{h}}=\sum _{I,J\in \{0\}\times \{L,T\}}<\xi |\xi _{I}>t^{IJ}\xi _{J}$.

$<|>$ is a non degenerate invariant bilinear form on $\mathfrak{g}$
defined in the appendix. We introduce the Casimir tensor $t=t^{IJ}\xi _{I}\otimes \xi _{J}$
such that $t^{IJ}<\xi _{J}|\xi _{K}>=\delta _{K}^{I}$ and the dual
basis $\xi ^{I}=t^{IJ}\xi _{J}$.

An element $\chi ^{a,b}=a\xi _{L,0}+b\xi _{T,0}$ of $\mathfrak{g}$
being given, it will be useful to introduce $\kappa ^{a,b}\in \mathfrak{g}$
defined by \begin{eqnarray}
\kappa ^{a,b} & = & \frac{-a\xi _{L,0}+b\xi _{T,0}}{a^{2}+b^{2}}\; .
\end{eqnarray}
 From this definition, it is immediate to show that the following
operator: \begin{eqnarray}
\iota :\mathfrak{g}\longrightarrow \mathfrak{g}\; \; ,\; \; \; \;  &  & \xi \longmapsto [\kappa ^{a,b},[\chi ^{a,b},\xi ]]\label{projector}
\end{eqnarray}
 is such that $\iota (\xi )=\xi -\xi _{\mathfrak{h}}$.

For any representation $\Pi $ of $\mathcal{U}(\mathfrak{g})$, $V_{\Pi }$
its module, $\Pi ^{*}$ will denote its contragredient representation,
and $\iota $ will denote the trivial representation.

\subsection{\label{sub:Dirac-free}Dirac bracket for the relativistic particle}

From the expression (\ref{Poissonbracket}), it is straightforward
to compute strongly the Poisson algebra between first class and second
class constraints as follows: \begin{eqnarray}
\{C{}_{R}^{f},C{}_{S}^{f}\} & = & 0\\
\{C{}_{R}^{f},C{}_{A}^{s}\} & = & \epsilon _{RA}{}^{B}C{}_{B}^{s}\\
\{C{}_{A}^{s},C{}_{B}^{s}\} & \equiv  & \Delta _{AB}=-\epsilon _{AB}{}^{R}(<\chi ^{b_{+},b_{-}}|\xi _{R}>-C{}_{R}^{f})\; .\label{Deltamatrix}
\end{eqnarray}
 The Dirac bracket between two functions on the phase space $f$ and
$g$, defined, as usual, by the expression: \begin{eqnarray}
\{f,g\}_{D}=\{f,g\}-\{f,C{}_{A}^{s}\}(\Delta ^{-1})^{AB}\{C{}_{B}^{s},g\}\; , &  & 
\end{eqnarray}
 with $\Delta ^{-1}$ being the strong inverse of $\Delta $. Due
to the previous properties, it verifies the usual axiomatic, i.e.
antisymetry, Leibniz rule, strong Jacobi identity and \begin{eqnarray}
\{f,C{}_{A}^{s}\}_{D}=0\; \; ,\; \; \{f,C{}_{R}^{f}\}_{D}\thickapprox \{f,C{}_{R}^{f}\}\; \; ,\; \; \{f,\{C{}_{R}^{f},C{}_{S}^{f}\}_{D}\}_{D}\thickapprox \{f,\{C{}_{R}^{f},C{}_{S}^{f}\}\}\; . &  & 
\end{eqnarray}

In order to compute the Dirac bracket, we have to invert strongly
the matrix $\Delta $. From the expression (\ref{Deltamatrix}), we
have: \begin{eqnarray}
\Delta _{AB}=-\epsilon _{AB}{}^{R}<\chi ^{\tilde{b}_{+},\tilde{b}_{-}}|\xi _{R}> &  & \label{DeltaAB}
\end{eqnarray}
 where we have introduced $\tilde{b}_{+}=b_{+}+C{}_{0,T}^{f},$ $\tilde{b}_{-}=b_{-}+C{}_{0,L}^{f}$
and we have to notice that $\chi ^{\tilde{b}_{+},\tilde{b}_{-}}=\chi ^{b_{+},b_{-}}-C[M,P]_{\mathfrak{h}}=-(M^{-1}PM)_{\mathfrak{h}}$.
It is easy to show that the inverse matrix $(\Delta ^{-1})^{AB}$
is given by: \begin{eqnarray}
(\Delta ^{-1})^{AB} & = & -\epsilon ^{AB}{}_{R}<\kappa ^{\tilde{b}_{+},\tilde{b}_{-}}|\xi ^{R}>\; .
\end{eqnarray}
 As $P$ commutes with the second class constraints, the Poisson brackets
(\ref{Poissonbracket}) involving $P$ are not modified by the Dirac
reduction. However, the Poisson bracket (\ref{Poissonbracket}) between
matrix elements of $M$ is modified into the following quadratic bracket:
\begin{eqnarray}
\{M_{1},M_{2}\}_{D}=M_{1}M_{2}r_{12}^{(0)}(\tilde{b}_{+},\tilde{b}_{-}) &  & 
\end{eqnarray}
 where $r$ is defined as follows: \begin{eqnarray}
r^{(0)}:\mathbb{R}^{2}\longrightarrow \mathfrak{g}\otimes \mathfrak{g}\; \; ,\; \; (b_{+},b_{-})\longmapsto r_{12}^{(0)}(b_{+},b_{-})=-[\kappa _{2}^{b_{+},b_{-}},t_{12}]\; . &  & 
\end{eqnarray}

The basic properties of the Dirac bracket can be mapped to the following
properties: \begin{eqnarray}
r_{12}^{(0)}+r_{21}^{(0)} & = & 0\label{rmatrixcoupling0}\\{}
[r_{13}^{(0)},r_{12}^{(0)}]+[r_{23}^{(0)},r_{12}^{(0)}]+[r_{23}^{(0)},r_{13}^{(0)}] & = & -\frac{\partial r_{12}^{(0)}}{\partial m}(\xi _{0,T})_{3}+\frac{\partial r_{13}^{(0)}}{\partial m}(\xi _{0,T})_{2}-\frac{\partial r_{23}^{(0)}}{\partial m}(\xi _{0,T})_{1}\nonumber \\
 &  & -\frac{\partial r_{12}^{(0)}}{\partial s}(\xi _{0,L})_{3}+\frac{\partial r_{13}^{(0)}}{\partial s}(\xi _{0,L})_{2}-\frac{\partial r_{23}^{(0)}}{\partial s}(\xi _{0,L})_{1}\; \; .\nonumber \\
 &  & \label{CDYBE}
\end{eqnarray}
 As a result, $r^{(0)}$ is a solution of the dynamical classical
Yang-Baxter equation with zero coupling constant associated to the
algebra $\mathfrak{g}$.

\subsection{Operator $K_{\phi }^{X}$\label{sub:Operator}}

The aim of this appendix is to study the following operator: \begin{eqnarray}
K_{\phi }^{X}:C^{\infty }(\mathcal{B},\mathfrak{g}) & \longrightarrow  & C^{\infty }(\mathcal{B},\mathfrak{g})\label{operatorK}\\
\bar{u} & \longmapsto  & K_{\phi }^{X}\bar{u}=\partial _{\phi }(\bar{u})+[\breve{X},\bar{u}-\breve{u}]\; ,\nonumber 
\end{eqnarray}
where $\breve{X}$ is the mapping from $\mathcal{B}$ to $\mathfrak{g}$
defined by \ref{definitionX}. In fact, this operator can be rewritten
as follows: \begin{eqnarray*}
(K_{\phi }^{X}u)(x_{(n)}(t,\phi )) & = & M_{(n)}(t)\left(\mathbf{K}_{\phi }^{\frac{-1}{2\pi }\chi _{n}}\left(M_{(n)}^{-1}(t)u(x_{(n)}(t,\phi ))M_{(n)}(t)\right)\right)M_{(n)}^{-1}(t)\; 
\end{eqnarray*}

$\forall \; u\in C^{\infty }(\mathcal{B},\mathfrak{g})\; ,$ where
\begin{equation}
\mathbf{K}_{\phi }^{\chi }:C^{\infty }(S^{1},\mathfrak{g})\longrightarrow C^{\infty }(S^{1},\mathfrak{g}),\: f\mapsto \partial _{\phi }(f)+[\chi ,f-f^{av.}]\label{operatorKbold}\end{equation}
 for any $\chi =\alpha _{T}\xi _{0,T}+\alpha _{L}\xi _{0,L}\in \mathfrak{h}.$
Properties of (\ref{operatorK})will be linked back to that of (\ref{operatorKbold}).

It is immediate to see that (\ref{operatorKbold}) is not injective
and its kernel consists in the subspace of constant functions. This
operator will be invertible on the subspace $C_{s}^{\infty }(S^{1},\mathfrak{g})$
of functions having zero as mean value. Any function $u\in C_{s}^{\infty }(S^{1},\mathfrak{g})$
can be decomposed in terms of Fourier modes as follows: $u(\phi )=\sum _{k\neq 0,I\in I_{\mathfrak{g}}}u_{k}^{I}e^{ik\phi }\xi _{I}\; ,$
and the action of the operator (\ref{operatorKbold}) on $f$ reduces
to an action on each component of the Fourier decomposition: \begin{eqnarray}
\mathbf{K}_{\phi }^{\chi }u=\sum _{k\neq 0,I,J\in I_{\mathfrak{g}}}\left(\mathbf{K}_{k\, J}^{\chi \, I}u_{k}^{J}\right)e^{ik\phi }\xi _{I}\; , &  & 
\end{eqnarray}
 where the coefficients $\mathbf{K}_{k,J}^{\chi \, I}$ are given
by $\mathbf{K}_{k,S}^{\chi \, R}=ik\delta _{S}^{R}+<\alpha _{T}\xi _{0,T}+\alpha _{L}\xi _{0,L},[\xi _{S},\xi ^{R}]>,$
i.e. \begin{eqnarray}
\mathbf{K}_{k\, 0,T}^{\chi \, 0,T} & = & \mathbf{K}_{k\, 0,L}^{\chi \, 0,L}=ik\label{coeffKchi}\\
\mathbf{K}_{k\, b,L}^{\chi \, a,L} & = & \mathbf{K}_{k\, b,T}^{\chi \, a,T}=ik\delta _{b}^{a}+\alpha _{L}\varepsilon ^{0a}\, _{b}\nonumber \\
\mathbf{K}_{k\, b,L}^{\chi \, a,T} & = & -\mathbf{K}_{k\, b,T}^{\chi \, a,L}=\alpha _{T}\varepsilon ^{0a}\, _{b}.\nonumber 
\end{eqnarray}

Let us compute the inverse operator. We denote by $(\mathbf{K}^{\chi \, -1})_{k\, B}^{A}$
the coefficients of the inverse operator. 

In order to compute the inverse operator explicitely, it will be useful
to introduce the complex number $\underline{\alpha }=(\alpha _{L}+i\alpha _{T})$
and, the functions $c_{k}:\mathbb{C}\rightarrow \mathbb{C}\; \; \; \; \; \; z\mapsto \frac{1}{k+z}\; .$
A straightforward computation, shows that: \begin{eqnarray}
(\mathbf{K}^{\chi \, -1})_{k\, 0,T}^{0,T} & = & (\mathbf{K}^{\chi \, -1})_{k\, 0,L}^{0,L}=\frac{1}{ik}\label{coeffKchimoins1}\\
(\mathbf{K}^{\chi \, -1})_{k\, b,L}^{a,L} & = & (\mathbf{K}^{\chi \, -1})_{k\, b,T}^{a,T}=Re(c_{k}(-\underline{\alpha }))\frac{-i\delta _{b}^{a}+\varepsilon ^{0a}\, _{b}}{2}-Re(c_{k}(\underline{\alpha }))\frac{i\delta _{b}^{a}+\varepsilon ^{0a}\, _{b}}{2}\nonumber \\
(\mathbf{K}^{\chi \, -1})_{k\, b,L}^{a,T} & = & -(\mathbf{K}^{\chi \, -1})_{k\, b,T}^{a,L}=Im(c_{k}(-\underline{\alpha }))\frac{-i\delta _{b}^{a}+\varepsilon ^{0a}\, _{b}}{2}-Im(c_{k}(\underline{\alpha }))\frac{i\delta _{b}^{a}+\varepsilon ^{0a}\, _{b}}{2}.\nonumber 
\end{eqnarray}

\subsection{Dirac bracket for spin-networks\label{sub:Dirac-gravit}}

Let us compute the following equal-time Poisson bracket$\left\{ (U_{[yz]}^{\Pi _{[yz]}}M_{(k)})_{1},(U_{[y'z']}^{\Pi _{[y'z']}}M_{(k)})_{2}\right\} _{D}$
where $z=x_{(k)}(t,\phi ),z'=x_{(k)}(t,\phi ')$ (we will denote $\varphi =\phi -\phi '$).In
order to apply formula (\ref{definitiveDiracbracket}), we recall
that, for any $v\in \check{\mathbb{g}}$, we have 

\begin{eqnarray*}
\left\{ \tilde{\Omega }(v),U_{[yz]}^{\Pi _{[yz]}}\right\} _{2}(A) & =\frac{\vartheta }{2l_{c}} & \int _{y}^{z}ds\: U_{[ys]}^{\Pi _{[yz]}}(A)(D_{s}^{(A)}v)U_{[sz]}^{\Pi _{[yz]}}(A)\\
 & = & \frac{\vartheta }{2l_{c}}(-v(y)\: U_{[yz]}^{\Pi _{[yz]}}(A)+U_{[yz]}^{\Pi _{[yz]}}(A)\: v(z))\\
 & = & \frac{\vartheta }{2l_{c}}(U_{[yz]}^{\Pi _{[yz]}}(A)\: v(z))\\
\left\{ U_{[yz]}^{\Pi _{[yz]}}{}_{1},U_{[y'z']}^{\Pi _{[y'z']}}{}_{2}\right\} _{2} & = & 0\\
\left\{ U_{[yz]}^{\Pi _{[yz]}}{}_{1},M_{(k)}{}_{2}\right\} _{2} & = & \left\{ U_{[yz]}^{\Pi _{[yz]}}{}_{1},M_{(k)}{}_{2}\right\} _{2}=0\\
\left\{ \tilde{\Omega }(v),M_{(k)}\right\} _{2} & = & 0
\end{eqnarray*}

hence,\begin{eqnarray*}
\left\{ (U_{[yz]}^{\Pi _{[yz]}}M_{(k)})_{1},(U_{[y'z']}^{\Pi _{[y'z']}}M_{(k)})_{2}\right\} _{D}(A) & = & U_{[yz]}^{\Pi _{[yz]}}(A)_{1}U_{[y'z']}^{\Pi _{[y'z']}}(A)_{2}\left\{ M_{(k)1},M_{(k)2}\right\} _{2}+\\
+\frac{\vartheta }{2l_{c}}U_{[yz]}^{\Pi _{[yz]}}(A)_{1}U_{[y'z']}^{\Pi _{[y'z']}}(A)_{2} & \times  & \left(\int \! \! \! \! \! \int _{\check{\mathbb{g}}\times \check{\mathbb{g}}}[\mathcal{D}u][\mathcal{D}v]\: \bar{u}(z)_{1}\bar{v}(z')_{2}\tilde{K}{}^{-1}(u,v)\right)M_{(k)1},M_{(k)2}\\
 & = & (U_{[yz]}^{\Pi _{[yz]}}M_{(k)})_{1},(U_{[y'z']}^{\Pi _{[y'z']}}M_{(k)})_{2}r_{12}^{(0)}(\tilde{b}{}_{(+)}^{n},\tilde{b}{}_{(-)}^{n})+\\
 &  & +\frac{\vartheta }{2l_{c}}U_{[yz]}^{\Pi _{[yz]}}(A)_{1}U_{[y'z']}^{\Pi _{[y'z']}}(A)_{2}(\xi _{I})_{1}(w_{z}^{I}(z'))_{2}M_{(k)1},M_{(k)2}
\end{eqnarray*}

where we have introduced certain functions $(w_{z}^{I})_{I\in I_{\mathfrak{g}}}$
defined on the boundary, with value in $\mathfrak{g}$, smooth everywhere
except at $z$ (where they are allowed to have a step) and such that
for any $u\in \check{\mathbb{g}},$ $\bar{u}(z)=\tilde{K}(w_{z}^{I},u)\xi _{I}$
.

We can easily verify, that the solution of this equation is given
by \[
(\xi _{I})_{1}(w_{z}^{I}(z'))_{2}=-M_{(k)1}M_{(k)2}(\sum _{n\neq 0}(\mathbf{K}^{\frac{-1}{2\pi }\tilde{\chi }_{k}\, -1})_{n\, J}^{I}\frac{e^{in\varphi }}{2\pi }\xi _{I1}\xi _{2}^{J})M_{(k)1}^{-1}M_{(k)2}^{-1}.\]

Using the results of the previous appendix we have \begin{eqnarray*}
\mathfrak{R}_{12}^{(1)}(M_{(k)},P_{(k)};\varphi ) & \equiv  & \frac{-\vartheta }{2l_{c}}\sum _{n}(\mathbf{K}^{\frac{-1}{2\pi }\tilde{\chi }_{k}\, -1})_{n\, J}^{I}e^{in\varphi }\xi _{I1}\xi _{2}^{J}\\
 & = & \frac{-\vartheta }{4\pi l_{c}}\left(A(\varphi )(\xi _{0,T}\otimes \xi _{0,L}+\xi _{0,L}\otimes \xi _{0,T})+\right.\\
 &  & -Im(B(\varphi ))(\xi _{2,T}\otimes \xi _{1,T}-\xi _{1,T}\otimes \xi _{2,T}-\xi _{2,L}\otimes \xi _{1,L}+\xi _{1,L}\otimes \xi _{2,L})+\\
 &  & -Re(B(\varphi ))(\xi _{2,T}\otimes \xi _{1,L}+\xi _{2,L}\otimes \xi _{1,T}-\xi _{1,T}\otimes \xi _{2,L}-\xi _{1,L}\otimes \xi _{2,T})+\\
 &  & +Im(C(\varphi ))(\xi _{1,T}\otimes \xi _{1,T}-\xi _{1,L}\otimes \xi _{1,L}+\xi _{2,T}\otimes \xi _{2,T}-\xi _{2,L}\otimes \xi _{2,L})+\\
 &  & \left.+Re(C(\varphi ))(\xi _{1,T}\otimes \xi _{1,L}+\xi _{1,L}\otimes \xi _{1,T}+\xi _{2,T}\otimes \xi _{2,L}+\xi _{2,L}\otimes \xi _{2,T})\right)\\
\mathfrak{R}_{12}^{(0)}(M_{(k)},P_{(k)}) & \equiv  & \frac{-\vartheta }{2l_{c}}(\mathbf{K}^{\frac{-1}{2\pi }\tilde{\chi }_{k}\, -1})_{0\, J}^{I}\xi _{I1}\xi _{2}^{J}\\
 & = & r_{12}^{(0)}(\tilde{b}{}_{(+)}^{n},\tilde{b}{}_{(-)}^{n})\\
 & = & \frac{\vartheta }{4\pi l_{c}}\left(Im(\frac{1}{\tilde{\underline{b}}})(\xi _{2,T}\otimes \xi _{1,T}-\xi _{1,T}\otimes \xi _{2,T}-\xi _{2,L}\otimes \xi _{1,L}+\xi _{1,L}\otimes \xi _{2,L})+\right.\\
 &  & \left.+Re(\frac{1}{\underline{\tilde{b}}})(\xi _{2,T}\otimes \xi _{1,L}+\xi _{2,L}\otimes \xi _{1,T}-\xi _{1,T}\otimes \xi _{2,L}-\xi _{1,L}\otimes \xi _{2,T})\right)
\end{eqnarray*}

where we have defined \begin{eqnarray*}
\tilde{\underline{b}} & \equiv  & \frac{\vartheta }{4\pi l_{c}}(\tilde{b}{}_{+}^{(n)}+i\tilde{b}{}_{-}^{(n)})\\
A(\varphi ) & \equiv  & (\varphi -\pi )_{[0,2\pi ]}\\
B(\varphi ) & \equiv  & \frac{\pi cos(\tilde{\underline{b}}(\pi -\varphi )_{[0,2\pi ]})}{sin(\pi \tilde{\underline{b}})}\\
C(\varphi ) & \equiv  & \frac{\pi \: sin(\tilde{\underline{b}}(\pi -\varphi )_{[0,2\pi ]})}{sin(\pi \tilde{\underline{b}})}
\end{eqnarray*}

and the symbol $(f(\varphi ))_{[0,2\pi ]}$ denotes the continuation
by $2\pi -$periodicity of the function $f$ defined in $[0,2\pi ].$

The r-matrix $r_{12}^{(1)}(\frac{\vartheta }{4\pi l_{c}}\tilde{b}_{+},\frac{\vartheta }{4\pi l_{c}}\tilde{b}_{-})\equiv \mathfrak{R}_{12}^{(1)}(M_{(k)},P_{(k)};0^{+})$
is solution of classical dynamical Yang-Baxter equation and is such
that $-r_{21}^{(1)}(\frac{\vartheta }{4\pi l_{c}}\tilde{b}_{+},\frac{\vartheta }{4\pi l_{c}}\tilde{b}_{-})=\mathfrak{R}_{12}^{(1)}(M_{(k)},P_{(k)};2\pi ^{-})=\frac{\vartheta }{2l_{c}}t_{12}+r_{12}^{(1)}(\frac{\vartheta }{4\pi l_{c}}\tilde{b}_{+},\frac{\vartheta }{4\pi l_{c}}\tilde{b}_{-}).$
As a result its coupling constant is $\frac{-\vartheta }{2l_{c}}.$
\end{document}